\documentclass[%
    11pt,
    letterpaper,
    reprint,
    notitlepage,
    superscriptaddress,
    aps,
    usenames,dvipsnames
]{revtex4-2}

\usepackage[pdftex]{graphicx}
\usepackage{svg}
\usepackage{amsmath,amssymb,amsfonts,amsthm} 
\usepackage{empheq}
\usepackage{physics}
\usepackage{placeins}
\usepackage{adjustbox}
\usepackage{lipsum}
\usepackage{hyperref}
\usepackage{newunicodechar}
\newunicodechar{π}{$\pi$}
\hypersetup{colorlinks=true,citecolor=blue,linkcolor=blue,urlcolor=blue}

\newcommand{\suppoptbayesian}{Supplementary Note 1}
\newcommand{\suppclocks}{\text{Supplementary Note 2}}
\newcommand{\suppallandev}{\text{Supplementary Note 3}}
\newcommand{\supplukin}{\text{Supplementary Note 4}}
\newcommand{\suppwiseman}{\text{Supplementary Note 5}}
\newcommand{\suppqubitnumbers}{\text{Supplementary Note 6}}
\newcommand{\suppsinestate}{\text{Supplementary Note 7}}
\newcommand{\suppmthreeproof}{\text{Supplementary Note 8}}
\newcommand{\suppadaptivemeas}{\text{Supplementary Note 9}}
\newcommand{\supptuviaproof}{\text{Supplementary Note 10}}
\newcommand{\suppunwinding}{\text{Supplementary Note 11}}
\newcommand{\suppuniformprior}{\text{Supplementary Note 12}}
\newcommand{\suppamplitudedamping}{\text{Supplementary Note 13}}

\usepackage{dsfont}
\usepackage[normalem]{ulem}
\usepackage{lineno}
\newcommand{\Caltech}{California Institute of Technology, Pasadena, CA 91125, USA}
\newcommand{\huji}{Racah Institute of Physics, The Hebrew University of Jerusalem, Jerusalem 91904, Givat Ram, Israel}
\newcommand{\telaviv}{School of Physics and Astronomy, Tel Aviv University, 69978, Israel}

\begin{document}

\title{Heisenberg-limited Bayesian phase estimation with low-depth digital quantum circuits }

\author{Su Direkci}
\email{sdirekci@caltech.edu}
\affiliation{\Caltech}
\author{Ran Finkelstein}
\affiliation{\Caltech}
\affiliation{\telaviv}
\author{Manuel Endres}
\affiliation{\Caltech}
\author{Tuvia Gefen}
\affiliation{\Caltech}
\affiliation{\huji}

\date{\today}

\begin{abstract}
    Optimal phase estimation protocols 
    require complex state preparation and readout schemes, generally unavailable or unscalable in many quantum platforms. We develop a scheme that achieves near-optimal precision up to a constant overhead for Bayesian phase estimation, using simple digital quantum circuits with depths scaling logarithmically with the number of qubits. This is done by approximating the optimal initial states with products of Greenberger-Horne-Zeilinger states for Gaussian prior phase distributions with arbitrary widths.
    We study various protocols that employ this class of states with different levels of measurement and post-processing complexities, and obtain improvement compared to previously proposed schemes.
    We then use our scheme to address phase slip errors and laser noise, which impose a major limitation in Bayesian phase estimation and atomic clocks.  
     Based on our scheme, we develop an efficient protocol to suppress this noise
     that 
     outperforms existing methods.
\end{abstract}

\maketitle

\section*{Introduction}

Quantum metrology studies fundamental precision limits in physical measurements imposed by quantum physics. 
Recent advances in this field have led to novel protocols and improved precision for a variety of sensing devices:  magnetometers \cite{brask_improved_2015, baumgratz_quantum_2016}, atomic clocks \cite{borregaard_near-heisenberg-limited_2013, rosenband2013exponential,kaubruegger_quantum_2021,marciniak2022optimal, Robinson2024, Eckner2023, finkelstein2024universal,cao2024multiqubit}, nano-NMR \cite{schmitt_submillihertz_2017, boss_quantum_2017, aharon_nv_2019, schmitt_optimal_2021,cohen_confined_2020}, and atom interferometers \cite{salvi_squeezing_2018,ockeloen_quantum_2013} to name a few. Many of these sensing applications can effectively be described as a single-shot phase estimation with an ensemble of $N$ qubits, where an unknown phase $\phi,$
due to e.g. an electromagnetic field is imprinted on the qubits and is estimated after performing a measurement on the system.

A conventional phase estimation protocol is Ramsey interferometry, where the qubits are set to an initial state that is a superposition of the eigenstates of the computational basis, and are subjected to a projective measurement in the same basis after sensing $\phi$. 
The precision limits of this scheme are well established: the estimated phase standard deviation (STD) scales as $1/\sqrt{N}$ when uncorrelated qubits are used as the initial state, also known as the standard quantum limit (SQL) \cite{pezze_quantum_2018}. Employing a suitable entangled state as the initial state, it is possible to achieve an improved estimation uncertainty that scales as $1/N$, referred to as the Heisenberg limit (HL) or Heisenberg scaling (HS) in the literature \cite{giovannetti_advances_2011}.
The initial state that achieves the minimum possible estimation uncertainty is referred to as the \textit{optimal initial state}.

The two approaches to the phase estimation problem in the literature are (i) the frequentist approach, where the unknown phase $\phi$ is assumed to be a \textit{deterministic} parameter and it can be measured many times.
(ii) The Bayesian approach,  where $\phi$ is assumed to be a \textit{stochastic} variable and a single-shot phase estimation is performed \cite{kolodynski2015thesis}. In the Bayesian approach, the prior information about $\phi$ is encoded in a prior distribution $\mathcal{P}_{\delta\phi}(\phi)$, with a standard deviation of $\delta\phi$, also referred to as the prior width. In this work, we 
consider the Bayesian approach, which is relevant to various applications, e.g. atomic clocks \cite{kaubruegger_quantum_2021}.

For the frequentist approach, it is well known that the optimal initial state is the Greenberger-Horne-Zeilinger (GHZ) state. Such states were analyzed in the literature extensively, in the context of Ramsey spectroscopy \cite{bollinger_optimal_1996, huelga_improvement_1997, meyer_experimental_2001}, quantum interferometry \cite{lee_quantum_2002, nagata_beating_2007}, quantum lithography \cite{dangelo_two-photon_2001} and imaging \cite{nasr_demonstration_2003, israel_supersensitive_2014}.
Conversely, for the Bayesian approach, the optimal initial state depends on the shape of the prior distribution: specifically, the prior width $\delta\phi$. For a small prior width (\mbox{$\delta\phi < 1/N$}, where $N$ is the number of qubits), the GHZ state, combined with local readout, keeps on performing optimally \cite{macieszczak_bayesian_2014}. However, for prior widths larger than the dynamical range of the GHZ states (\mbox{$\delta\phi > 1/N$}), they become sub-optimal. 
In this regime of wide prior phase distribution, the optimal initial state is the so-called \textit{sine state}, a spin-squeezed state \cite{Kitagawa_spin_squeezed} that can be generated by the two-axis counter-twisting Hamiltonian \cite{tact_states}.
Furthermore, the optimal measurement involves a Quantum Fourier Transform (QFT) followed by a projective measurement on the computational basis \cite{buzek_optimal_quantum_clocks,berry_optimal_2000, summy_phase_1990, macieszczak_bayesian_2014}. Lastly, the ultimate estimation precision in this limit is given by the $\pi$-corrected HL, $\Delta\phi \approx \pi/N$ \cite{jarzyna_true_2015, gorecki__2020}.

\begin{figure*}
\raggedright
\includegraphics[width=1.\linewidth]{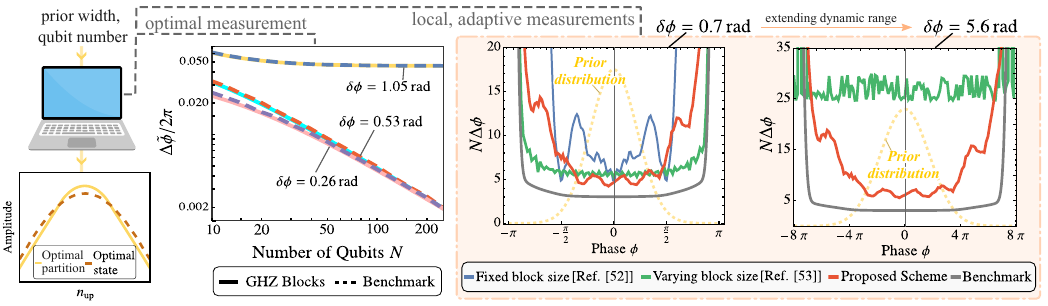}
\caption{\textbf{Summary of contributions.} We propose a scheme that, given the prior width $\delta\phi$ and the total number of qubits $N$, computes the optimal partition over blocks of GHZ states, as well as optimal local, adaptive measurements. Partitions of blocks of GHZ states can be used to approximate the optimal initial states for phase estimation. If optimal measurements are available, blocks of GHZ states approximately reach the benchmark sensitivity for all prior widths. If local, adaptive measurements are used instead, we obtain Heisenberg scaling up to a constant overhead. Furthermore,  optimal partitions can be rescaled to include slow atoms (atoms that accumulate phases of $\phi/2, \phi/4, \dots$), which enable an extended dynamic range beyond $\delta\phi \gg \pi$. We show that the proposed scheme surpasses other phase unwinding protocols in the literature that extend the dynamic range. }
\label{fig:summary_fig}
\end{figure*}

There is an ongoing experimental effort to achieve below SQL precision levels \cite{Hosten2016} and to reach the optimal limit in the wide prior regime \cite{kaubruegger_quantum_2021}.
A recent experiment achieved nearly optimal precision levels with trapped ions by optimizing over global squeezing operations \cite{marciniak2022optimal}.
However, such operations are not available (or scalable) in many quantum information platforms, e.g. atoms in tweezer arrays \cite{finkelstein2024universal, Kaufman2021}, superconducting qubits \cite{Bao_superconducting, Place2021Superconducting, Verjauw2022_superconducting}, and more generally quantum computing devices that rely on circuits of few-qubit gates \cite{nielsen2001quantum}.
That is because generating a 
sine state 
and implementing the QFT measurement both require polynomial depth circuits \cite{barenco_approximate_1996, cleve2000fast}.

This raises the question: is it possible to saturate the fundamental precision limits, for any prior width, with constant to logarithmic depth digital quantum circuits?
One particularly practical motivation behind this question is tweezer optical clocks \cite{Norcia2019, Majdarov2019}.
This technology emerges as a promising platform for quantum enhanced clocks, thanks to high fidelity quantum control and readout \cite{finkelstein2024universal, cao2024multiqubit}. However, since the quantum control in these clocks is limited to digital circuits, it is unknown whether the optimal precision can be obtained with 
state-of-the-art capabilities.
Another motivation for this question is quantum phase estimation algorithms, which are performed using digital quantum devices. It is highly desirable to understand the fundamental precision limits of 
these algorithms when implemented with low-depth circuits \cite{ni2023low, li2023adaptive, smith2024adaptive}.

One set of initial states that can be prepared efficiently by digital 
circuits is an ensemble of GHZ states with varying numbers of qubits. Such a state contains $m_i$ copies of GHZ states with $2^{k_i}$ qubits, where $k_i, m_i \in \mathbb{N}$. We also refer to such states as \textit{blocks of GHZ states}, and $m_i$ as the block size. Several protocols using such initial states have been proposed in the literature \cite{berry_how_2009, kaftal2014usefulness, kessler_heisenberg-limited_2014, higgins_demonstrating_2009}. However, to our knowledge, the performance of these schemes in the Bayesian setting has not been investigated (i.e. different prior distributions or prior widths have not been considered).

Here, we analyze how such initial states perform and compare various measurement schemes with an increasing level of control.
Specifically, we consider a scheme with a \emph{fixed block size} adopted from Ref. \cite{kessler_heisenberg-limited_2014} that employs non-adaptive, qubit-resolved (local) measurements, as well as a scheme with a \emph{varying block size}~\cite{higgins_demonstrating_2009} which further employs single-qubit rotations before measurement.
We show that the scheme with a varying block size surpasses the scheme with a fixed block size and achieves HS up to a constant overhead, in the large prior width within the dynamic range regime ($\delta\phi = 0.7$ rad). 

However, both schemes have several drawbacks. First, they offer exact solutions only for specific values of the number of qubits. For example, the scheme with a varying block size offers a solution only for qubit numbers $N = 2, 9, 26, 63, 140, \dots$ Extending these schemes to any number of qubits is highly desirable for near-term quantum devices that operate with a small number of qubits. Second, they consider a uniform prior phase distribution in the interval of $[- \pi, \, \pi]$, which is not realistic for devices such as atomic clocks. This motivates extension to an arbitrary prior width, and Gaussian prior distributions, which are more relevant for atomic clocks. 

We thus propose and analyze the following protocol: 
we numerically optimize over all possible partitions of a total number of $N$ qubits into blocks of GHZ states with varying numbers of qubits, assuming a Gaussian prior distribution with an arbitrary prior width $\delta\phi$. As for the measurement strategy, we utilize local, adaptive measurements, instead of local, non-adaptive measurements used in Refs. \cite{kessler_heisenberg-limited_2014,higgins_demonstrating_2009}. These measurements were introduced in \cite{berry_how_2009}, but the performance for Bayesian phase estimation was not studied. Given the initial state, we optimize over the local, adaptive measurements and show that this protocol achieves HS for any given prior width, and surpasses the previously suggested protocols.

To answer our initial question, we show that it is possible to approximately saturate fundamental precision limits for any prior width, with constant to logarithmic depth digital quantum circuits. We observe that the initial states used in the proposed scheme perform almost optimally and reach the fundamental limit of sensitivity for Bayesian phase estimation for all prior widths, if arbitrary level of control in measurement schemes is allowed.

Lastly, we discuss the application of these schemes in the context of atomic clocks, and propose a method to reach the ultimate limit of sensitivity. We do so by increasing the dynamic range of the proposed scheme, i.e. the interval for which the phase $\phi$ can be unambiguously estimated, by introducing \textit{slow} atoms that accumulate fractional phases. We present a method to allocate the atoms into slow atoms and GHZ states efficiently, in order to extend the dynamic range with a negligible amount of additional atoms. We show that our method outperforms existing schemes that utilize slow atoms by comparing the number of atoms needed to obtain the same level of sensitivity.

\section*{Results}

\subsection*{Formulation}
\label{sec:formulation}

We consider a Ramsey interferometry experiment where $N$ 
qubits are initialized in the state $\ket{\psi_i}$, and undergo a unitary $\phi$-encoding transformation $U(\phi)$ such that
\begin{align}
    U(\phi) &= e^{-i\phi J_z} \nonumber \\
    J_{z} &= \frac{1}{2}\underset{i}{\sum}\sigma_z^{i}, \;\; i = 1, 2, \dots N,
\end{align}
where $\sigma_z^i$ is the Pauli $Z$ operator for the $i^\text{th}$ qubit, we set $\hbar = 1$, and $\phi \in \mathbb{R}$ is unknown. Let us denote the state after the encoding as $|\psi_{f} \rangle$, with $|\psi_{f} \rangle =U(\phi) |\psi_{i}\rangle.$  
Information about $\phi$ is obtained through performing positive operator valued measures (POVMs) on $|\psi_{f} \rangle,$
 i.e. a set of positive Hermitian operators $\{\Pi_{x}\}_{x}$ such that $\sum_x \Pi_{x} = \mathbf{1}$, where $\mathbf{1}$ is the identity operator.
The probabilities of the measurement outcomes are given by the distribution $\{p(x|\phi)\}_{x}$, with $p(x|\phi) = \bra{\psi_f}\Pi_x\ket{\psi_f}$. Given a measurement result, the phase is estimated by an estimator $\phi_{est}(x)$, and the accuracy of the phase estimation for any $\phi$ is defined by the mean-squared error (MSE),
\begin{align}
\label{eqn:mse}
    (\Delta \phi)^2 = \sum_{x} p(x|\phi)(\phi-\phi_{est}(x))^2 \,.
\end{align}
Given the MSE as a function of $\phi$, we seek to minimize a cost function that quantifies the overall performance of the phase estimation, for all $\phi$. For this purpose, we choose to utilize the Bayesian mean squared error (BMSE). Given a prior distribution of $\phi,$ $\mathcal{P}_{\delta\phi}(\phi),$ the BMSE is defined as
\begin{align}
\label{eqn:bmse}
(\Delta \widetilde{\phi})^2 = \int (\Delta \phi)^2 \mathcal{P}_{\delta\phi}(\phi) d\phi \,.
\end{align}
We refer to the square-root of the BMSE, $\Delta \widetilde{\phi}$, as root Bayesian mean squared error (RBMSE), or the posterior width. We assume
a Gaussian prior distribution with a standard deviation of $\delta\phi$, defined as
\begin{align}
\label{eqn:gaussian_prior}
    \mathcal{P}_{\delta\phi}(\phi) = \frac{1}{\sqrt{2\pi (\delta\phi)^2}} \text{exp} \left[ -\frac{\phi^2}{2 (\delta \phi)^2} \right] \,.
\end{align}
$\delta\phi$ is also referred to as the prior width. Then, given a prior width, the problem of minimizing the BMSE reduces to optimizing over $\ket{\psi_i}$, $\{\Pi_x\}$, and $\{\phi_{est}(x)\}$.
An efficient numerical algorithm for 
this task was given in \cite{macieszczak_bayesian_2014} (see \suppoptbayesian\ for details).
This algorithm provides the minimal BMSE along with the optimal initial state, the POVM, and the optimal estimators that saturate it. 
We refer to this minimal BMSE as the optimal quantum interferometer (OQI).
The OQI constitutes a benchmark for the maximum attainable sensitivity in Bayesian phase estimation.

\subsection*{Blocks of GHZ States for Phase Estimation}


It is desirable to use GHZ states as building blocks to construct initial states for phase estimation, as they are relatively easy to prepare. 
$N$-qubit GHZ states can be prepared using various short depth circuits: (i) a circuit depth of $O(\log N)$ assuming full-connectivity \cite{Mooney_2021}, (ii) constant depth if measurement and feedback strategies \cite{tantivasadakarn_long-range_2022} or multi-qubit gates \cite{cao2024multiqubit} are used, (iii) linear depth if only nearest-neighbor interactions are allowed \cite{finkelstein2024universal}.
GHZ states are optimal initial states for Bayesian phase estimation given a vanishing prior phase width and for the frequentist case. To show this, let us first define the $J_x,J_y$ operators: $J_x=1/2\sum_{i=1}^{N} \sigma_x^i$, $J_y=1/2\sum_{i=1}^{N} \sigma_y^i$, where $\sigma_x^i,\,\sigma_y^i$ are the Pauli $X,\,Y$ Pauli operators on the $i^\text{th}$ qubit, respectively. Consider a $J_y$ measurement on the $N$-qubit GHZ state. The first and second moments of the parity operator, defined as $\Pi=\left(-1\right)^{J_y+N/2}$, are given by $\langle\Pi\rangle=\sin\left(N\phi\right)$, $\langle\Pi^2\rangle=\cos\left(N\phi\right)^2$. Given $M$ repetitions of the measurement, 
the MSE is
\begin{align}
(\Delta\phi)^2 =\frac{(\Delta\Pi)^2}{M(\partial_{\phi}\langle\Pi\rangle)^{2}}\approx\frac{1}{MN^{2}} \quad\text{for}\quad\phi \ll 1\,,    
\end{align}
where $(\Delta\Pi)^2 = \langle\Pi^2\rangle - \langle\Pi\rangle^2$. Thus, GHZ states attain the HL in the frequentist approach. 
However, this limit is unachievable in Bayesian quantum metrology, where only a single measurement is allowed ($M = 1$), and $\phi$ has a prior distribution with a non-zero width. Bayesian phase estimation resembles the frequentist phase estimation in the limit of small prior width, where 
$\delta \phi \ll \pi/N$, for which bimodal states close to GHZ states are the optimal initial states \cite{macieszczak_bayesian_2014}.

For $\delta \phi > \pi/N$, GHZ states cannot attain the HL due to their limited dynamic range. Measurement outcomes repeat themselves after a period of $2\pi/N$, causing phase wrap errors to appear outside of this period, and leading to a poor performance. In order to have both a high precision in estimation, and a large dynamic range, we look for ways to approximate the optimal initial states using a combination of GHZ states with varying numbers of qubits, i.e. blocks of GHZ states. Such a state reads
\begin{align}
\label{eqn:general_ghz}
\ket{\psi} = \frac{1}{2^{\sum_i m_i/2}} \prod_i \left(\ket{0}^{\otimes 2^{k_i}} + \ket{1}^{\otimes 2^{k_i}} \right)^{\otimes m_i},
\end{align}
where $\ket{0},\ket{1}$ are the eigenstates of $\sigma_z$ with a negative and positive eigenvalue, respectively, $m_i$ is the size of the block that contains $2^{k_i}$-qubit GHZ states, $k_i, m_i \in \mathbb{N}$, and the total number of qubits is given by \mbox{$N = \sum_i m_i 2^{k_i}$}. 
We want to find optimal partitions, i.e. $\{k_i\}$,$\{m_i\},$ for every $N$ and optimal readout schemes and estimators.
In Refs. \cite{berry_how_2009, kaftal2014usefulness}, it was shown that blocks of GHZ states can attain the HL up to a constant overhead given a flat prior in $[-\pi, \pi]$. This was achieved by setting $m_i \geq 3 \, \forall \, i$, and performing optimal measurements. Before outlining our protocol, let us first review existing schemes that utilized blocks of GHZ states.

\subsection*{Review of existing schemes}
\label{sec:review_lukin_wiseman}

Protocols using blocks of GHZ states with different partitions (i.e. different sets of $\{k_i\}$,$\{m_i\}$ in Eq. (\ref{eqn:general_ghz})) have been studied in the literature \cite{berry_how_2009, higgins_demonstrating_2009,kessler_heisenberg-limited_2014}.
We review and further analyze the performance of these schemes, focusing on the scheme in Ref. \cite{kessler_heisenberg-limited_2014} to which we refer as the scheme with a fixed block size ($m_i = M$), and the scheme in Ref. \cite{higgins_demonstrating_2009}
to which we refer as the scheme with a varying block size.

In the scheme with a fixed block size, all of the GHZ states (with $\left\{ 2^{k_{i}}\right\} _{k_{i}=0}^{k_{\text{max}}}$ qubits) are repeated by a constant $m_i = M$. This constant is found from the implicit equation $M = \frac{16}{\pi^2}\,\text{log}(N)$, where $N$ is the total number of qubits.
For example, for a state with 1, 2, and 4-qubit GHZ states, $M = 6$.
For each block of $2^{k}$-atoms GHZ states,
$M/2$ of the GHZ states are measured in the local Pauli $\sigma_{x}$ and the other $M/2$ in the $\sigma_{y}$ basis,
i.e. a dual quadrature readout is performed to increase the dynamic range. Ref. \cite{kessler_heisenberg-limited_2014} assumes a uniform phase prior, which leads to an RBMSE
of $\Delta\tilde\phi = \frac{8}{\pi} \sqrt{\text{log}(N)}/N$, i.e. HS up to a logarithmic correction.
This scheme is analyzed in 
\supplukin, where we numerically calculate the RBMSE with two different estimation strategies: a bit-by-bit estimation strategy that was developed in \cite{kessler_heisenberg-limited_2014} and the optimal Bayesian estimation.
In the bit-by-bit estimation, we write $\phi$ (mod $2 \pi$) as $\phi = 2\pi(0.Z_1 Z_{2} Z_{3} \dots),$
where $Z_{j} \in \{0, 1\}$ are the binary digits.
By combining the results of the parity measurements of the GHZ states we estimate the binary digits $Z_j$ of the phase $\phi$. We find that while the Bayesian estimation achieves a sub-SQL BMSE, the bit-by-bit estimation does not achieve the expected precision (see \supplukin\ and Ref. \cite{kessler_heisenberg-limited_2014} for more details).

\begin{figure*}
    \raggedright\includegraphics[width=2.\columnwidth]{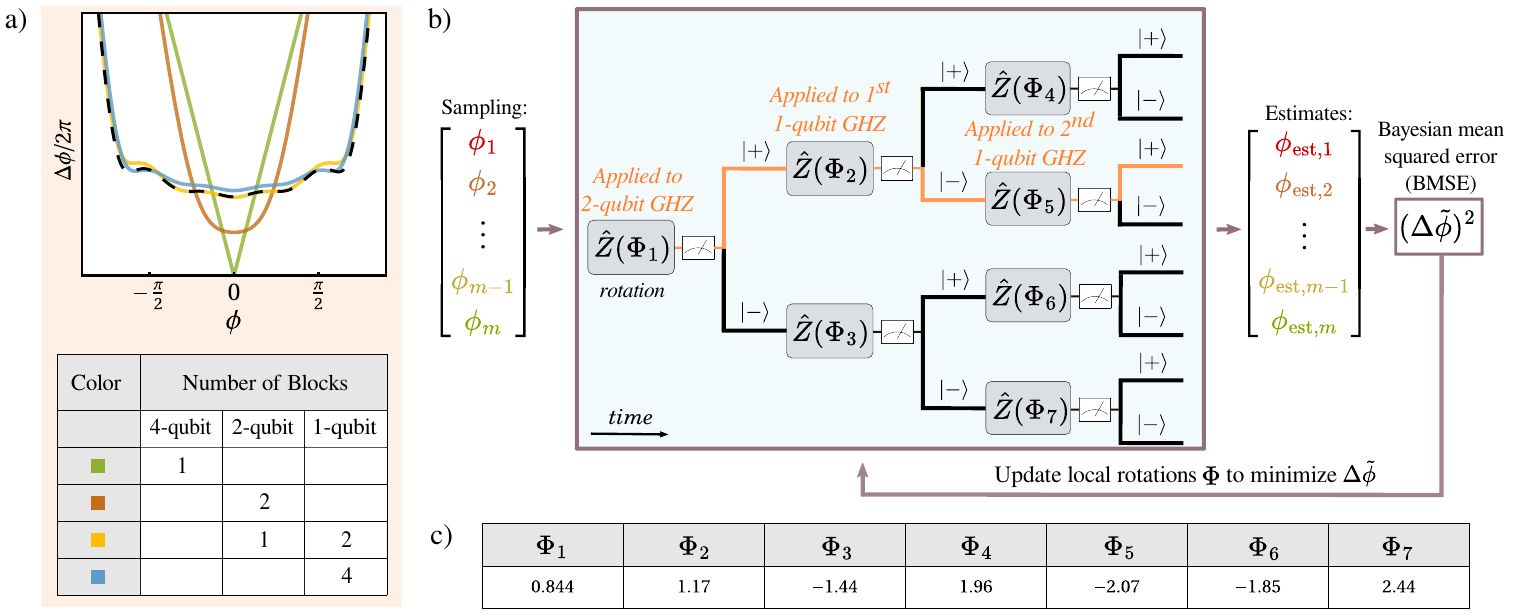}
    \caption{ {\bf{Illustration of the proposed scheme.}} Summary of the introduced protocol for $N=4$ qubits, and a prior width of $\delta\phi = \pi/3 \approx 1.05$ rad. (a) The possible partitions of the blocks of GHZ states for $N=4$ qubits are given in the Table, and the MSE as a function of $\phi$ is plotted. For this prior width, the best-performing partition is plotted in yellow, highlighted in the Table.  (b) Optimization over local, adaptive measurements. The measurement consists of successive $\hat Z(\Phi_i)$ operations and projective measurements. The $\hat Z(\Phi_i)$ operation corresponds to a rotation around the $z$-axis by $\Phi_i$ in the Bloch sphere picture, such that $\phi-\Phi_i$ is measured. Based on the measurement outcome, denoted with $\ket{+}$ and $\ket{-}$, the next rotation angle is determined. During the optimization, $m$ phases are sampled uniformly from the interval $[-6 \delta\phi,\, 6\delta\phi]$, which are then used to compute the Bayesian mean squared error (BMSE) analytically. Gradient descent is used to tune the single-qubit rotations $\Phi_i$ in the adaptive measurement. This process is iterated until the BMSE converges to a minimum. The path highlighted with orange corresponds to the measurement outcome $\{+,-,+\}$, for example. (c) Numerical values (in units of radians) of the single-qubit rotations $\Phi_i$ at the end of the optimization. 
    } 
\label{fig:intro_fig}
\end{figure*}

In the scheme with a varying block size, the state consists of GHZ states with sizes of $2^{k_i}$, where $k_i = 0, 1, \dots, k_\text{max}$, and \mbox{$m_i = m_{k_{\text{max}}} + \mu\,(k_\text{max}-k_i)$}, $m_{k_\text{max}},\,\mu \in \mathbb{N}^{+}$. Then, the largest GHZ state has $2^{k_\text{max}}$ qubits, and is repeated $m_{ k_{\text{max}} }$ times. Similarly,  $2^{k_\text{max}-1}$-qubit GHZ states are repeated $m_{ k_{ \text{max} } } + \mu$ times, etc. 
Before a measurement, the $m_i$ copies of GHZ states are subjected to a set of single-qubit rotations $\theta_j$, with $\theta_j = \pi \frac{j}{m_i}$, $j = 0, 1, \dots, m_i - 1$. Finally, they undergo parity measurements. 
Ref. \cite{higgins_demonstrating_2009} observed numerically that the maximum sensitivity is obtained for $m_{ k_{ \text{max} } } = 2$, and $\mu = 3$. For example, to prepare an initial state with 2-qubit and 1-qubit GHZ states, this protocol employs two copies of 2-qubit GHZ states, and five copies of 1-qubit GHZ states. Furthermore, the 2-qubit (1-qubit) GHZ states are subjected to single-qubit rotations of $0, \pi/2$ ($0, \pi/5, \dots, 4\pi/5$). 
Ref. \cite{higgins_demonstrating_2009} showed that this scheme achieves HL up to a constant overhead  of $2.03$ \footnote{Note that the cost function quantifying the performance of phase estimation in Ref. \cite{higgins_demonstrating_2009} is the Holevo variance, defined as $(\Delta \tilde \phi)^2 = \int_0^{2\pi} 4 \sin{\left(\frac{\phi-\bar \phi}{2} \right)}\,d\phi $, where $\bar \phi$ is the estimator of $\phi$.}. 
We implemented this scheme numerically with Gaussian prior distributions and observed that it achieves the sensitivity of the OQI up to a constant overhead that depends on the prior width. For example, for a prior width of $\delta\phi = 0.7$ rad, the overhead is calculated to be 1.66 (see \suppwiseman). 

The schemes mentioned above raise several issues: First, they are not immediately applicable to atomic clocks since they assume a uniform prior distribution on the phase. To our knowledge, they have not been analyzed for Gaussian prior distributions, which are the relevant phase distributions for the regime where current atomic clocks operate in.

Second, they correspond to very specific values of $N,$
e.g. for the scheme with a varying block size, the considered states have a total number of $N = 2, 9, 26$ qubits for $k_{\text{max}} = 0,1, 2.$
For intermediate numbers of qubits, it is necessary to modify these states by removing or adding GHZ states, which leads to sub-optimal states. Furthermore, the density of the qubit numbers suitable to these schemes decreases with increasing $N$, further complicating the use of them for large $N$ (see \suppqubitnumbers\ for a plot of the suitable qubit numbers for these schemes). We therefore want to devise a scheme that will achieve optimal precision for every $N$. 

These issues motivate us to find better schemes for Bayesian phase estimation that are applicable to all qubit numbers, $N$, and to Gaussian prior distributions. To this end, we use an adaptive protocol, inspired by Ref. \cite{berry_how_2009}.
We extend this protocol to Bayesian phase estimation and generalize it to all phase prior widths and all qubit numbers, $N.$ 

\subsection*{Proposed Adaptive Scheme}
\label{sec:scheme}

We now turn to outline our protocol. We design the protocol assuming a relatively small phase slip probability, i.e. $\delta\phi < \pi$, where $\delta\phi$ is the prior width. In the context of atomic clocks, this regime corresponds to a Ramsey time smaller than the local oscillator (LO) coherence time. In other words, $T \gamma_{\text{LO}} < 1$, where $T$ is the Ramsey time and $\gamma_{\text{LO}}$ is the linewidth of the laser (see \suppclocks). We extend the proposed scheme in the regime where $T \gamma_{\text{LO}} > 1$ through the introduction of phase unwinding.

Our goal is to get as close as possible to the performance of the OQI with 
low-depth circuits. The initial states and the POVMs of the OQI can be efficiently calculated, but they are difficult to implement: for a large prior width within the dynamic range ($\delta\phi \approx$ 1 rad), the optimal state resembles a sine state \cite{berry_optimal_2000}, and the optimal measurements have high 
complexity \cite{macieszczak_bayesian_2014} (for a derivation of the optimality of the sine state, see \suppsinestate\ and Ref. \cite{friis_flexible_2017}). Furthermore, optimizing over all possible digital circuits is quite formidable. Hence, we pursue a two-step approach. First, for a given atom number and prior width, we ``mimic" the optimal initial state using simpler states, i.e. blocks of GHZ states. Then, given this approximated state, we optimize over the local measurements to minimize the BMSE.

A summary and illustration of the proposed adaptive scheme is given in Fig. \ref{fig:intro_fig} for $N=4$ qubits,
and a prior width $\delta\phi = \pi/3 \approx 1.05$ rad (such that phase slips are three-sigma events).
We list all the possible partitions of $N=4$ qubits into blocks of GHZ states, and plot the MSE as a function of $\phi$ of all such partitions as a function of $\phi$ in Fig. \ref{fig:intro_fig}a.
We remark that their MSE depends on the prior width through the measurement optimization that is based on it.
The MSE of the OQI as a function of $\phi$ is given by the black, dashed line in Fig. \ref{fig:intro_fig}a. 
The smallest BMSE with our scheme is obtained by the partition whose MSE is closest to the OQI, which corresponds here to the yellow line.

\begin{figure*}
\center
\includegraphics[width=2\columnwidth]{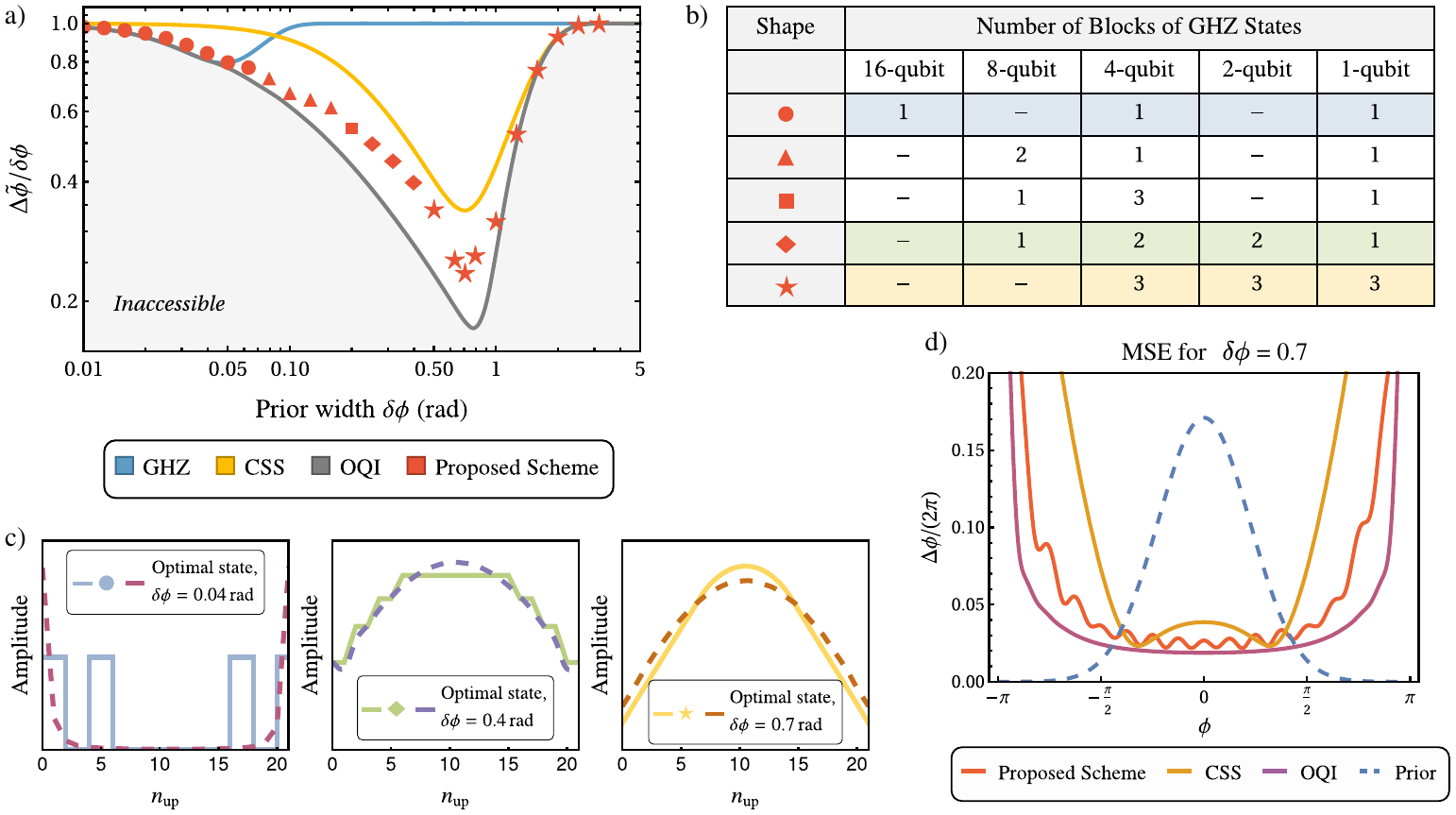}
\caption{\textbf{Performance of the proposed scheme for $N = 21$ qubits.} (a) Root Bayesian mean squared error (RBMSE) $\Delta\tilde{\phi}$ as a function of the prior width, $\delta\phi$, for $N = 21$. The performance of the phase estimation with a 21-qubit GHZ state, 21-qubit CSS, the OQI, and the proposed scheme is plotted in blue, yellow, gray, and red, respectively. The different shapes of the red markers correspond to different ways of partitioning the 21 qubits into blocks of GHZ states, listed in the table in b. (b) Optimal partitions of 21 qubits into blocks of GHZ states for different prior widths. For a small prior width (first row), we see that the optimal strategy is to use the GHZ state with the maximum number of atoms possible, which is the 16-atom GHZ state for $N = 21$. The remaining atoms are distributed to smaller GHZ states. In contrast, for a large prior width within the dynamic range (last row), it is more advantageous to use GHZ states with smaller numbers of atoms and distribute them equally, i.e. use 3 blocks of each. (c) The amplitudes of states with $n_{\text{up}}$ excitations, $0 \leq n_{\text{up}} \leq 21$, for the optimal initial states of the OQI and the initial states obtained with blocks of GHZ states, for different prior widths. $n_{\text{up}} = 0$ ($n_{\text{up}} = 21$) corresponds to the state $\ket{0}^{\otimes N}$($\ket{1}^{\otimes N}$). We see that for $\delta\phi \approx 1$ rad, the optimal initial state of the OQI resembles a sine state. (d) MSE of the OQI, CSS, and the proposed scheme for $N = 21$ and $\delta \phi = 0.7$ rad, plotted in purple, yellow, and red, respectively. The proposed scheme achieves a metrological gain of $g=2.29$ (3.59 dB).}
    \label{fig:21_atom_bmse}
\end{figure*}

In Fig. \ref{fig:intro_fig}b, local, adaptive measurements are depicted: a single-qubit rotation of $\Phi_i$ around the $z$-axis is applied before each measurement, such that $\phi-\Phi_i$ is measured. Depending on the measurement result, the single-qubit rotation that will be applied to the next GHZ state is chosen. This process is initiated with the largest GHZ state in the partition in terms of the number of qubits, and ends with the smallest GHZ state (see Fig. \ref{fig:intro_fig}b, where an example to a measurement result is highlighted in orange).
To find the optimal set of single-qubit rotations $\{\Phi \}$, $m$ phases $\phi$ are uniformly sampled from the interval $[-6\,\delta\phi,\, 6\,\delta\phi]$, and the BMSE is calculated analytically, which is used to tune $\{\Phi \}$. This process is iterated until the BMSE converges to a minimum. In Fig. \ref{fig:intro_fig}c, we list the numerical values of the single-qubit rotations obtained after performing the optimization for this case.

The proposed scheme can be realized experimentally with current quantum processors, where computing and sensing can be interleaved. State preparation and sequential feed-forward measurements can be implemented using e.g. Yb-171 nuclear spin qubits, where midcircuit measurements without entanglement were shown \cite{Norcia2023_midcircuit_Yb, omg_architecture_2023}. We lay out an experimental procedure in the subsection ``Experimental Realization".

\subsection*{Performance of the schemes}
\label{sec:results}

First, for a fixed qubit number, we analyze how the proposed scheme performs as we change the prior width, $\delta\phi$. Then, we assume a fixed prior width and observe the performance of all of the schemes as a function of the number of qubits, $N$.
To quantify the increase in the sensitivity of phase estimation when the proposed scheme is used, we define the metrological gain $g = (\Delta\phi_{\text{CSS}})^2/(\Delta\Tilde{\phi})^2$ as our figure of merit, where $(\Delta\phi_{\text{CSS}})^2$ and $(\Delta\Tilde{\phi})^2$ are the BMSE of the phase estimation performed using an $N$-qubit coherent spin state (CSS) and the proposed scheme, respectively. Note that the prior width and the number of qubits are the same for both schemes when calculating the gain, and we assume a single-quadrature readout for the CSS.


We study how the performance of our scheme, as well as the optimal initial state and partition depend on the prior width. The results for $N=21$ qubits can be found in Fig. \ref{fig:21_atom_bmse}a. The performance of the OQI is indicated in gray, and the gray filling below this line is used to emphasize the fact that it is not possible to perform better than the OQI, i.e., this part of the parameter space is \textit{inaccessible}. We also plot the performance of a GHZ state and a CSS for $N = 21$. We see that the GHZ state performs optimally for a small prior width, however, the performance quickly deteriorates as the prior width grows. We plot the performance proposed scheme with red points, and see that the proposed scheme surpasses the performance of the CSS with 21 qubits for any prior width. The optimal performance, i.e. the minimum value of $\Delta \Tilde{\phi}/\delta\phi$ for the proposed scheme occurs at $\delta\phi \approx 0.7$ rad. For each of the points of the proposed scheme, we (i) find the best partition for the given prior width, (ii) optimize over the local, adaptive measurements. The changing shapes of the red points indicate the different partitions of the 21 atoms into blocks of GHZ states: for example, the stars correspond to the partition with three 4-qubit, 2-qubit, and 1-qubit GHZ states, whereas the circles correspond to one 16-qubit, 4-qubit, and 1-qubit GHZ state. The full list of partitions corresponding to each shape can be found in Fig. \ref{fig:21_atom_bmse}b. We also plot the initial states of the proposed scheme and the initial states of the OQI for some of these shapes in Fig. \ref{fig:21_atom_bmse}c.

We then compare the MSE obtained with the proposed scheme, as a function of $\phi$, to the performance of the OQI and the CSS for $\delta\phi = 0.7$ rad in Fig. \ref{fig:21_atom_bmse}d. As can be seen from the Figure, the OQI surpasses the CSS for all $\phi$, performing optimally over almost the entire $\left[ -\pi, \pi \right]$ range. We see that the proposed scheme also surpasses the CSS for almost all $\phi$ except a very small region around $\phi = \pm 3\pi/8$ rad, where the CSS performs slightly better. However, the proposed scheme has a much larger bandwidth and a better performance around $\phi = 0$ rad. We calculate the metrological gain at $\delta\phi = 0.7$ rad as $g=2.29$ (3.59 dB).

\begin{figure}
    \centering\includegraphics[width=0.95\columnwidth]{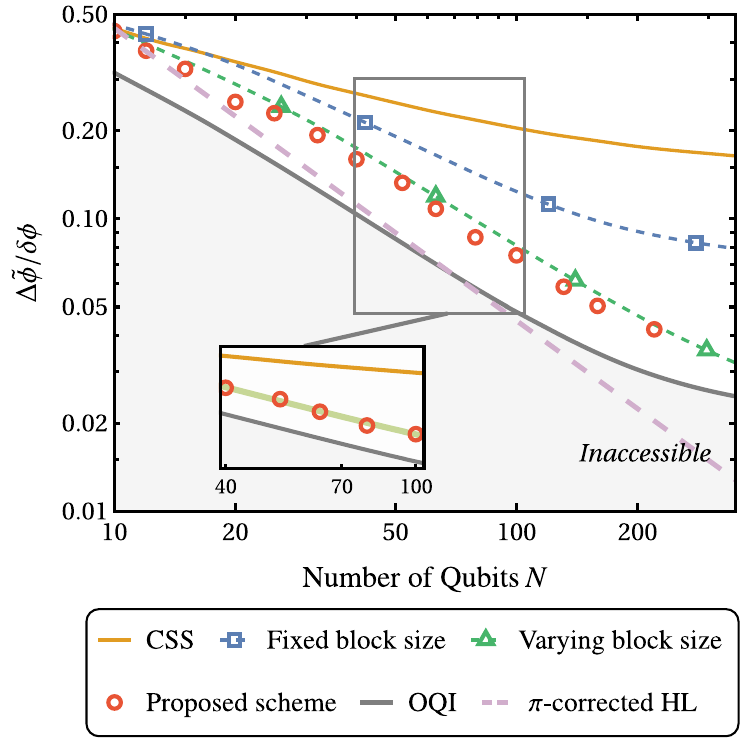}
    \caption{{\bf{Performance as a function of the number of qubits $N$.}} The root Bayesian mean squared error (RBMSE) as a function of number of qubits $N$, normalized by the prior width, for $\delta\phi = 0.7$ rad. We see that the CSS and the OQI constitute the SQL and HL, respectively. For larger qubit numbers, the BMSE of all of the schemes converge to a constant independent of $N$. We also plot the $\pi$-corrected HL in absence of phase slip errors, which touches the OQI curve at $N \approx 70$. We observe that all of the schemes that use blocks of GHZ states show sub-SQL sensitivity. Furthermore, the scheme with a varying block size \cite{higgins_demonstrating_2009} and the proposed scheme obtain HS. For all $N$, the proposed scheme obtains the maximal sensitivity, among existing schemes. \textit{Inset}. We plot the RBMSE of the OQI, CSS, and the proposed scheme in the range $40 \leq N \leq 100$, with gray, yellow, and red, respectively. We see that the proposed scheme shows HS with an overhead of $\approx 1.56$, shown with the light green line.}
    \label{fig:bmse_vs_N_no_noise}
\end{figure}


To see how the performance of the different schemes scale with the qubit number $N$, we fix $\delta\phi = 0.7$ rad, from the observation that the proposed scheme works optimally around that point in Fig. \ref{fig:21_atom_bmse} \footnote{Ref \cite{kaubruegger_quantum_2021} observed that this point depends on the qubit number, $N$. However, $\delta\phi = 0.7$ rad is still close to the optimal point, hence relevant for qubit numbers that we work with.}. In Fig. \ref{fig:bmse_vs_N_no_noise}, we plot the performance of these schemes, as well as the sensitivities of the OQI and the CSS for $10 \leq N \leq 350$.
We note that for our numerical simulations, we employed optimal Bayesian estimators. The blue, green, and red points in Fig. \ref{fig:bmse_vs_N_no_noise} correspond to the scheme with a fixed block size, the scheme with a varying block size, and the proposed scheme respectively. The implementation complexity of the proposed measurements also increase in this order: the scheme with a fixed block requires a double-quadrature readout and local measurements only, whereas the other schemes require single-qubit rotations. Moreover, for the proposed scheme, the single-qubit rotations are determined adaptively, which thus requires mid-circuit measurements.

However, there is a significant difference in the discontinuous nature of the scheme with a fixed block size and the scheme with a varying block size, compared to the scheme of this paper. For these schemes, the optimization over the blocks of GHZ states (Eq. (\ref{eqn:general_ghz})) is performed with respect to block sizes $\{m_i\}$, instead of $N$. Therefore, such schemes perform optimally only at certain $N$ where $N = \sum_i m_i 2^{k_i}$ holds, which are the points that were used to show their performance in Fig. \ref{fig:bmse_vs_N_no_noise}. For a general qubit number, such schemes must be modified by reducing or increasing the number of repetitions of certain blocks in order to fit the constraint of a fixed qubit number, which causes them to behave sub-optimally. In contrast, since we perform our optimization by fixing the number of qubits, the proposed scheme performs optimally for each $N$.

We observe from Fig. \ref{fig:bmse_vs_N_no_noise} that as the implementation complexity of the measurement proposed by a protocol increases, so does its sensitivity. Accordingly, the proposed scheme and the scheme with a varying qubit number scale as the OQI, surpassing the scheme with a fixed block size for every $N$. These schemes are Heisenberg-limited, and their overheads are calculated to be $1.56$ and $1.66$, respectively. The smaller overhead belongs to the proposed scheme, and the fit is shown in the inset of Fig. \ref{fig:bmse_vs_N_no_noise}. For example, the proposed scheme obtains a smaller posterior variance than that of the scheme with a varying block size by a factor of 1.22 (0.85 dB) for $N=63$. 
We also compare these schemes for a smaller prior width ($\delta\phi = 0.2$ rad) in \suppwiseman, and observe that the schemes continue to show HS. Furthermore, we observe that the proposed scheme surpasses the scheme with a varying block size by a larger margin. For example, for $N=63$, the gain (in variance) obtained by the former over the latter is calculated to be 1.57 (1.97 dB).

\subsection*{Fundamental limits due to phase slip errors}
\label{sec:fundamental_limits}

Our analysis so far focused on achieving quantum-enhanced precision with low-depth circuits.
However, another crucial limitation is phase slip errors, which arguably received less attention in the literature. Here, we first derive fundamental limits in precision due to phase slip errors. Then, we describe how to overcome such errors in the context of atomic clocks by proposing an efficient phase unwinding scheme.

Without phase slip errors, i.e. for any prior distribution with a  negligible probability outside of the range of $\left[ -\pi, \pi \right]$, it is well established that the minimal achievable RBMSE in the limit of large $N$ is $\Delta \tilde \phi \approx\pi/N$ \cite{jarzyna_true_2015, gorecki__2020}, where $N$ is the number of qubits. This is the relevant, $\pi$-corrected HL. Furthermore, the RBMSE achieved with $N$ uncorrelated qubits is given by $\Delta \tilde \phi \approx 1/\sqrt{N}$, which is the SQL. In the limit of an infinite qubit number $N$, both the HL and the SQL approach zero.

However, any prior distribution with a support outside of the range $\left[ -\pi, \pi \right]$ will give rise to an imperfect estimation, even in the limit of $N \rightarrow \infty$, since the phase can be estimated only up to modulo $2\pi$. This means that if a phase $\phi$ is outside of this range, its estimation error is given by $2\pi k$ in the limit of a large qubit number $N \gg 1$, where $k = \lfloor (|\phi|+\pi)/2\pi \rfloor$. The effect of these errors is shown in Fig. \ref{fig:21_atom_bmse}a, where the RBMSE increases with increasing prior width $\delta\phi$ due to phase slip errors, in the regime where $\delta\phi > 1$ rad. For $\delta\phi \gg 1$, the RBMSE becomes approximately equal to $\delta\phi$, signifying that the phase estimation scheme does not provide any information about $\phi$. Furthermore, in the presence of these errors, the RBMSE does not vanish in the limit of $N \rightarrow \infty.$ This is shown in Fig. \ref{fig:bmse_vs_N_no_noise}: the RBMSE converges to a finite value in the limit of $N \rightarrow \infty,$ which is determined by the statistics of these errors.  

We now explicitly find this limit, i.e. we find the OQI BMSE and the CSS BMSE with a single quadrature readout in the limit of $N \rightarrow \infty$.
These bounds are given in the following claim, in which we use the following expression of the BMSE (equivalent to Eq. (\ref{eqn:mse})):
\begin{align}
\begin{split}
\label{eqn:bmse_2}
&(\Delta\tilde{\phi})^{2}=(\delta\phi)^{2}-\sum_{x}\frac{\left(\int\phi\,p(x|\phi)\mathcal{P}_{\delta\phi}(\phi)d\phi\right)^{2}}{\int p(x|\phi)\mathcal{P}_{\delta\phi}(\phi)d\phi}\\
&=\left(\delta\phi\right)^{2}-\sum_{x}\phi_{\text{est}}\left(x\right)^{2}p\left(x\right).
\end{split}
\end{align}

\textit{Claim I.} The OQI
BMSE in the limit of $N \rightarrow \infty$ for any prior distribution $\mathcal{P}_{\delta \phi}(\phi)$ is \begin{align}
\label{eqn:plateau_hl}
(\Delta \Tilde{\phi})^2_{\text{OQI}} =\left(\delta\phi\right)^{2}-\overset{\pi}{\underset{-\pi}{\int}}\frac{\left(\underset{k}{\sum}\phi_{k}\mathcal{P}_{\delta\phi}\left(\phi_{k}\right)\right)^{2}}{\underset{k}{\sum}\mathcal{P}_{\delta\phi}\left(\phi_{k}\right)}d\phi,
\end{align}
where $\phi_k = \phi + 2\pi k, \; k \in \mathbb{Z}.$\\
Assuming a $J_{\theta} \coloneqq \cos\left(\theta\right)J_{x}-\sin\left(\theta\right)J_{y}$ single quadrature readout, the CSS BMSE in the limit of $N \rightarrow \infty$ for any prior distribution $\mathcal{P}_{\delta \phi}(\phi)$ is
\begin{align}
\begin{split}
&(\Delta\Tilde{\phi})_{\text{SQL}}^{2}=\left(\delta\phi\right)^{2}-
\overset{\pi}{\underset{-\pi}{\int}}d\phi \, \cdot\\
&\frac{\left(\underset{k}{\sum}\left(\phi_{k}-\theta\right)\mathcal{P}_{\delta\phi}\left(\phi_{k}-\theta\right)-\left(\phi_{k}+\theta\right)\mathcal{P}_{\delta\phi}\left(-\phi_{k}-\theta\right)\right)^{2}}{\underset{k}{\sum}\mathcal{P}_{\delta\phi}\left(\phi_{k}-\theta\right)+\mathcal{P}_{\delta\phi}\left(-\phi_{k}-\theta\right)}.
\label{eqn:plateau_sql}
\end{split} 
\end{align}

We defer the detailed proof to \supptuviaproof, and provide here a sketch of the proof.
To demonstrate that Eq. (\ref{eqn:plateau_hl}) is the OQI BMSE in the limit of $N \rightarrow \infty$, we first show that it is a lower bound of the BMSE for any possible scheme. Then, we show that there exists a scheme that saturates it.
The following minimization problem lower-bounds the OQI for any $N$:
\begin{align}
\text{\ensuremath{\underset{\phi_{\text{est}}\left(\phi\right)}{\text{min}}}}\int_{-\pi}^{\pi}\sum_{k}\left(\phi_{\text{est}}\left(\phi\right)-\left(\phi+2\pi k\right)\right)^{2}p_{\delta\phi}\left(\phi+2\pi k\right)\;d\phi,    
\end{align}
where $\phi_{\text{est}}\left(\phi\right):\left[-\pi,\pi\right]\rightarrow\mathbb{R}.$
The BMSE given in Eq. (\ref{eqn:plateau_hl}) is a solution of this minimization problem.
For attainability, we show in \supptuviaproof\ that a CSS interrogation with dual quadrature readout saturates this lower bound in the limit of $N \rightarrow \infty.$

For the bound assuming a CSS $J_\theta$  quadrature readout, Eq. (\ref{eqn:plateau_sql}), we find the measurement probabilities $p\left(x_{\theta}|\phi\right),$
where $x_{\theta}$ is the outcome of $2J_\theta/N$ measurement, i.e. a measurement outcome of $J_{\theta}$ normalized with respect to $N/2$.
In the limit of $N\rightarrow \infty$, $p\left(x_{\theta}|\phi\right)\rightarrow\delta\left(x_{\theta}-\cos\left(\phi+\theta\right)\right)$. Inserting this into Eq. (\ref{eqn:bmse_2}) leads to Eq. (\ref{eqn:plateau_sql}).

These bounds are the ``plateaus" of the BMSE observed in Fig. \ref{fig:bmse_vs_N_no_noise}: this is the remaining uncertainty that cannot be suppressed by increasing $N$. They increase as the prior width, $\delta\phi$, increases.
We remark that the CSS bound shown in Fig. \ref{fig:bmse_vs_N_no_noise}
corresponds to a $J_{y}$ measurement.
The difference between the bounds in Eqs. (\ref{eqn:plateau_hl}), (\ref{eqn:plateau_sql}) correspond to the gap between the uncertainty of the OQI and the CSS in the limit of $N \gg 1$, as observed in Fig. \ref{fig:bmse_vs_N_no_noise}.

\begin{figure*}
    \raggedright
    \includegraphics[width=2.05\columnwidth]{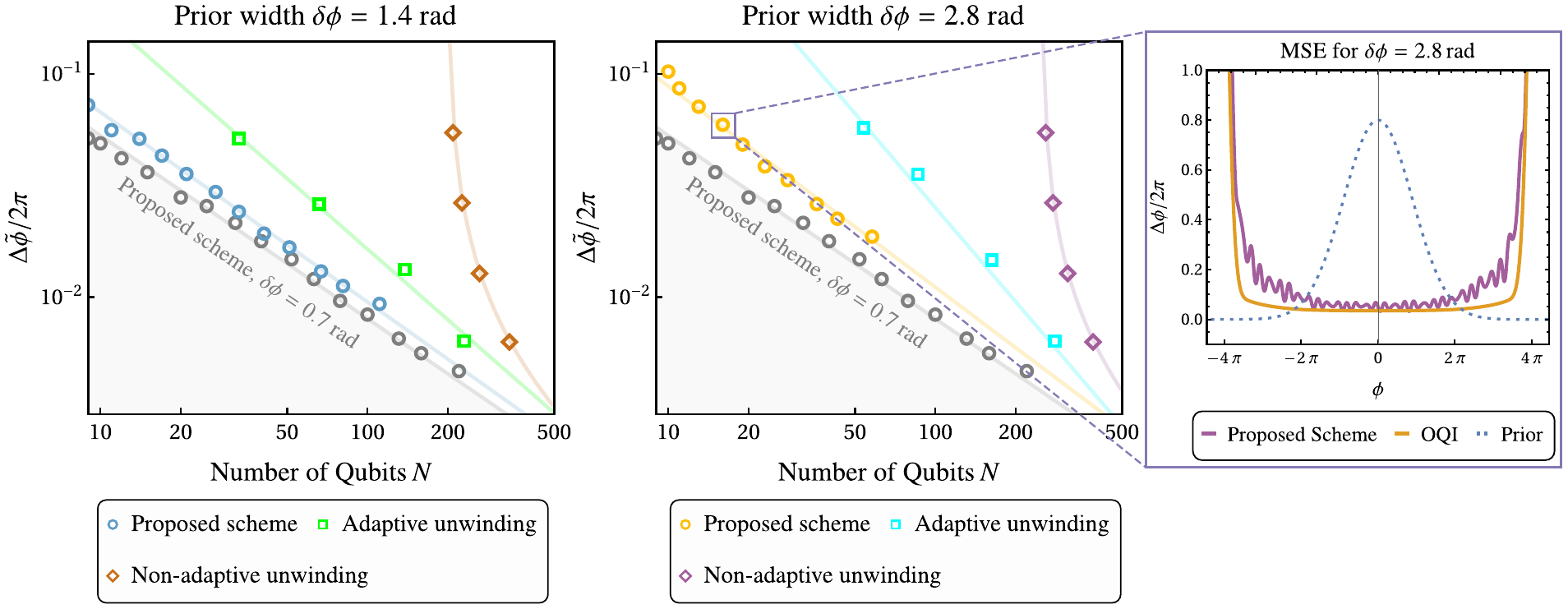}
    \caption{\textbf{Extending the dynamic range with slow atoms.} Root Bayesian mean squared error (RBMSE) $\Delta\tilde\phi$, normalized by $2\pi$, as a function of the total qubit number \mbox{$N$}, for different prior widths.
    For small enough prior widths ($\delta\phi = 0.7$ rad), no slow atoms are needed. The RBMSE of the proposed scheme in this regime is plotted in gray. For larger prior widths, the number of atoms needed to extend the dynamic range depends on the \textit{phase unwinding protocol}. These protocols estimate $\phi = 2\pi P + \theta$, $P \in \mathbb{Z}, \, \theta = \phi \,\text{mod}\,2\pi$. We define non-adaptive (estimates $P$ with slow atoms and $\theta$ with the scheme with a varying block size \cite{higgins_demonstrating_2009}) and adaptive (estimates both $P$ and $\theta$ with slow atoms with, then estimates $\phi-\phi_\text{est}$ with the scheme with a varying block size) as phase unwinding. We also propose a more efficient phase unwinding protocol that extends the proposed scheme to large prior widths. Furthermore for the proposed scheme, in the limit of $N \gg 1$, the RBMSE for different prior widths converge to the RBMSE for $\delta\phi = 0.7$ rad. 
    Therefore, in this limit, we can obtain a clock stability that scales as the total interrogation time $\tau$ even in the regime where $\tau \gamma_{\text{LO}} > 1$, where $\gamma_{\text{LO}}$ is the linewidth of the free running laser. \textit{Inset}. MSE for the proposed phase unwinding scheme for $N = 16$ atoms and a prior width of $\delta\phi = 2.8$ rad. We see that the MSE is close to that of the OQI, and that the proposed scheme has a dynamic range of $\approx 8\pi$. }
    \label{fig:phase_unwinding_ours}
\end{figure*}

\subsection*{Overcoming phase slip errors: phase unwinding schemes}
\label{sec:phase_unwinding_scheme}

For a wide phase prior distribution, phase slip errors are the main factor limiting the precision, preventing HS or even SQL for large $N.$
In the context of atomic clocks, this regime of a wide prior corresponds to a Ramsey time longer than the LO coherence time. In other words, $T \gamma_\text{LO} >1$, where $T$ is the Ramsey time, or the total interrogation time, and $\gamma_\text{LO}$ is the linewidth of the laser (see \suppclocks). 
In this regime, phase slip errors dominate, and further Ramsey interrogations do not lead to any improvement in phase sensitivity.
To suppress these phase slip errors, and to extend the Ramsey time beyond the LO noise limit, some techniques were devised in Refs. \cite{rosenband2013exponential, Borregaard_efficient_atomic}.
These schemes introduce additional CSS interrogations with \textit{slow} atoms, i.e. atoms that accumulate phases $\phi/2, \phi/4 \dots$ during the interrogation time, instead of accumulating a phase $\phi$. We refer to such schemes as phase unwinding schemes.

These slow atoms that accumulate only a fractional phase can be obtained in several different ways, depending on the platform and the relevant phase noise. In the simplified case, where the laser frequency $\omega_L(t)$ is almost constant throughout the Ramsey time, they can be obtained through evolving the qubits for the relevant fraction of the Ramsey time $T.$
In this case, a single qubit can accumulate a phase of $\phi/2^{l}$ for $2^{l}$ times.  In other words, a single atom can be re-used as $2^{l}$ slow atoms with a phase of $\phi/2^{l}$.
More generally, given an arbitrary stochastic process of the frequency, different methods should be employed and re-using of the qubits may not be possible. A reduced evolution rate in the presence of a general noise can be achieved using dynamical decoupling, as was shown in Refs. \cite{rosenband2013exponential, shaw_multi-ensemble_2023}. The idea is to divide the Ramsey time into short time intervals of $\tau',$ where $\tau'$ is shorter than the correlation time of the frequency noise. In each $\tau'$ interval we can apply a $\pi$-pulse after a period of $\tau'\left(1/2+1/2^{l+1}\right),$ this would lead to an accumulated phase of $\phi/2^l.$ Alternatively, midcircuit measurement and reset can be used \cite{rosenband2013exponential}.
With this method, each qubit corresponds to a single slow atom, and re-using of the atoms is not possible.
An alternative method requires memory qubits, but allows re-using of sensing qubits: each memory qubit encodes a different slow atom, such that the sensing qubit will be entangled to different memory qubits at different periods of the $\tau'$ interval. This method requires multiple memory qubits and fast entangling gates between the sensors and memory qubits, but it allows more efficient use of the sensing qubits.

Let us first analyze and extend some existing phase unwinding methods that use slow atoms.
The phase unwinding protocol in Ref. \cite{rosenband2013exponential} defines \mbox{$\phi = 2\pi P + \theta$}, where $P \in \mathbb{Z}$, and $\theta = \phi \text{ mod } 2\pi$. Assuming that $m$ copies of slow atoms that accumulate phases of $\phi/2^{l}$, $l = 0, 1, \dots, l_{\text{max}}$ are available, it estimates $P$ iteratively from the following equation
\begin{align}
    P_{j-1} = 2P_{j} + \frac{2\beta_j -\beta_{j-1}}{2\pi}
\end{align}
where $j = 0, 1, \dots, l$, $P = P_0$, $P_{l_{\text{max}}} = 0$, and $\beta_j$ is the estimate of $\phi$ after performing dual quadrature readout on the $j^\text{th}$ block of slow atoms.
After estimating $P$, the posterior distribution on the phase resembles a uniform distribution in $\left[-\pi, \pi\right]$, in the limit of a large initial prior width $\delta\phi$ (see \suppunwinding\ for a derivation). The protocol in Ref. \cite{rosenband2013exponential} concludes by estimating $\theta$ from the slow atoms that accumulate a phase $\phi$.

Alternatively, phase unwinding protocols can be combined with phase estimation using entangled atoms to obtain a higher precision.
We can use the schemes in Refs. \cite{kessler_heisenberg-limited_2014, higgins_demonstrating_2009}, or the proposed scheme, for this purpose. We illustrate two possible ways, which we refer to as \textit{adaptive} and \textit{non-adaptive} phase unwinding. For non-adaptive phase unwinding, one can estimate $P$ from the protocol in Ref. \cite{rosenband2013exponential}, and estimate $\theta$ with schemes that use GHZ blocks. Here, we assume that one can further constrain the posterior distribution on $\phi \text{ mod } 2\pi$ after estimating $P$ to be approximately a uniform distribution in the interval $[-\pi/2, \, \pi/2]$ instead of the interval $[-\pi, \, \pi]$, by including 2-qubit GHZ states in the protocol in Ref. \cite{rosenband2013exponential}. We choose to do so since for a uniform prior distribution in $[-\pi, \, \pi]$, the BMSE is large due to the fact that small errors for phases close to $-\pi, \, \pi$ lead to large errors in estimating $\phi$. In other words, the phase estimation error jumps to $2\pi$ for $|\phi|$ close to $\phi$, which is not suppressed when a uniform prior distribution in $[-\pi, \, \pi]$ is used.
We plot the performance of all of the schemes that use GHZ blocks, for a uniform prior in $[-\pi/2, \, \pi/2]$, in \suppuniformprior. 

In Fig. \ref{fig:phase_unwinding_ours}, we plot the RBMSE obtained by combining non-adaptive phase unwinding with the scheme with a varying block size (Ref. \cite{higgins_demonstrating_2009}). We analyze two prior widths, $\delta\phi = 1.4$ rad (plotted in brown), and $\delta\phi = 2.8$ rad (plotted in purple), where we plot the sensitivity as a function of the total number of qubits $N$. $N$ includes both the atoms in the GHZ states, and the atoms used to extend the Ramsey time by the protocol in Ref. \cite{rosenband2013exponential}. HS is not observed for the regime plotted in Fig. \ref{fig:phase_unwinding_ours}, due to the large overhead of additional atoms needed for this type of phase unwinding.

For adaptive phase unwinding, one can estimate both $P$ and $\theta$, thus estimate $\phi$, written as $\phi_\text{est}$. After the estimation, the posterior distribution of $\phi$ will approximately be a Gaussian distribution centered at $\phi_\text{est}$, with a width of $\Delta\tilde\phi$, i.e. the RBMSE. Then, one can apply a global phase shift $-\phi_\text{est}$, and estimate $\phi-\phi_\text{est}$ with schemes that use GHZ blocks, assuming an effective prior width of $\Delta\tilde\phi$. We also plot the RBMSE obtained by adaptive phase unwinding in Fig. \ref{fig:phase_unwinding_ours} for $\delta\phi = 1.4, 2.8$ rad in green and light blue, respectively. We observe that adaptive phase unwinding is more efficient than non-adaptive phase unwinding, meaning that adaptive phase unwinding obtains the same sensitivity with a smaller number of atoms. We can explain this increase in performance by noticing that adaptive unwinding makes use of both $P$ and $\theta$ before estimating $\phi$ with schemes that use GHZ blocks. Conversely, non-adaptive phase unwinding only computes $P$. Therefore, the posterior distribution on $\phi$ before doing the phase estimation with GHZ blocks is more localized in the former phase unwinding protocol, giving rise to a better sensitivity. In our simulations, we observed that when adaptive phase unwinding was optimized with respect to the number of slow atoms, the posterior width was in the vicinity of $\Delta\tilde\phi \approx 0.5$ rad.

We now outline a more efficient phase unwinding protocol that optimizes over all partitions of blocks of slow atoms and GHZ states. 
Assuming a large $\delta \phi$ such that phase slips errors dominate, we first choose an integer $l_\text{max}$ such that $\delta\phi/2^{l_\text{max}}$ is small enough. We now introduce slow atoms that accumulate phases $\phi/2^{l_\text{max}},\phi/2^{l_\text{max}-1},...,\phi/2.$ 
For a given partition into blocks of slow atoms and GHZ states, we can write the relevant quantum state as:
\begin{align}
\begin{split}
&|\psi\rangle \propto \underset{l=1}{\overset{l_\text{max}}{\Pi}}\left(|0\rangle+|1\rangle e^{i\phi \,2^{-l}}\right)^{\otimes m_{-l}}\otimes\\
&\underset{k=0}{\overset{k_{\text{max}}}{\Pi}}\left(|0\rangle^{\otimes2^{k}}+|1\rangle^{\otimes2^{k}}e^{ i\phi \, 2^k }\right)^{\otimes m_{k}}.
\end{split}
\end{align}
The first term of the product is the slow or classical part, while the second term is the fast rotating part that contains GHZ states.
The relation between $N$ and the number of atoms in the different blocks requires special attention. Assuming that the same qubit can be used as a slow atom several times, e.g. a single qubit can accumulate a phase of $\phi/2^{l}$ for $2^{l}$ times, then the constraint on $N$ is given by: $ N=\lceil\sum_{l=-l_{\text{max}}}^{k_{\text{max}}} 2^{l} m_{l} \rceil.$
If we cannot re-use qubits, then each slow atom corresponds to a single qubit, and the constraint on $N$ is given by: $N=\sum_{l=-l_{\text{max}}}^{-1} m_{l} + \sum_{k=0}^{k_{\text{max}}} 2^{k} m_{k}.$

Given a total number of qubits $N,$
we want to find an optimal partition of these $N$ qubits into slow atoms and GHZ states. 
To do this we use a rescaling transformation that reverts the problem back to partitions of GHZ states only.
We define $\phi'=\phi/2^{l_\text{max}},$ and we use the following equality:
\begin{align}
\begin{split}
&\underset{\vec{n}}{\sum}\underset{-\infty}{\overset{\infty}{\int}}\left(\phi-\phi_{e}\right)^{2}p\left(\vec{n}|\phi\right)\mathcal{P}_{\delta\phi}\left(\phi\right)\;d\phi=\\
&2^{2 l_\text{max}}\underset{\vec{n}}{\sum}\underset{-\infty}{\overset{\infty}{\int}}\left(\phi'-\phi'_{e}\right)^{2}p\left(\vec{n}|\phi'\right)\mathcal{P}_{\delta\phi/2^{l_\text{max}}}\left(\phi'\right)\;d\phi',
\label{eq:rescaling_equivalence}
\end{split}
\end{align}
which is obtained by changing the integration variable to $\phi'=\phi/2^{l_\text{max}}.$
This equality can be written as:
\begin{align}
(\Delta\widetilde{\phi})_{|\psi\rangle,\delta\phi}^{2}=2^{2 l_\text{max}}(\Delta\widetilde{\phi}')_{|\psi'\rangle,\delta\phi/2^{l_\text{max}}}^{2},
\label{eq:rescaling_equivalence_2}
\end{align}
where the l.h.s is our usual figure of merit: the BMSE in estimating $\phi$ with the original state, $|\psi\rangle$, and the original prior width $\delta \phi.$ The r.h.s corresponds to the rescaled frame: it is $2^{2 l_\text{max}}$ times the BMSE in estimating $\phi'=\phi/2^{l_\text{max}}$ with a prior width of $\delta\phi/2^{l_\text{max}}$ and a rescaled state $|\psi'\rangle.$
The rescaled state is: 
\begin{align}
|\psi'\rangle=\underset{k=0}{\overset{l_{\text{max}}+k_{\text{max}}}{\Pi}}\left(|0\rangle^{\otimes2^{k}}+|1\rangle^{\otimes2^{k}}e^{i\phi'2^{k}}\right)^{\otimes m_{k-l_{\text{max}}}}.    
\end{align}
Hence based on Eq. (\ref{eq:rescaling_equivalence}), we can optimize the BMSE of $\phi'$ in the rescaled frame using the techniques developed here, in which $|\psi'\rangle$ contains no slow atoms and the prior width is shrunk to $\delta \phi/2^{l_{\text{max}}}.$

The rescaling transformation maps $\phi,\delta\phi,|\psi\rangle$ into $\phi',\delta\phi/2^{l_{\text{max}}},|\psi'\rangle$.
Since the number of qubits in $|\psi' \rangle$ is not equal to the number of qubits in $|\psi \rangle$, $N$ is also transformed into $N'$, given by \mbox{$N'= 2^{l_{\text{max}}} \sum_{l = -l_{\text{max}}}^{k_{\text{max}}} 2^{l} m_l$}.
The transformation of $N$ depends on whether qubits can be re-used
as several slow atoms or not. Assuming that re-using is possible, we see that $N' \approx 2^{l_{\text{max}}} N.$ 
However if re-using is not possible, \mbox{$N' = 2^{l_{\text{max}}} N - \sum_{l = 0}^{l_{\text{max}}} (2^{l_{\text{max}}}-2^l) m_{l-l_{\text{max}}}$}. Here, we similarly optimize the BMSE in the $\phi'$ frame for the relevant values of $N'$, and transform back to the BMSE in the $\phi$ frame using Eq. (\ref{eq:rescaling_equivalence_2}). 

For example, we have calculated the optimal partition for $N' = 159, \delta\phi/2^{l_{\text{max}}} = 0.7$ rad  as $m_{k-l_{\text{max}}} = 3, 2, 4, 3, 3, 2$, $k = 0, 1, \dots, 5$. We can rescale this partition by setting $l_{\text{max}} = 3$, which corresponds to a prior width of $\delta\phi = 5.6$ rad in the rescaled frame. In this frame, the largest GHZ state is a 4-qubit GHZ state.
Furthermore, assuming that re-using qubits is not possible, the rescaled frame contains $N = 12 + 2\cdot 3 + 4\cdot 2 = 26$ atoms. In summary, we were able to map the problem of finding the optimal partition of slow atoms and GHZ states into finding the optimal partition in presence of GHZ states only, which was already addressed and solved. In this way, we incorporated the slow atoms into the proposed scheme. The BMSE in presence of slow atoms can therefore be found using Eq. (\ref{eq:rescaling_equivalence_2}).

Finally, we plot the RBMSE obtained by the proposed phase unwinding scheme for $\delta\phi = 1.4, 2.8$ rad with blue and yellow, respectively, assuming that a qubit cannot be re-used as several slow atoms. We also plot the RBMSE of the proposed scheme for $\delta\phi = 0.7$ rad in gray as a reference since no slow atoms are used to extend the dynamic range for this prior width. 
We observe that the proposed phase unwinding achieves the same precision as adaptive and non-adaptive phase unwinding with a much smaller number of slow atoms. Furthermore for large $N$, the RBMSE of the proposed scheme for $\delta\phi = 1.4, 2.8$ rad start converging to the reference RBMSE ($\delta\phi = 0.7$), meaning that the number of slow atoms becomes negligible compared to the number of atoms of the GHZ blocks.

Then, in this limit, the interrogation time can be extended by the addition of a negligible number of atoms, resulting in the Allan deviation to scale as $\sigma_y(\tau) \propto \tau^{-1}$, where $\tau$ is the total interrogation time (see  \suppclocks). Furthermore, the proposed phase unwinding protocol reduces the number of slow atoms needed to extend the interrogation time significantly compared to other phase unwinding protocols, as seen from Figure \ref{fig:phase_unwinding_ours}.

While the slow atoms methods presented above mitigate phase slip errors, they still do not recover the noiseless limit. Since the noiseless ultimate precision limit is the $\pi$-corrected HL \footnote{The $\pi$-corrected HL is relevant when $\delta \phi N>1$. If $\delta \phi N \ll 1$, the conventional HL, $1/N$, can be attained asymptotically \cite{gorecki__2020}.}, the relevant question is whether there exists a method that achieves this limit for arbitrarily large prior widths and arbitrary N.
The following claim provides an indication that the answer is positive: assuming ancillas and coherent control, we can construct slow replicas of optimal states that achieve HS.

\textit{Claim II.}
For every phase prior distribution $\mathcal{P}_{\delta\phi}\left(\phi\right)$, and for every number of probes, $N,$ there exists a protocol that achieves a BMSE, $(\Delta\tilde{\phi})_{\mathcal{P}_{\delta\phi}\left(\phi\right),\,N}^{2},$ with
\begin{align}
\label{eq:oqi_proof}
(\Delta\tilde{\phi})_{\mathcal{P}_{\delta\phi}\left(\phi\right),\,N}^{2}=l^{2}(\Delta\tilde{\phi})_{\text{OQI},\mathcal{P}_{\delta\phi/l}\left(\phi\right),\,lN}^{2},    
\end{align}
where $(\Delta\tilde{\phi})_{\text{OQI},\mathcal{P}_{\delta\phi/l}\left(\phi\right),\,lN}^{2}$ is the OQI BMSE with
a rescaled prior distribution $\mathcal{P}_{\delta\phi/l}\left(\phi\right)$ and with $lN$ probes, where $l\in\mathbb{N}_{+}.$ This protocol assumes the use of ancillas and time dependent control.

Before proving the claim, let us first 
elaborate on its meaning. This protocol should allow us to achieve the $\pi$-corrected HL, $\pi^2/N^2$, with arbitrarily large prior width. If we can find small enough $\delta \phi'=\delta\phi/l$ such that 
$(\Delta\tilde{\phi})_{\text{OQI},\mathcal{P}_{\delta\phi/l}\left(\phi\right),\, lN}^{2}=\pi^2/(Nl)^2,$
then this claim implies that the same scaling is achievable for $\delta \phi:$
$(\Delta\tilde{\phi})_{\mathcal{P}_{\delta\phi}\left(\phi\right),\,N}^{2}=\pi^2/N^2$.

\textit{Proof.}
We denote the OQI with $\mathcal{P}_{\delta\phi/l}\left(\phi\right)$ and $Nl$ probes as $(\Delta\tilde{\phi})_{\text{OQI},\mathcal{P}_{\delta\phi/l}\left(\phi\right),\,lN}^{2}.$
The optimal state that achieves it is denoted as $|\psi\rangle=\sum_{j=0}^{Nl} c_{j}|j\rangle$ . After the phase encoding, this state is given by $|\psi_{\phi}\rangle=\sum_{j=0}^{Nl} c_{j}e^{-ij\phi}|j\rangle$.
We have shown in Eq. (\ref{eq:rescaling_equivalence}) that the following rescaling transformation
\begin{subequations}
\begin{align}
& \phi\mapsto\phi/l,\;  \mathcal{P}_{\delta\phi/l}\left(\phi\right)\mapsto\mathcal{P}_{\delta\phi}\left(\phi\right),\\
&|\psi_{\phi}\rangle\mapsto|\psi_{\phi/l}\rangle=\sum_{j=0}^{Nl} c_{j}e^{-ij\phi/l}|j\rangle. 
\end{align}
\end{subequations}
transforms the BMSE, $(\Delta\tilde{\phi})_{\mathcal{P}_{\delta\phi}\left(\phi\right),N}^{2}$, as given in Eq. (\ref{eq:oqi_proof}).
This is the BMSE we want attain with $N$ qubits, hence the only thing that is needed to show is that we can construct the following state (and phase encoding) $|\psi_{\phi/l}\rangle=\sum_{j=0}^{Nl} c_{j}e^{-ij\phi/l}|j\rangle$, and perform the optimal measurements using $N$ qubits.

We provide a construction that uses ancillas and pulses that depend on the state of the ancillas.
We first construct the state $|\psi\rangle= \sum_{j=0}^{Nl} c_{j}|j\rangle$  on the ancillas.
In principle, $\log\left(Nl+1\right)$ ancillas are sufficient to generate this state. It can also be generated by using a symmetric or a unary encoding, but this requires $Nl$ ancillas. 
The $N$ probes are initialized to the state $|N\rangle$ , such that the initial state is $|N\rangle\left(\sum_{j=0}^{Nl} c_{j}|j\rangle\right).$
For every $|j\rangle$,  we want to obtain the following phase accumulation: $|N\rangle|j\rangle\mapsto e^{-ij\phi/l}|N\rangle|j\rangle$ .
This can be achieved with $|j\rangle$-dependent pulses that slow down the evolution, e.g.
by adding a control Hamiltonian of $H_{c}=\sum_{j=0}^{Nl} f_{j}\left(t\right)J_{x}\otimes|j\rangle\langle j|$,
with 
$f_{j}\left(t\right)=\pi\delta\left(t-T\frac{Nl+j}{2Nl}\right),$
where $T$ is the Ramsey time ($\omega T = \phi$ \footnote{We assumed in the proof the simplest case of a fixed $\omega$ during the time evolution, i.e. $H=\omega J_{z}.$
In most realistic cases, $\omega(t)$ fluctuates with time.
In these cases, to obtain an effective Hamiltonian of $H=\sum_{j=0}^{Nl}\frac{j}{Nl}\omega\left(t\right)J_{z}\otimes|j\rangle\langle j|$, we need to apply $\pi$-pulses during time periods shorter than the correlation time of $\omega(t).$ Denoting such time intervals as $\tau,$ the ancilla-dependent control function $f_{j}(t)$ is then given by  $f_{j}\left(t\right)=\ensuremath{\sum_{k=0}^{T/\tau-1}}\pi\left(\delta\left(t-\tau\left(k+\frac{Nl+j}{2Nl}\right)\right)+\delta\left(t-\left(k+1\right)\tau\right)\right)$}). These ancilla-dependent pulses lead to the effective Hamiltonian of (after moving to the interaction picture with respect to $H_c$) $H= \sum_{j=0}^{Nl} \frac{j}{Nl}\omega J_{z}\otimes|j\rangle\langle j|$, and thus to the final state of $|N\rangle(\sum_{j=0}^{Nl} c_{j}e^{-ij\phi/l}|j\rangle).$
The state of the ancillas is now $|\psi_{\phi/l} \rangle,$ hence by implementing the optimal measurement on the ancillas the desired BMSE is obtained. \qedsymbol 

\subsection*{Experimental Realization}

Here, we provide a roadmap for the experimental implementation of the proposed scheme in existing quantum platforms. Specifically, neutral atom arrays have recently emerged as a promising platform for high fidelity control of qubits, where entanglement is mediated through Rydberg interactions \cite{levine_high-fidelity_2018, levine_multiqubit_2019, Evered2023_highfidelity, omg_architecture_2023,shaw_multi-ensemble_2023,finkelstein2024universal, cao2024multiqubit}.

To implement the proposed scheme, 
the following experimental stages need to be carried out:
(1) State preparation: high fidelity preparation of GHZ states of optical clock qubits. (2) Evolution period: LO phase is imprinted on the qubits. The evolution time should be shorter than the qubit coherence time (determined by amplitude damping or dephasing noise when slow atoms are used to extend the dynamic range).
(3) Measurement: sequential local measurements should be performed, combined with adaptive single-qubit rotations. As the measurement is sequential and adaptive, it is crucial that when a qubit is measured, the remaining qubits to-be-measured are ``frozen", i.e. decoupled from the LO phase and remain coherent during the measurement period, in order to achieve high sensitivity.

Due to the constraints above, we propose an implementation that would make use of 
interleaving sensing and computing with optical clock qubits and non-optical qubits with long coherence times, respectively.
For the computing qubits, we can use nuclear spins in a metastable state (i.e. Yb-171). This system allows for long coherence times, high fidelity gates and protection from imaging light \cite{Jenkins_Ytterbium}. Or, we can use a hyperfine alkali qubit in a dual-species architecture. Let us explain what these implementations entail for state preparation and readout in the following.


In order to prepare the blocks of GHZ states in neutral atom array platforms, one can use high-fidelity entangling gates obtained through e.g. phase-modulated Rydberg pulses. Such gates have recently been implemented for optical \cite{finkelstein2024universal, cao2024multiqubit}, as well as hyperfine and nuclear qubits  \cite{Evered2023_highfidelity, Ma2023_mid-circuit_atomic}.

For fast and high-fidelity preparation of GHZ states, we can use nuclear qubits in a metastable state \cite{PhysRevX_Yb_2022, Jenkins_Ytterbium}. For the clock operation, these qubits need to be mapped to optical clock qubits, which can be realized via a global $\pi$-pulse. This implementation allows for a faster state preparation that is robust to the clock laser noise. Another possible implementation is via a dual-species architecture, where state preparation is realized with e.g. the hyperfine states in Rubidium \cite{PhysRevX.12.011040}. In order to map these qubits to optical clock qubits, an interspecies SWAP gate is needed. Lastly, if state preparation is realized using optical clock qubits, a high-coherence laser (e.g. the clock laser systems of Refs. \cite{bothwell_resolving_2022,young_half-minute-scale_2020}) can be used to achieve a high fidelity \cite{finkelstein2024universal, cao2024multiqubit}.

Experimentally, the circuit to prepare an $N$-qubit GHZ state can be implemented through combining single-qubit rotations with controlled-Z (CZ) gates, which can be realized with two-qubit \cite{finkelstein2024universal} or multi-qubit gates \cite{levine_multiqubit_2019, Graham2022_Multi-qubit, cao2024multiqubit}. High-fidelity CZ gates allow for the construction of high-fidelity GHZ states. The state-of-the-art fidelities are cited as 99.72(3)\% for two-qubit CZ gates, and  99.963(2)\% for single-qubit rotations, for systems that use Yb-171 nuclear spin qubits \cite{muniz2024highfidelityuniversalgates171yb}.


The proposed scheme requires the implementation of single-qubit rotations and sequential midcircuit measurements with feed-forward. As the initial states of the proposed scheme contain multiple copies of GHZ states with different sizes, the remaining states should be protected from the imaging light during the measurement of a GHZ state. This can be achieved by, for example, transporting the qubits to a separate readout zone before a measurement \cite{Bluvstein2024_reconfigurable}, or, by shelving the other qubits such that they are dark to the imaging light \cite{omg_architecture_2023, graham_midcircuit, Ma2023_mid-circuit_atomic}.

Let us describe how nuclear qubits can be used to achieve adaptive readout. After the accumulation of the LO phase, optical qubits can be mapped back to metastable nuclear qubits via another global $\pi$-pulse, so that they are decoupled from the LO and additional decoherence effects, as well as from the imaging light for the readout. Finally, a local $\pi$-pulse can be used to map one-by-one the GHZ states of nuclear qubits into GHZ states of optical or ground state qubits for sequential, local readout.

Finally, any experimental realization would suffer from decoherence, due to e.g. amplitude damping, state preparation errors and detection noise. We study the effect of these noises on the proposed scheme in Supplementary Notes 13, 14. In Supplementary Note 13, we analyze the effect of amplitude damping on blocks of GHZ states, and plot the BMSE for different levels of noise. A similar analysis is done in Supplementary Note 14 for state preparation and measurement errors.

\section*{Discussion}

In this paper, we analyzed phase estimation schemes that utilize state preparation with low-depth circuits in the Bayesian framework. Specifically, we focused on protocols that employ blocks of GHZ states with varying numbers of qubits, such as the protocols in Refs. \cite{kessler_heisenberg-limited_2014, higgins_demonstrating_2009}. Inspired by these protocols, we proposed a scheme 
that combines such initial states
with local, adaptive measurements, where we approximate optimal initial states for any prior width using such blocks of GHZ states. Constraining our optimization by fixing the number of qubits, $N$, we were able to achieve a good approximation for all $N$, which is one of the distinguishing features of the proposed scheme compared to other protocols that we have analyzed. 

After obtaining the approximated optimal initial state, we optimized over local, adaptive measurements. These are implemented by applying single-qubit rotations (selected adaptively based on the previous measurement result) before a projective measurement.
We analyzed the performance of the proposed scheme 
using Gaussian prior distributions. However, our optimization algorithm is completely general and can be applied to any prior distribution function.

Working with a large prior width within the dynamic range, we showed that it is possible to obtain Heisenberg scaling by using the protocol of Ref. \cite{higgins_demonstrating_2009} that combines non-adaptive measurements and single-qubit rotations with blocks of GHZ states. Moreover, we showed that the proposed protocol also exhibits Heisenberg scaling with a constant overhead of 1.56 for a prior width of $\delta\phi = 0.7$ rad, and outperforms the scheme of Ref. \cite{higgins_demonstrating_2009}. Since these schemes require logarithmic depth circuits, they may pave the way to a large-scale Heisenberg-limited atomic clock.
A non-trivial challenge, however, is the adaptive local measurements, which require fast and non-destructive detection.
This raises a question about 
how much we lose in sensitivity by restricting ourselves to non-adaptive measurements, for a general qubit number. We leave this question to future investigation.

Lastly, we proposed an efficient method to extend the dynamic range of the proposed scheme in order to obtain an enhanced precision. We showed that the optimal initial states of our scheme can be rescaled in order to obtain partitions that include slow atoms that accumulate fractional phases, as well as GHZ states. Compared to existing phase unwinding schemes, we showed that our method uses a smaller number of slow atoms.

Further future directions include a more detailed study of the effects of noise and decoherence.
Our study of these effects (in Supplementary Notes 13,14) can be improved by a closer examination of the optimal initial states and measurements given relevant noise processes. Such optimal measurements may involve error detection that mitigates the effect of noise, as in Ref. \cite{kielinski2024ghz}. 
Lastly, inquiring whether modifying the partitioning into GHZ states adaptively can improve the performance is also an intriguing future direction.

\section*{Methods}

\subsection*{Approximating optimal initial states}
\label{sec:approx_initial_state}

First, we explain how the optimally performing partition is computed for a given qubit number $N$, and prior width $\delta\phi$. For this purpose, we fix $N$ and write down all possible partitions into $\{k_i\}$,$\{m_i\}$, given the constraint $N = \sum_i m_i 2^{k_i}$. For all the partitions, we numerically calculate the minimum possible BMSE using the algorithm outlined in Ref. \cite{macieszczak_bayesian_2014}, assuming that optimal measurements (for example, the QFT operation followed by projective measurements) are available.
The partition that achieves this minimum possible BMSE, for a given qubit number $N$ and prior width $\delta\phi$, is referred to as the optimal partition. The optimal partition depends on the prior width: for a narrow prior distribution, it is desirable to use most of the qubits to construct a GHZ state with a high number of qubits, whereas for a wider prior distribution, we expect the optimal partition to include more blocks of GHZ states with smaller numbers of qubits. 

\begin{figure}
    \centering
\includegraphics[width=0.95\columnwidth]{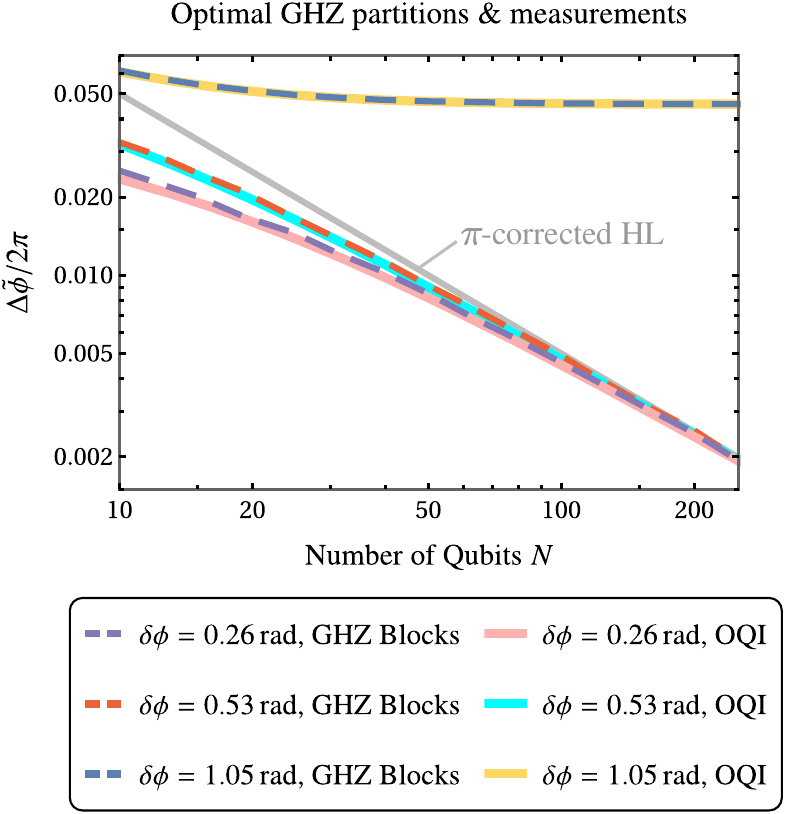}
    \caption{{\bf{Performance of optimal partitions given optimal measurements.}} Root Bayesian mean squared error (RBMSE) normalized by $2\pi$, as a function of the number of qubits $N$, for the optimal partition of the blocks of GHZ states, and the OQI.
    The OQI constitutes a benchmark for Bayesian phase estimation, and we assume that arbitrarily complex optimal measurements are available for the optimal partitions. 
    The comparison to the OQI shows that the BMSE of the optimal partitions perform almost optimally for all prior widths. We also plot the $\pi$-corrected HL, defined as $\Delta\tilde\phi = \pi/N$ in the limit of $N \gg 1$ \cite{jarzyna_true_2015}, with a gray line. The BMSE of the OQI (as well as the optimal partition) converges to the HL for all prior widths except $\delta\phi \approx 1.05$ rad, due to phase slip errors.}
\label{fig:bmse_comparison_oqi_ghz_block}
\end{figure}

It is interesting to study how well these optimal partitions of blocks of GHZ states perform given that optimal measurements are applied, and how close their minimal BMSE is to the BMSE of the OQI.
This analysis is shown in Fig. \ref{fig:bmse_comparison_oqi_ghz_block}, where we compare the BMSE obtained by the optimal partitions with their optimal measurements to the BMSE of the OQI for different prior widths ($\delta\phi = \pi/12 \approx 0.26$ rad, $\; \pi/6 \approx 0.53$ rad, $\; \pi/3 \approx 1.05$ rad). We observe that optimal partitions achieve the performance of the OQI almost exactly for all qubit numbers. Therefore, given optimal measurements, approximating the optimal initial state with blocks of GHZ states causes almost no loss of sensitivity.

To get a better analytical understanding of how close the sensitivity obtained with blocks of GHZ states can get to the OQI, we calculate the estimation error
obtained by the state containing $M=3$ copies of GHZ states with sizes $k = 0,1, \dots 2^{k_{\text{max}}}$. Such a state can be written as
\begin{align}
    \ket{\psi} = \frac{1}{2^{3(k_\text{max}+1)/2}} \sum_{n=0}^N \sqrt{f(n)} \ket{n},
\end{align}
where
$\ket{n}$ denotes the state with $n$ excited qubits, i.e. $n$ $\ket{1}$'s, that is symmetric under qubit-exchange operations.
For this state, $f(n)$ is the coefficient multiplying $x^n$ in the polynomial $\prod_{i=0}^{k_\text{max}} ( 1+ x^{2^i})^{3}$.
To simplify the analytical calculation we take a uniform prior distribution in
$[-\pi, \pi]$ and a cost function of $4\sin^2{((\phi-\phi_\text{est})/2)}.$
This cost function is commonly used for this prior distribution, due to the $\phi=\phi+2\pi$ symmetry, and it is equivalent to the standard
$(\phi-\phi_\text{est})^2$
for small $\phi-\phi_{\text{est}}.$
In Ref. \cite{kaftal2014usefulness}, 
it was shown that the estimation error in this case can be written as
\begin{align}
    (\Delta \widetilde{\phi})^2 = 2 \left[1 - \frac{1}{2^{3(k_\text{max}+1)}} \sum_{n=0}^{N-1} \sqrt{f(n) f(n+1)} \right].
\end{align}
Using this expression, we explicitly calculate $(\Delta \widetilde{\phi})^2$ and find that it reduces to
\begin{align}
    (\Delta \widetilde{\phi})^2 = \frac{9\sqrt{3} \ln{(2+\sqrt{3})}}{2N^2} + O(N^{-3}) \approx \frac{10.26}{N^2}.
\end{align}
The detailed calculation can be found in \suppmthreeproof.
Comparing this expression 
to the
OQI with this prior and cost function,
 $\pi^2/N^2$ \cite{berry_optimal_2000, gorecki2023heisenberg}, 
we get a very small overhead
of $\approx 1.04$. Then, in accordance with the numerical results in Fig. \ref{fig:bmse_comparison_oqi_ghz_block}, we observe that blocks of GHZ states can approach the optimal sensitivity limits significantly, if optimal measurements are available. 

We note that the optimal partitions differ from the ones used by previous schemes. The latter are tailored for non-adaptive, local measurements, whereas the partitions of this work assume optimal measurements. We observed numerically that in general these optimal measurements cannot be replicated with non-adaptive, local measurement protocols. 

\subsection*{Optimizing over local, adaptive measurements}
\label{sec:opt_adaptive_measurements}

Given the optimal partition, the optimal measurement strategy might be 
complex. For example, the optimal measurement includes performing a QFT in the limit of a large prior width within the dynamic range \cite{friis_flexible_2017}. We therefore look for ways to perform a measurement that approximates the optimal measurement as much as possible. It has been shown that the QFT operation, followed by projective measurements, can be approximated by performing adaptive measurements \cite{berry_optimal_2000}, where one starts measuring from the GHZ state with the highest $k_i$ in Eq. (\ref{eqn:general_ghz}), and performs single-qubit rotations to the rest of the states adaptively before their respective measurements. 

Taking the optimal partitions of the previous section as our initial states, the adaptive measurement can be optimized with respect to the initial state and the prior width to minimize the BMSE in Eq. (\ref{eqn:bmse}). First, defining $ M = \sum_i m_i$, note that one needs to perform $M$ measurements in total. For a parity measurement 
performed on a $k$-qubit GHZ state, preceded by
a single-qubit rotation
of $\Phi,$ the probabilities of obtaining an even or odd parity (denoted by $+,-$ respectively) are \mbox{$P(\pm) = (1 \pm \cos{(k(\phi-\Phi))})/2$}. Furthermore, since there are $M$ measurements, the number of possible outcomes is $2^M$. Denoting the probability of each outcome by $p(\vec{n}, \vec{\Phi} |\phi)$, we write the total BMSE as
\begin{align}
    (\Delta \tilde \phi)^2 = \int d\phi \, \mathcal{P}_{\delta\phi}(\phi) \sum_{\vec{n}} p(\vec{n}, \vec{\Phi} |\phi)(\phi-\phi_{est}(\vec{n}))^2
\end{align}
where the sum is over all outcomes, or branches $\vec{n}$. Note that the single-qubit rotations before measurements, $\vec{\Phi}$, and the estimators, $\phi_{est}(\vec{n})$, are functions of the branches. We use the optimal Bayesian estimators, defined as
\begin{align}
\label{eqn:opt_bayesian_est_mt}
\phi_{est}(\vec{n}) = \frac{\int d\phi \, \phi \, p(\vec{n}, \vec{\Phi} |\phi) \, \mathcal{P}_{\delta\phi}(\phi)}{\int d\phi \, p(\vec{n}, \vec{\Phi} |\phi) \, \mathcal{P}_{\delta\phi}(\phi)}
\end{align}
We employ gradient descent techniques to find the optimal single-qubit rotations $\vec{\Phi}$ for a given initial state and prior width (see \suppadaptivemeas\ for more details about the optimization procedure).

\section*{Data Availability}
Data is available from the corresponding author upon reasonable request.
\section*{Code Availability}
The codes used to generate the results are available from the corresponding
author upon reasonable request.

\section*{Acknowledgments}
We thank Elie Bataille, Yanbei Chen, Rafał Demkowicz-Dobrzański, Klemens Hammerer, and Raphael Kaubruegger for their helpful discussions. We acknowledge funding from the Army Research Office MURI program (W911NF2010136) and from the Institute for Quantum Information and Matter, an NSF Physics Frontiers Center (NSF Grant PHY-1733907). RF acknowledges support from the Troesh postdoctoral fellowship. TG acknowledges funding provided by the Institute for Quantum Information and Matter.

\section*{Author Contributions}
Conceptualization: S.D., T.G. Funding acquisition: R.F., M.E., T.G. Methodology: S.D., R.F., M.E., T.G. Writing---
original draft: S.D., T.G. Writing—review \& editing: S.D., R.F., M.E., T.G.
\section*{Competing Interests}
The authors declare no competing interests.

\appendix

\renewcommand{\figurename}{Supplementary Figure}
\renewcommand{\thefigure}{S\arabic{figure}}
\renewcommand{\theequation}{S\arabic{equation}}

\section*{Supplementary Note 1. Optimal quantum Bayesian protocol}
\label{app:opt_quantum}

In the main text, we calculate the minimum possible BMSE for phase estimation using the method presented in \cite{macieszczak_bayesian_2014}, which we refer to as the optimal quantum interferometer (OQI). For completeness, we present the main idea. During the interferometry experiment, an initial state $\rho=|\psi\rangle \langle \psi|,$
undergoes an evolution under the channel $\Lambda_{\phi}$ such that the final state is $\rho_{\phi}=\Lambda_{\phi}\left(\rho\right) =  U(\phi) \rho U^{\dagger}(\phi).$ In the context of phase estimation, the unitary transformation $U(\phi)$ is defined as $U(\phi) = e^{-i \phi J_z}.$
The BMSE, given a prior distribution $\mathcal{P}_{\delta\phi} \left( \phi \right)$,
reads
\begin{align}
\label{eqn:bmse_oqi_proof}
(\Delta \tilde{\phi})^2=\int d\phi\;dx\;\mathcal{P}_{\delta\phi}\left(\phi\right)\text{Tr}\left(\rho_{\phi}\Pi_{x}\right)\left(\hat{\phi}_{x}-\phi\right)^{2}\,    
\end{align}
where $\hat{\phi}_{x}$ is the estimator of $\phi$ given an outcome $x,$ and $\Pi_{x}$ is the corresponding projection operator. Note that it was proven in \cite{macieszczak_bayesian_2014} that we can restrict ourselves to projection operators. We can then define the operator $L=\int dx\Pi_{x}\hat{\phi}_{x}$ and $L^{2}=\int dx\Pi_{x}\hat{\phi}_{x}^{2}$. Plugging these operators in Eq. (\ref{eqn:bmse_oqi_proof}), we obtain
\begin{align}
(\Delta \tilde{\phi})^2&=(\delta \phi)^2+\text{Tr}\left(\Lambda_{\phi}\left(\rho\right)\int d\phi\;\mathcal{P}\left(\phi\right)\left(L^{2}-2\phi L\right)\right) \nonumber \\
&=(\delta \phi)^2+\text{Tr}\left[\rho\Lambda_{\phi}^{\dagger}\left(\int d\phi\;\mathcal{P}\left(\phi\right)\left(L^{2}-2\phi L\right)\right)\right],
\end{align}
where $(\delta \phi)^2$ is the variance of the prior distribution. To find the minimum possible BMSE, we need to optimize over the initial state, $\rho$, and the measurement and estimator operator, $L$. The optimization is done iteratively. For a given initial state $\rho$, the optimal $L$ is
\begin{align}
\left\{ L,\bar{\rho}\right\} =2\bar{\rho}', 
\end{align}
with
\begin{align}
    \bar{\rho}&=\int d\phi\,\mathcal{P}_{\delta\phi} \left(\phi\right)\Lambda_{\phi}\left(\rho\right) \nonumber \\
    \bar{\rho}'&=\int d\phi\,\mathcal{P}_{\delta\phi} \left(\phi\right)\Lambda_{\phi}\left(\rho\right)\phi.
\end{align}
Furthermore, for any $L$, the optimal $\rho$ is the eigenstate of
$\Lambda_{\phi}^{\dagger}\left(\int d\phi\;\mathcal{P}\left(\phi\right)\left(L^{2}-2\phi L\right)\right)$ with the maximal eigenvalue. Therefore, starting from a random initial state, the optimal $\rho$ and $L$ can be achieved by iteratively calculating (i) the optimal $L$ given $\rho$, (ii) the optimal $\rho$ given $L$. The iteration stops after the desired precision is achieved. In the end, the minimal BMSE is then
\begin{align}
(\Delta \tilde{\phi})^2 = \delta\phi^2-\text{Tr}(\bar\rho L^2)\,.
\label{eq:minimal_bmse}
\end{align}

\section*{Supplementary Note 2. Phase estimation for atomic clocks}
\label{sec:atomic_clocks}

\begin{figure*}
\raggedright\includegraphics[width=2.05\columnwidth]{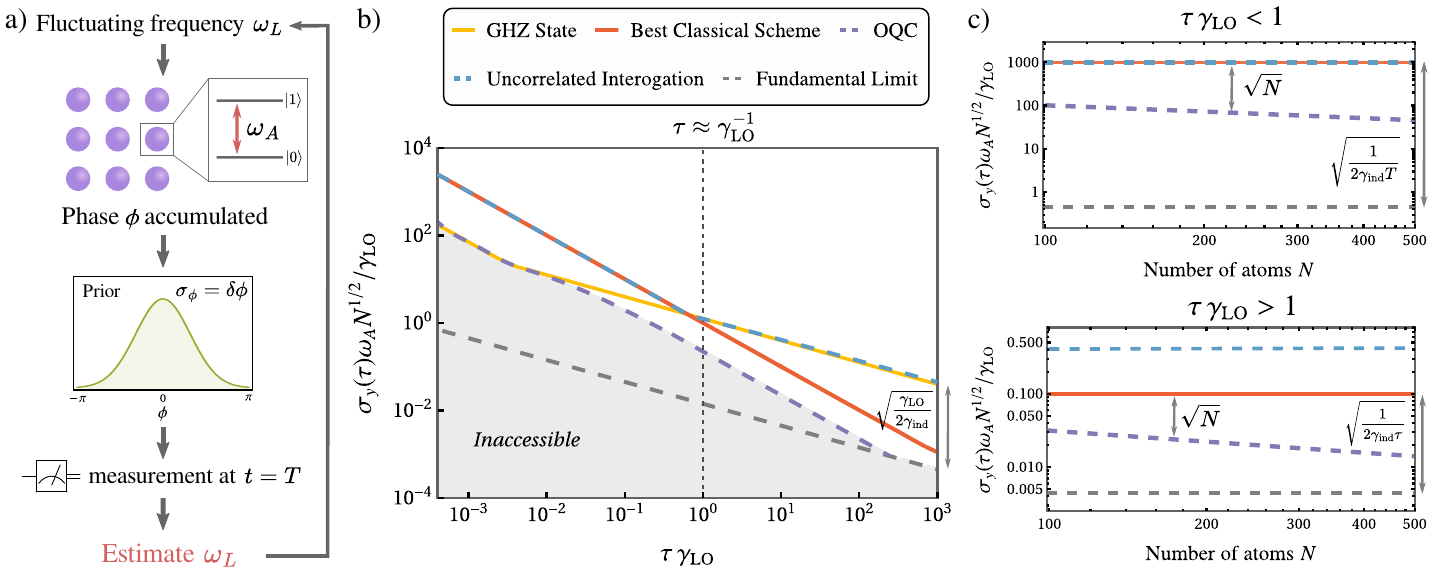}
    \caption{{Operation of an optical atomic clock in the Bayesian context.} (a) The fluctuating laser frequency $\omega_L(t)$ causes the atoms to accumulate a phase $\phi$. The phase is measured at the end of the interrogation cycle, and is used to estimate the laser frequency.
(b) Allan deviation $\sigma_y(\tau)$, normalized by $\gamma_{\text{LO}}/\omega_A N^{1/2}$, as a function of the total interrogation time $\tau$, for various interrogation schemes, and $N = 200$ atoms. 
    The vertical dashed line represents the coherence time limit due to local oscillator noise \cite{schulte_prospects_2020}, which determines the maximum Ramsey time $T$. The Allan deviation decreases as $\tau^{1/2}$ after this point for the uncorrelated and GHZ interrogation. If phase slip errors are suppressed, 
    it is possible for the stability to scale as $\tau^{-1}$, as seen from the stabilities of the best classical scheme and the optimal quantum clock (OQC). The fundamental limit of sensitivity is set by the individual particle decoherence (Eq. (\ref{eqn:fundamental_limit})), which is plotted with a gray, dashed line.
    (c) Allan deviation $\sigma_y(\tau)$, normalized by $\gamma_{\text{LO}}/\omega_A N^{1/2}$, as a function of the total number of atoms $N$, for $\tau = 10^{-3} \gamma_{\text{LO}}^{-1}$ (top) and $\tau = 10\,\gamma_{\text{LO}}^{-1}$ (bottom). Note that the legend for the plots is shown in b. For $\tau\,\gamma_{\text{LO}} < 1$, the stabilities of all the clocks scale as $N^{1/2}$. Protocols that show Heisenberg scaling (HS) can increase the clock stability by an additional factor of $N^{1/2}$. For $\tau\,\gamma_{\text{LO}} > 1$, $T$ takes its largest value of $T \approx \gamma_{\text{LO}}^{-1}$ for the uncorrelated interrogation, causing the stability to scale as $\tau^{1/2}$. Extending the Ramsey time with the best classical scheme enables to obtain a stability that scales as $\tau$, and using protocols that show HS, such as the OQC, can further boost the stability by a factor of $N^{1/2}$.}
    \label{fig:clock_fig}
\end{figure*}

An important application of phase estimation is in the context of optical atomic clocks \cite{Ludlow_opt_atomic_clocks}, where a local oscillator (LO) with a frequency $\omega_L(t)$, specifically a clock laser, is locked to an atomic transition with a frequency $\omega_A$. Due to the fluctuations in the laser frequency, $\omega_L(t)$ deviates from $\omega_A$, 
and a phase $\phi\coloneqq \int_0^{T} (\omega_A-\omega_L(t)) \, dt$ is accumulated. Here, $T$ is the Ramsey time, and it will also be referred to as
the interrogation time. 
Due to the fluctuating laser frequency, the accumulated phase $\phi$ is a stochastic variable, distributed according to a prior distribution with a prior width $\delta\phi$. For example, $\delta\phi$ grows with the Ramsey time as $\delta\phi \propto T$ for a laser frequency noise spectrum in the form of $S_\omega(f) \propto 1/f$ \cite{Leroux_2017}. At the end of the interrogation, the accumulated phase $\phi$ is estimated from a population measurement of the atoms. This estimate is consequently used to correct the deviations in the laser frequency. After the measurement, the uncertainty on $\phi$ is reduced from $\delta\phi$ to $\Delta\tilde\phi$, the posterior width, given in Eq. (\ref{eqn:bmse}).
Finally, the atoms are reinitialized
to perform a new interrogation. This process is described in Fig. \ref{fig:clock_fig}a. 
During the detection and initialization time, the laser frequency is not interrogated, and this interval is referred to as dead time. Phase diffusion during the dead time of atomic clocks gives rise to the Dick effect \cite{dick_effect}. 
This effect however can be circumvented by
introducing an additional ensemble of atoms that enables implementation of a zero-dead time (ZDT) clock \cite{biedermann_zero-dead-time_2013}. We ignore this effect in the following discussion.
The clock stability is characterized by the Allan deviation of the frequency, $\sigma_y(\tau)$. For a clock operation with $N_T$ interrogations, each having a Ramsey time $T$, the Allan deviation ideally follows \cite{kessler_heisenberg-limited_2014}
\begin{align}
\label{eqn:allan_dev}
    \sigma_y(\tau) = \frac{\Delta\Tilde{\phi}}{\omega_A \sqrt{\tau T}}  
\end{align}
where $\omega_A$ is the frequency of the atomic transition as mentioned above, $\tau = N_T \cdot T$ is the total duration of the clock operation, or the total interrogation time, and $\Delta\Tilde{\phi}$ is the posterior width \footnote{Here, we assume that $\Delta\Tilde\phi \ll \delta\phi$. If this assumption fails, the effective uncertainty defined in Eq. (\ref{eqn:effective_uncertainty}) in \suppallandev\ should be used instead of the posterior width as the numerator of Eq. (\ref{eqn:allan_dev}).}. Stable optical clocks are obtained through the minimization of $\sigma_y(\tau)$, therefore it is desirable to have long interrogation times $T$, and small posterior widths $\Delta\tilde\phi$. 
The posterior width $\Delta\tilde\phi$ depends on the initial state of the atoms, and the measurement strategy. For example, a coherent spin state (CSS), i.e. uncorrelated atoms, can be used as the initial state: an $N$-qubit CSS is defined as $\ket{\psi_{\text{CSS}}} = \left[(\ket{0}+\ket{1})/\sqrt{2}\right]^{\otimes N}$. Phase estimation with a CSS results in the SQL, where the posterior variance scales as $\Delta\tilde\phi \propto 1/\sqrt{N}$. Using entanglement, it is possible to achieve a better scaling with $N$. For example, a GHZ state can be used as an initial state, where an $N$-qubit GHZ state reads $\ket{\psi_{\text{GHZ}}} = (\ket{0}^{\otimes N} + \ket{1}^{\otimes N})/\sqrt{2}$. An N-qubit GHZ state achieves a posterior variance of \mbox{$\Delta\tilde\phi \propto1/N$}, for $\delta\phi \approx \pi/N$ or smaller.

As for the interrogation time $T$, it is typically limited by the LO noise, i.e.
by $\gamma_{\text{LO}},$
the linewidth of the free-running laser.
For $T$ longer than
$\gamma_{\text{LO}}^{-1}$, $\delta \phi \approx \pi$,
leading to phase slip errors, i.e. phase estimation error arising from $\phi$ being outside of the interval $\left[-\pi, \,\pi \right]$. Since $\phi$ can be estimated up to modulo $2\pi$, the phase estimation results in a $2\pi k$ error, $k \geq 1,\,k \in \mathbb{Z}$.
These errors degrade the accuracy in phase estimation and consequently the Allan deviation. Therefore, for a naive clock operation using uncorrelated atoms, the maximum allowable Ramsey time is $T \approx \gamma_{\text{LO}}^{-1}$. Consequently, we see two scalings of the Allan deviation emerge as a function of $\tau$: for $\tau < \gamma_{\text{LO}}^{-1}$, one can set 
$T = \tau$, such that $\sigma_y(\tau) \propto \tau^{-1}$. For $\tau > \gamma_{\text{LO}}^{-1}$, the Ramsey time is fixed to its maximum value, and $\sigma_y(\tau) \propto \tau^{-1/2}$ due to averaging over the many independent interrogations. 

In a practical setting, optical atomic clocks are operated in the regime of $\tau > \gamma_{\text{LO}}^{-1}$. Since GHZ states perform optimally for $\delta\phi < \pi/N$, their maximum allowable Ramsey time is $T \approx (\gamma_{\text{LO}}N)^{-1}$, which is a factor of $N$ shorter compared to the case of an $N$-qubit CSS. Then, the clock that uses an $N$-qubit GHZ state achieves a similar stability as the clock that uses an $N$-qubit CSS \footnote{GHZ states achieve slightly smaller posterior variances, see Ref. \cite{kaubruegger_quantum_2021} or Fig. \ref{fig:clock_fig}b}, providing no metrological gain.
Therefore, optimal metrology seeks (i) protocols that extend the Ramsey time $T$ beyond the LO noise limit, $\gamma_{\text{LO}}^{-1}$, (ii) protocols that show Heisenberg-scaling with respect to the number of atoms $N$, given a wide prior width, to achieve metrological advantage.

The first objective can be accomplished by introducing additional classical interrogations with uncorrelated atoms and performing phase unwinding \cite{rosenband2013exponential, Borregaard_efficient_atomic}. Phase unwinding is achieved by using
ensembles of atoms that acquire smaller phases (which can be achieved by e.g. mid-circuit measurement and reset). These ensembles 
provide the ability to correct for phase slip errors, and thus, to extend the interrogation time such that $T = \tau$. We refer to the classical scheme that extends the Ramsey time in this way as the \textit{best classical scheme}, and the clock that uses this scheme as the \textit{best classical clock}.

Furthermore, the second objective can be achieved with the OQI. The OQI achieves the minimum possible posterior variance given any interrogation time (see \suppoptbayesian). We refer to the clock that combines the OQI with phase unwinding techniques beyond the LO noise limit as the optimal quantum clock (OQC): the OQC achieves the maximum possible clock stability given $N$ qubits, for any interrogation time.

Ultimately, the clock stability is limited by the individual particle decoherence arising from amplitude damping (comprised of scattering from trapping light and radiative decay), which restricts the linewidth of the atomic clock transition. The fundamental limit to phase sensitivity is given by \cite{Escher2011, demkowicz2014using}
\begin{align}
\label{eqn:fundamental_limit}
    \sigma_{y, \text{min}}(\tau) \approx \frac{1}{\omega_A}\sqrt{\frac{2\gamma_{\text{ind}}}{\tau N}}
\end{align}
where $\gamma_{\text{ind}}$ is the rate of amplitude damping. Note that the fundamental limit scales as $N^{-1/2},$
where neither entanglement nor any classical protocol can retrieve the Heisenberg scaling \cite{Escher2011, demkowicz2014using}.

We illustrate the different regimes and the possible gain from quantum metrology in Fig. \ref{fig:clock_fig}. For a fixed $\gamma_{\text{LO}}/\gamma_{\text{ind}} = 10^4$, we plot in Fig. \ref{fig:clock_fig}b the Allan deviation $\sigma_y(\tau)$, with respect to the total interrogation time $\tau$, for different protocols (see \suppallandev for more details). From the Figure, the stability of all of the protocols scales as $\tau^{-1}$ for small total interrogation times $\tau$. However,
the Allan deviation
scales as $\tau^{-1/2}$
for $\tau > \gamma_{\text{LO}}^{-1}$, for uncorrelated atoms and the GHZ state. This leads to a factor of $\gamma_{\text{LO}}^{1/2}/(2\gamma_{\text{ind}})^{1/2}$ gap from the fundamental sensitivity limit given in Eq. (\ref{eqn:fundamental_limit}), and plotted with gray dashed lines in Fig. \ref{fig:clock_fig}b. 
The stability of
the best classical clock, and the OQC, continue to scale as $\tau^{-1}$
for times shorter than the single-qubit decoherence limit $(\tau<\gamma_{\text{ind}}^{-1})$, as they assume the use of phase unwinding. Furthermore, the OQC surpasses the best classical clock since it shows HS in this time scale. The area below the OQC is marked as \textit{inaccessible}, since it sets the benchmark for optimal sensitivity. Note that we assumed $\gamma_{\text{ind}} = 0$ when computing the performance of the OQC to simplify the calculations.
As the total interrogation time $\tau$ approaches $\gamma_{\text{ind}}^{-1}$, the Allan deviation for all of the schemes start scaling as $\tau^{-1/2}$ due to the single-qubit noise. 

In Fig. \ref{fig:clock_fig}c, we plot the Allan deviation $\sigma_y(\tau)$, normalized by $\gamma_{\text{LO}}/\omega_A N^{1/2}$, with respect to the total number of atoms $N$, for $\tau = 10^{-3} \gamma_{\text{LO}}^{-1}$ (top) and $\tau = 10\,\gamma_{\text{LO}}^{-1}$ (bottom). For $\tau < \gamma_{\text{LO}}^{-1}$, the uncorrelated interrogation has the same stability as the best classical scheme. Since the OQC shows HS, the stability increases faster by a factor of $N^{1/2}$ compared to the other schemes and gets closer to the fundamental limit of sensitivity. For $\tau > \gamma_{\text{LO}}^{-1}$, the best classical clock performs better than the clock using uncorrelated interrogations by a factor of $(\tau \gamma_{\text{LO}})^{1/2}$ due to its extended Ramsey time. On top of this, the OQC surpasses the best classical clock by a factor of $N^{1/2}$, due to its extended Ramsey time, and its HS. In summary, the maximum gain in sensitivity obtained by the OQC with respect to the sensitivity of the uncorrelated interrogation is $\sqrt{N}$ in the regime where $\tau\gamma_{\text{LO}} < 1$, and $\sqrt{N} \sqrt{\tau\gamma_{\text{LO}}}$ in the regime where $\tau\gamma_{\text{LO}} > 1$, respectively.

\section*{Supplementary Note 3. Allan deviation for various clocks}
\label{app:allan_dev}

To calculate the clock stability for a given protocol, we need to find the optimal interrogation time $T$ to operate with. This optimal interrogation time is a function of the total interrogation time $\tau$. For an uncorrelated interrogation, $T = \tau$ for $\tau < \gamma_{\text{LO}}^{-1}$, and $T \approx \gamma_{\text{LO}}^{-1}$ for $\tau > \gamma_{\text{LO}}^{-1}$. Here, we calculate the exact value of $T$ numerically.

The unknown phase $\phi$ of the LO is a stochastic variable, sampled from a prior distribution with a variance of $(\delta\phi)^2$. The variance grows with the interrogation time $T$, in the form of $\delta\phi = \gamma_{\text{LO}} T$, for a $1/f$ laser noise frequency spectrum. Therefore, the probability of a phase slip occurring increases with $T$. However, the clock stability $\sigma_y(\tau)$ also increases with $T$, which results in an optimal $T$ to balance this trade-off. 

We assume a posterior variance $(\Delta\Tilde\phi)^2$ for a clock protocol, in absence of phase slips. The broadening of the posterior variance due to phase slips can be represented by 
\begin{align}
    (\Delta\phi)^2_{\text{slip}} &\approx 2\sum_{i=1}^\infty (2 k \pi)^2 \int_{(2 k-1)\pi}^{(2 k+1)\pi} d\phi \, \frac{1}{\sqrt{2\pi (\delta\phi)^2}} e^{- \frac{\phi^2}{2(\delta\phi)^2}} \nonumber \\
    &= \sum_{k=1}^\infty (2k\pi)^2 \left[ \text{erf}\left(\frac{(2k+1)\pi}{\sqrt{2}\delta\phi} \right) - \text{erf}\left(\frac{(2k-1)\pi}{\sqrt{2}\delta\phi} \right) \right]
\end{align}
for $(\Delta\Tilde\phi)^2 \ll 1$, or $N \gg 1$. Then, assuming that this effect is independent for each clock cycle, $\sigma_y(\tau)$ is calculated as 
\begin{align}
\label{supp_eq:allan_dev}
    \sigma_y(\tau) = \frac{1}{\omega_A \sqrt{\tau T}} \left[ (\Delta\Tilde\phi)^2 + \frac{\tau}{T} (\Delta\phi)^2_{\text{slip}}\right]^{1/2}
\end{align}
Then, the optimal interrogation time $T$ is found from minimizing Eq. (\ref{supp_eq:allan_dev}), resulting in the minimum Allan deviation
\begin{align}
\label{supp_eq:allan_dev_min}
 \sigma_y^*(\tau) = \min_{T} \sigma_y(\tau)
\end{align}
To obtain the Allan deviation curve in Fig. \ref{fig:clock_fig}b and Fig. \ref{fig:clock_fig}c for uncorrelated atoms, we plot Eq. (\ref{supp_eq:allan_dev_min}) for $N=200$. We find numerically that the Allan deviation, normalized by $N^{1/2}$ is approximately independent of $N$.

For a clock interrogation using an $N$-atom GHZ state, we take a different approach. First, we redefine Eq. (\ref{eqn:allan_dev}) in \suppclocks\ for short interrogation times, where the posterior variance is close to the prior width, i.e., $\Delta\Tilde\phi \approx \delta\phi$. In this limit, the assumption that we made in \suppclocks\ ($\Delta\Tilde\phi \ll \delta\phi$) breaks down, and the uncertainty of a stable clock is given by $\Delta\tilde\phi_{\text{eff}}$, with
\begin{align}
\label{eqn:effective_uncertainty}
    (\Delta\tilde\phi_{\text{eff}})^2 = \left[ (\Delta\tilde\phi)^{-2} - (\delta\phi)^{-2} \right]^{-1}
\end{align}
which we refer to as the \textit{effective uncertainty}.
The numerator in Eq. (\ref{eqn:allan_dev}) in \suppclocks\ is therefore given by this $\Delta\tilde\phi_{\text{eff}}$.
Now, assume a parity measurement on the $N$-atom GHZ state: the measurement has two outcomes, even or odd parity. The probabilities of obtaining these results are given by $P(\pm) = (1 \pm \cos{(N \phi)})/2$. Then, using optimal estimators (see Eq. (11) in the main text), the BMSE is calculated as
\begin{align}
\begin{split}
    &(\Delta\Tilde\phi)^2 = (\delta\phi)^{2}-2\frac{\left(1/2\underset{-\infty}{\overset{\infty}{\int}}\phi\sin\left(N\phi\right)\mathcal{P}_{\delta\phi}\left(\phi\right)\;\text{d}\phi\right)^{2}}{1/2} \\
    &=(\delta\phi)^2 - N^2 (\delta\phi)^4 e^{-N^2 (\delta\phi)^2}
\end{split}
\end{align}
Plugging in $\delta\phi = \gamma_{\text{LO}} T$, the effective uncertainty is given by
\begin{align}
\label{eqn:effective_unc_ghz}
    \Delta\tilde\phi_{\text{eff}} = \frac{\left(e^{N^2 (\gamma_{\text{LO}} T)^2}-N^2 (\gamma_{\text{LO}} T)^2 \right)^{1/2}}{N}
\end{align}
Then, the optimal Allan deviation for a GHZ interrogation can be obtained from minimizing $\sigma_y(\tau)$ with respect to the interrogation time $T$, given the effective uncertainty in Eq. (\ref{eqn:effective_unc_ghz}). In Fig. \ref{fig:clock_fig}b, we plot the Allan deviation, normalized up to some prefactors, for an $N = 200$ atom GHZ state. 

When slow atoms are introduced to a given clock protocol, we can extend the interrogation time $T$ beyond the LO noise limit, i.e. $T = \tau$ even in the regime where $\tau > \gamma_{\text{LO}}^{-1}$. Here, we assume that enough slow atoms are introduced such that the phase slip probability is small, and that the number of slow atoms is negligible compared to the initial number of atoms $N$. Therefore, in this regime, the Allan deviations of the best classical clock and the OQC are given by $1/\omega_A \sqrt{N} \tau$, and by $ \pi/\omega_A N \tau$, respectively. The stabilities of these clocks eventually reach the fundamental limit of sensitivity (up to a possible constant factor of order one, see Ref. \cite{Escher2011}), given in Eq. (\ref{eqn:fundamental_limit}) in \suppclocks.

\section*{Supplementary Note 4. Analysis of the scheme with a fixed block size}
\label{app:lukin_bit_by_bit}

\begin{figure}
    \centering
\includegraphics[width=0.95\columnwidth]{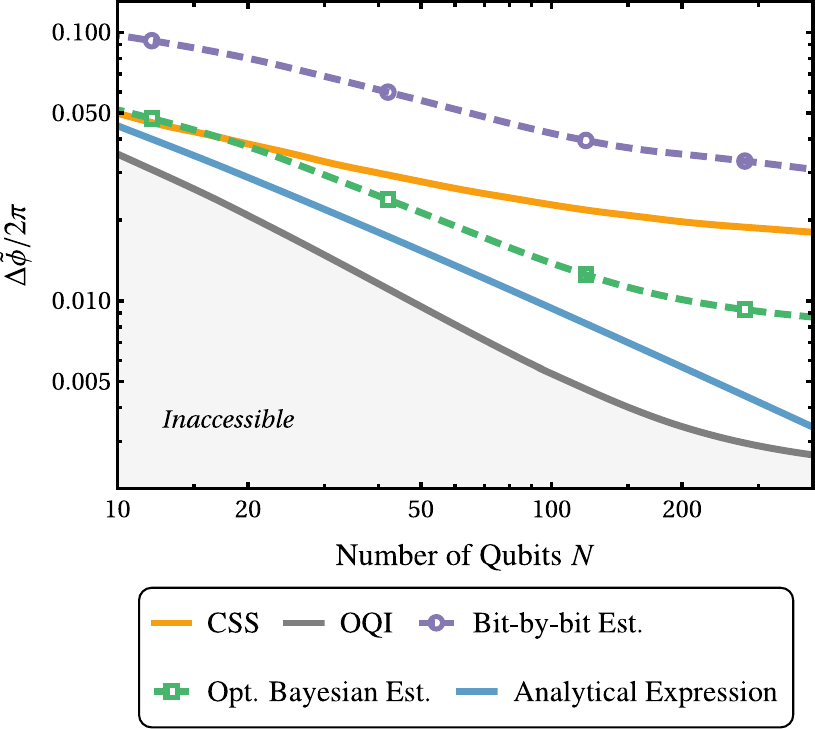}
    \caption{RBMSE $\Delta \tilde \phi$ of the OQI, the CSS, and the scheme with a fixed block size \cite{kessler_heisenberg-limited_2014} using different estimators, normalized by $2\pi$, as a function of the number of qubits, $N$. We work with a prior width of $\delta\phi = 0.7$ rad. We observe that the bit-by-bit estimator performs significantly worse than the optimal Bayesian estimator, failing to surpass the SQL. For intermediate qubit numbers, the optimal Bayesian estimator scales as the expected performance of the scheme calculated analytically in Ref. \cite{kessler_heisenberg-limited_2014}, shown with the blue curve. However, we observe a numerical overhead between the analytical expression and the observed performance of the scheme in this regime, which is calculated to be 1.38.}
    \label{fig:lukin_bitbybit}
\end{figure}

Here, we perform a detailed analysis of the scheme with a fixed block size \cite{kessler_heisenberg-limited_2014}. In this scheme, we have $M$ copies of GHZ states with $2^k$ qubits, $k = 0, 1, \dots k_{\text{max}}$, and $M$ is given implicitly by $M = \frac{16}{\pi^2}\, \text{log}\,(M(2^{k_\text{max}+1}-1))$. Note that $M$ is rounded to the closest integer to the solution of this implicit equation. Ref \cite{kessler_heisenberg-limited_2014} derives that this protocol leads to an RBMSE
of $\Delta\tilde\phi = \frac{8}{\pi} \sqrt{\text{log}(N)}/N$. To see why there is a logarithmic correction on the BMSE, we calculate the minimum achievable posterior variance using these initial states, from the Quantum Cramer-Rao Bound (QCRB). We find that this bound gives the following lower bound on the MSE
\begin{align}
    (\Delta \phi)^2 \geq \left[ M \sum_{k=0}^{k_{\text{max}}} 2^{2k} \right]^{-1} \approx \frac{3 M}{N^2} = \frac{48}{\pi^2} \frac{\text{log}(N)}{N^2}
\end{align}
where $N \approx M\,2^{k_{\text{max}}+1}$ is the total number of qubits, in the limit of $2^{k_{\text{max}}+1} \gg 1$. Hence, one can see that the logarithmic correction to the MSE arises from the fact that the number of copies of the GHZ states, $M$, is logarithmic in the total number of qubits, $N$.
Now, we analyze two types of estimators
and compare their performance in phase estimation.

The first estimator is outlined in Ref. \cite{kessler_heisenberg-limited_2014}, which we refer to as the ``bit-by-bit estimator". For this estimator, we first note that for $\phi$ in $\left[-\pi, \, \pi \right]$, $\phi$ can be written in the binary basis as
\begin{align}
    \phi = 2\pi (0.Z_1 Z_2 Z_3 \dots) \;\; \text{mod}\; \left[-\pi, \, \pi \right]
\end{align}
where $Z_i \in \{0, 1\}$ are the binary digits. The bit-by-bit algorithm then proceeds as follows: starting from the $2^{k_\text{max}}$-qubit GHZ state, for all GHZ states with $2^k$ qubits, we first perform a dual quadrature parity measurement, where $M/2$ copies are subjected to a parity measurement in the $X$ or $Y$ basis. We denote $n_{k,x}$ and $n_{k,y}$ as the number of measurements where an even parity was obtained for the $X$ and the $Y$ basis respectively.
Then, the estimator for $\phi_k := 2^k \phi \;\text{mod}\; \left[-\pi, \, \pi \right]$ is given by \begin{align}
\label{app_eqn:dual_quad}
    \hat \phi_k = \text{arg}\left( \left(\frac{2n_{k,x}}{M} - \frac{1}{2} \right) + i \left(\frac{2n_{k,y}}{M} - \frac{1}{2} \right) \right)
\end{align}
Note that we define the probabilities of the outcomes of a parity measurement on a $2^k$-qubit GHZ state in the $X$ basis as \mbox{$P(\pm) = (1 \pm \cos{(2^k\phi)})/2$}, where $\{ +, -\}$ denote even, odd parity, hence the estimator in Eq. (\ref{app_eqn:dual_quad}) follows. After computing all such $\hat \phi_k$, we estimate the $j^{\text{th}}$ bit $Z_j$ of $\phi$ as
\begin{align}
    \hat Z_j = \frac{2\hat \phi_{j-1} - \hat \phi_{j}}{2\pi}
\end{align}
for $1 \leq j \leq k_{\text{max}}$. Combining all such bits, the bit-by-bit estimator is in the form of 
\begin{align}
    \hat \phi = \left( 2\pi (0.\hat Z_1 \hat Z_2 \dots \hat Z_{k_{\text{max}}}) + \frac{\hat \phi_{k_\text{max}}}{2^{k_\text{max}}} \right) \;\; \text{mod}\; \left[-\pi, \, \pi \right]
\end{align}

The second estimator is the optimal Bayesian estimator, defined for a measurement outcome $k$ as
\begin{align}
\label{app_eqn:opt_bayes_estimators}
\phi_{est}^*(k) = \frac{\int \phi \, p(n |\phi) \mathcal{P}_{\delta\phi}(\phi) d\phi}{\int p(n |\phi) \mathcal{P}_{\delta\phi}(\phi) d\phi}
\end{align}
where $p(n |\phi)$ is the probability of obtaining the corresponding measurement outcome, and $\mathcal{P}_{\delta\phi}(\phi)$ is the prior distribution. For this scheme, the probabilities of measurement outcomes are given for $n = \{n_{k,x},n_{k,y}\}$ as
\begin{align}
    p(\{n_{k,x},n_{k,y}\}|\phi) = \frac{1}{2^{M(k_\text{max}+1)}}\prod_{k=0}^{k_\text{max}} & (1 + \cos{(2^k\phi)})^{n_{k,x}} \nonumber \\ & (1 - \cos{(2^k\phi)})^{\frac{M}{2}-n_{k,x}} \nonumber \\ & (1 + \sin{(2^k\phi)})^{n_{k,y}} \nonumber \\ & (1 - \sin{(2^k\phi)})^{\frac{M}{2}-n_{k,y}}
\end{align}
where
$n_{k,x}$ ($n_{k,y}$) is the number of parity measurement outcomes of the $k$-qubit GHZ states that resulted in an even parity, when measured in the $X$($Y$) basis respectively.
Therefore $n_{k,x},n_{k,y} \leq M/2.$ 

In our simulations, we numerically computed the optimal Bayesian estimators and the BMSE for all qubit numbers. We used $ \approx 10^7-10^8$ phase samples drawn from the prior distribution. For the bit-by-bit estimation, we drew $\approx 200-300$ samples from the prior distribution and simulated the estimation scheme for these phases $\approx 10^5-10^6$ times with the bit-by-bit estimator. We saw that the BMSE converged with these numbers.

We compare the performances of the two estimators in Fig. \ref{fig:lukin_bitbybit} by plotting the RBMSE obtained by them as a function of the number of qubits $N$, for a prior width of $\delta\phi = 0.7$ rad. We also plot the OQI as the benchmark for sensitivity, and the RBMSE obtained by the CSS, which constitutes the SQL. We observe that the bit-by-bit estimator performs significantly worse than the optimal Bayesian estimator and the SQL, whereas the optimal Bayesian estimator achieves sub-SQL sensitivity. Ref. \cite{kessler_heisenberg-limited_2014} calculated the expected performance of the scheme analytically as $\Delta\tilde\phi = \frac{8}{\pi} \sqrt{\text{log}(N)}/N$, assuming a uniform prior distribution, and a large qubit number $N$. Since the $\pi$-corrected Heisenberg limit for a uniform prior distribution in the limit of a large qubit number is given by $\Delta\tilde\phi_{\text{HL}} = \pi/N$, comparing this with the analytical expression for the RBMSE of the scheme with a fixed block size, we assume that the scheme has an overhead of $\frac{8}{\pi^2} \sqrt{\text{log}(N)}$, compared to the RBMSE of the OQI. We plot this with a blue curve in Fig. \ref{fig:lukin_bitbybit}, where optimal fit to the performance of the OQI for this prior width is calculated as $\Delta\tilde\phi_{\text{OQI}} = 1.55/N^{0.83}$ for intermediate qubit numbers. We observe that the scheme with a fixed block size, combined with the optimal Bayesian estimator, scales similarly with the analytical expression for intermediate qubit numbers (see $N = 42,\,N = 120$ in the Figure). However, we observe that the analytical expression surpasses the observed performance by a fixed factor of $\approx 1.38$ in STD for these qubit numbers. For larger qubit numbers, the sensitivity of the scheme starts saturating for both estimators. Therefore, this analysis highlights the importance of the estimator for phase estimation with this scheme.

\section*{Supplementary Note 5. Analysis of the scheme with a varying
block size}
\label{app:wiseman}

Here, we perform a detailed analysis of the scheme with a varying block size \cite{higgins_demonstrating_2009}. As described in the main text, the initial states employed by this protocol are parametrized by two parameters $(m_{k_{\text{max}}},\,\mu)$. For an initial state of this protocol containing GHZ states with $2^k$ qubits, $k = 0, 1, \dots, k_{\text{max}}$, the $2^{k_{\text{max}}}$-GHZ state has $m_{k_{\text{max}}}$ copies, and the $2^{k_{\text{max}}-1}$-GHZ state has $m_{k_{\text{max}}}+\mu$ copies, etc. 
Searching over such states, Ref. \cite{higgins_demonstrating_2009} observed that the maximal sensitivity is obtained for the pair $(m_{k_{\text{max}}},\, \mu) = (2, 3)$. For the measurement strategy, the $m_k$ copies of the $2^k$-qubit GHZ states go under a set of single-qubit rotations $\theta_j = 0, \pi/m_i, \dots, \pi (m_i-1)/m_i$. The estimation strategy in Ref. \cite{higgins_demonstrating_2009} was chosen to minimize the Holevo variance (which implicitly assumes a uniform prior in $[-\pi, \pi]$). Since we study Bayesian estimation with a Gaussian prior, our optimal estimators are different and we compute these Bayesian estimators numerically. For this purpose, we write down the probability of obtaining a measurement outcome $n = \{n_k\}$ as
\begin{align}
    p(n |\phi) &= \prod_{k = 0}^{k_{\text{max}}} p(n_k |\phi) \nonumber \\
    p(n_k |\phi) &= \prod_{i = 0}^{m_k-1} \frac{1}{2} \left[1 + n_{k,i} \cos{\left(2^k \phi - \pi\frac{i}{m_k} \right)} \right]
\end{align}
with $m_k = m_{k_{\text{max}}} + \mu (k_{\text{max}}-k)$, and $n_{k,i} = \pm 1$ is the result of the parity measurement, with $+1$ ($-1$) denoting even (odd) parity, and $0 \leq k \leq k_{\text{max}},\, 0 \leq i \leq m_k$.

\begin{figure}
    \centering
    \includegraphics[width=0.95\columnwidth]{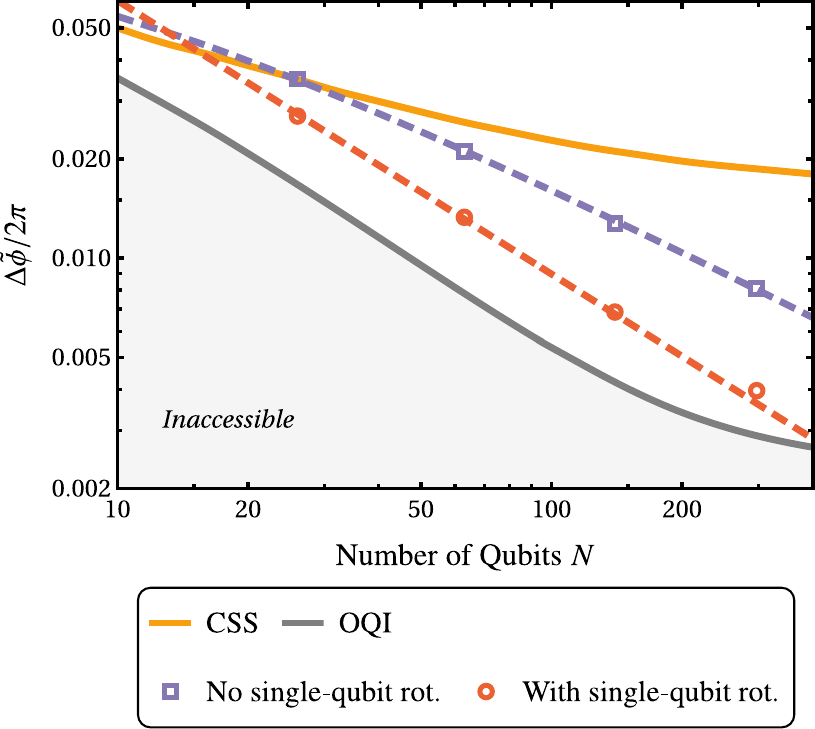}
    \caption{RBMSE $\Delta \tilde \phi$ of the OQI, the CSS, and two variations of the scheme with a varying block size \cite{higgins_demonstrating_2009}, normalized by $2\pi$, as a function of the number of qubits, $N$. We again work with a prior width of $\delta\phi = 0.7$ rad, and use the optimal Bayesian estimators for all of the schemes in the Figure. We plot the RBMSE of the scheme with a varying block size with and without single-qubit rotations with the red and purple data points respectively. We observe that the scheme with a varying block size performs better when single-qubit rotations are employed. Furthermore, we observe that the RBMSE of this scheme (with single-qubit rotations) scales as the OQI with an overhead of 1.66, plotted with the red, dashed line.}
    \label{fig:wiseman_scaling}
\end{figure}

\begin{figure}
    \centering
    \includegraphics[width=0.9\columnwidth]{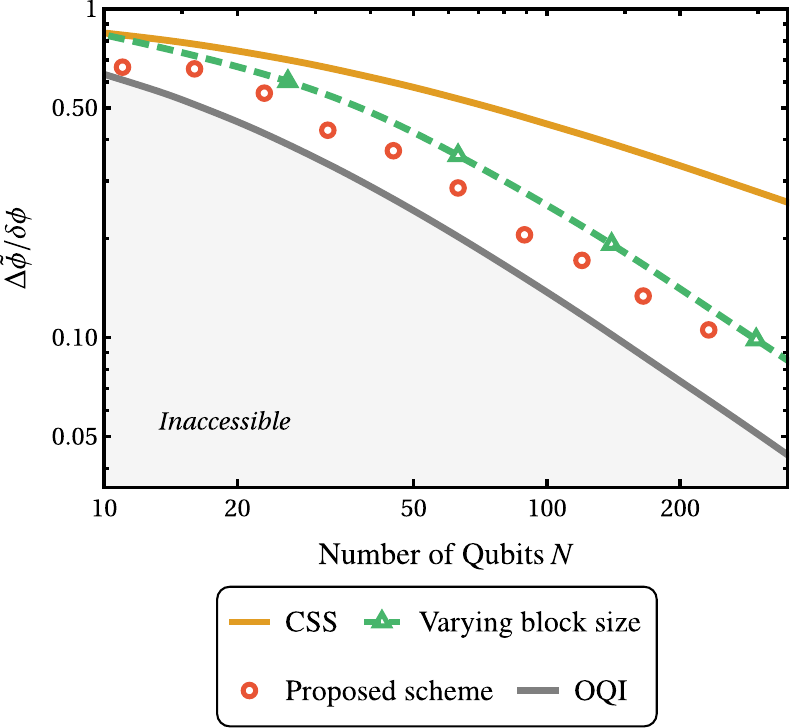}
    \caption{RBMSE $\Delta \tilde \phi$ of the OQI, the CSS, the scheme with a varying block size \cite{higgins_demonstrating_2009}, and the proposed scheme, normalized by the prior width $\delta\phi$, as a function of the number of qubits, $N$. We work with a prior width of $\delta\phi = 0.2$ rad, smaller than that of Fig. 5 in the main text, and observe that the sensitivity gap between the scheme with a varying block size and the proposed scheme grows as the prior width decreases. For example, for a qubit number of $N = 63$, the posterior variance of the proposed scheme is smaller than that of the scheme with a varying block size by a factor of 1.57 (1.97 dB), compared to a factor of 1.22 that we obtained for a prior width of $\delta\phi = 0.7$ rad . }
    \label{fig:wiseman_vs_us_0.2}
\end{figure}

We plot the RBMSE for $\delta\phi = 0.7$ rad of two variations of 
this scheme in Fig. \ref{fig:wiseman_scaling}: first, with red data points, we plot the scheme as it was described above. With purple data points, we plot a modified version of the scheme which uses the same partitions but no single-qubit rotations are performed in the measurement. The modified version was recently implemented experimentally in Ref. \cite{cao2024multiqubit}, using different partitions. 
We observe a gap between the two readout strategies, hence single-qubit rotations are needed for optimal precision. From our numerical results, the gap scales as $0.39\,\text{log}(N)$ in STD, where $N$ is the number of qubits. For $N = 297$, using single-qubit rotations in the readout provides a metrological gain of 4.16 (14.3 dB) in posterior variance over the readout without single-qubit rotations.
We also observe that the RBMSE of the scheme with single-qubit rotations scales as that of the OQI, with a constant overhead of 1.66.

The scheme with a varying block size performs similar to the proposed scheme in the large prior width within the dynamic range regime (see Fig. 5 in the main text, where we plot the RBMSE of both schemes for a prior width of $\delta\phi = 0.7$ rad). To see if this is the case for smaller prior widths, we compare the RBMSE of these schemes for a prior width of $\delta\phi = 0.2$ rad in Fig. \ref{fig:wiseman_vs_us_0.2}. We observe that the RBMSE of the proposed scheme and the scheme with a varying block size show similar scaling with respect to the number of qubits $N$ for this prior width, however, the sensitivity gap between the schemes is bigger compared to the gap that was observed for for a prior width of $\delta\phi = 0.7$ rad. For example, the proposed scheme surpasses the scheme with a varying block size in posterior variance by a factor of 1.57 (1.97 dB) for $N= 63$ qubits, compared to a factor of 1.22 (0.85 dB) that we observed for the same number of qubits, but for a prior width of $\delta\phi = 0.7$ rad. 

\section*{Supplementary Note 6. Suitable qubit numbers for the schemes with a fixed and a varying block size}
\label{app:suitable_qubit_numbers}

\begin{figure}
    \centering
    \includegraphics[width=0.9\columnwidth]{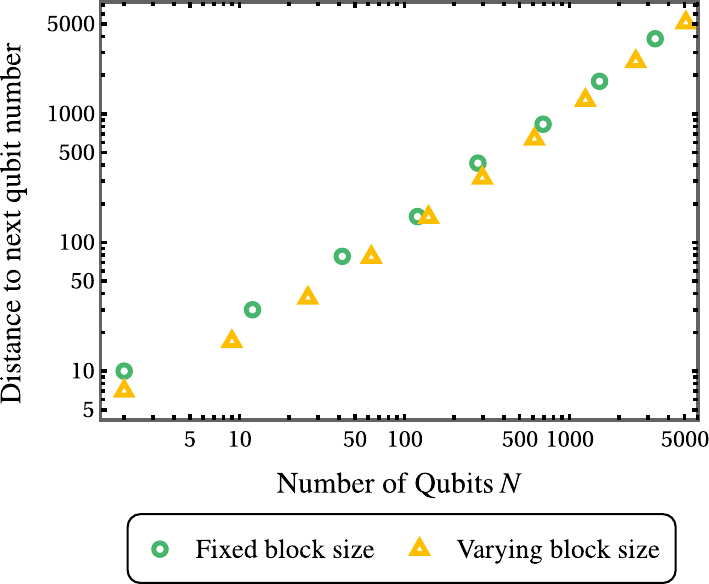}
    \caption{ 
    Difference between consecutive number of qubits, $N\left(k_{\text{max}}+1\right)-N\left(k_{\text{max}}\right)$
    for the two schemes (Refs. \cite{kessler_heisenberg-limited_2014}, \cite{higgins_demonstrating_2009}), as a function of $N.$ 
    This difference is the number of extra qubits needed to obtain the next initial state proposed by the given protocols.
    The differences grow approximately linearly with the total number of qubits for both schemes, signifying that with it gets more and more difficult to find an $N$ for which these schemes are defined, with increasing $N$.}
    \label{fig:suitable_qubit}
\end{figure}

As mentioned in the main text and in \supplukin\ and \suppwiseman, the schemes with a fixed and a varying block size are defined only for specific values of qubit numbers. In both schemes, there is a unique $N,$ total number of qubits, that corresponds to each $k_\text{max},$ let us denote it as $N \left( k_\text{max} \right).$  
For the scheme with a fixed block size, $N\left( k_\text{max} \right) = M\cdot (2^{k_\text{max}+1}-1)$, where $M$ is the nearest integer to the solution of the implicit equation $M=\frac{16}{\pi^2}\, \text{log}\,(M(2^{k_\text{max}+1}-1))$. For the scheme with a varying block size, the suitable qubit numbers are given by $N\left( k_\text{max} \right) = \sum_{k=0}^{k_\text{max}} [2 + 3(k_\text{max}-k)] 2^k = 5\cdot 2^{k_\text{max}+1} -3 k_\text{max} - 8$.
Therefore in both cases the difference between consecutive number of qubits, $N\left(k_{\text{max}}+1\right)-N\left(k_{\text{max}}\right)$, 
grows linearly with $N\left(k_{\text{max}}\right)$:
$N\left(k_{\text{max}}+1\right)-N\left(k_{\text{max}}\right)\approx N\left(k_{\text{max}}\right).$
This means that the density of the suitable qubit numbers decreases with increasing $N$.
This is illustrated in Fig. \ref{fig:suitable_qubit}, where we plot this difference
as a function of the total number of qubits $N$ for both of these schemes. 

\section*{Supplementary Note 7. Optimality of the sine state}
\label{app:sine_state}
Here, we show that the phase estimation protocol where the initial state is an $N$-qubit sine state, and the measurement is a QFT followed by a projective measurement on the basis states, attains the HL in the limit of large qubit number, $N \gg 1$. This regime is equivalent to having a large prior width within the dynamic range \cite{macieszczak_bayesian_2014}. The $N$-qubit sine state is given by
\begin{align}
    \ket{\psi_{\text{sine}}} = \sqrt{ \frac{2}{N+2}} \sum_{m=0}^{N} \sin{\left(\frac{\pi(m+1)}{N+2} \right)} \ket{m}
\end{align}
where the  states $\{\ket{m}\}$ are the symmetric-under-permutations eigenvectors of the angular momentum operator in the z-direction $J_{z}$, with
\begin{align}
    J_{z} &= \frac{1}{2}\underset{i}{\sum}\sigma^{i}_{z}, \quad i = 1, 2, \dots N
\end{align}
and $\sigma^{i}_{z}$ is the Pauli $Z$ operator for the $i^\text{th}$ qubit. After undergoing a unitary evolution $U(\phi) = e^{-i\phi J_z}$, the sine state transforms into
\begin{align}
    \ket{\psi} = \sqrt{ \frac{2}{N+2}} \sum_{m=0}^{N} e^{-im\phi} \sin{\left(\frac{\pi(m+1)}{N+2} \right)} \ket{m}
\end{align}
up to a phase factor. Performing a QFT, followed by a projective measurement on the basis $\{\ket{m}\}$ can be represented with a POVM $E_k = \ket{e_k}\bra{e_k}$, $k = 0, 1, \dots N$, where
\begin{align}
    \ket{e_k} = \sqrt{\frac{1}{N+1}} \sum_{m=0}^N e^{-im \frac{2\pi k}{N+1}} \ket{m}
\end{align}
And the probability of obtaining an outcome $k$ after the measurement is given by $p(k) = |\bra{\psi} E_k \ket{\psi}|^2$,
\begin{align}
    p(k) &= \frac{2}{(N+1)(N+2)} \left|\sum_{m=0}^N e^{im(\frac{2\pi k}{N+1}-\phi)} \sin{\left(\frac{\pi(m+1)}{N+2} \right)}\right|^2 \nonumber \\
    &\approx 2 \int_{0}^1 \int_{0}^1 dx \, dy \, e^{i(x-y)(2\pi k-N\phi)} \sin{\left( \pi x \right)} \sin{\left( \pi y \right)} \nonumber \\
    &= \frac{8\pi^2 \cos^2{(N\phi/2)}}{\left[ \pi^2-(2\pi k - N\phi)^2\right]^2}
\end{align}
for $N \gg 1$ and $k = 0, 1, \dots N$. Note that we assume no phase slips, $ \phi \in \left[ 0, \, 2\pi \right]$, such that $p(k)$ is non-zero only in this interval. Given the probabilities of measurement outcomes $p(k)$, we can derive the statistics of the measurement results.

Let us denote the random variable that corresponds to the measurement outcomes as $O_{\text{QFT}}.$ 
$O_{\text{QFT}}$ therefore takes the values $\left\{ k\right\} _{-N/2}^{N/2}$ with probabilities $p(k)$. The first and second moments of $O_{\text{QFT}}$ are given by
\begin{align}
    \left\langle O_{\text{QFT}} \right\rangle = \sum_{m = -N/2}^{N/2} \frac{8\pi^2 \cos^2{(N\phi/2)} \, m}{\left[ \pi^2-(2\pi m - N\phi)^2\right]^2} \nonumber \\
    \left\langle O_{\text{QFT}}^2 \right\rangle = \sum_{m = -N/2}^{N/2} \frac{8\pi^2 \cos^2{(N\phi/2)} \, m^2}{\left[ \pi^2-(2\pi m - N\phi)^2\right]^2}
\end{align}
where we define $\bar k = \lfloor N\phi/(2 \pi) \rfloor \approx N\phi/(2\pi)$. In the limit of $N \gg 1$, 
these expressions are approximately
\begin{align}
    &\left\langle O_{\text{QFT}} \right\rangle \approx \sum_{m = -\infty}^{\infty} \frac{8\pi^2 \cos^2{(N\phi/2)} \, m}{\left[ \pi^2-(2\pi m - N\phi)^2\right]^2} = \frac{N\phi}{2\pi} \nonumber \\
    &\left\langle O_{\text{QFT}}^2 \right\rangle \approx \sum_{m = -\infty}^{\infty} \frac{8\pi^2 \cos^2{(N\phi/2)} \, m^2}{\left[ \pi^2-(2\pi m - N\phi)^2\right]^2} = \left(\frac{N\phi}{2\pi}\right)^2 + \frac{1}{4} \nonumber \\
    &(\Delta O_{\text{QFT}})^2 = \left\langle O_{\text{QFT}}^2 \right\rangle - \left\langle O_{\text{QFT}} \right\rangle^2 = \frac{1}{4}
\end{align}
where $(\Delta O_{\text{QFT}})^2$ is the variance of $O_{\text{QFT}}$.
We can thus take the estimator of $\phi$ in the limit of $N \gg 1$ to be $\frac{2\pi}{N}O_{\text{QFT}}.$
The MSE of this estimator is then
\begin{align}
    (\Delta\phi)^2 =\frac{\left(2\pi\right)^{2}}{N^{2}}\left(\Delta O_{\text{QFT}}\right)^{2}=\frac{\pi^{2}}{N^{2}}.
\end{align}
We thus attain the $\pi$-corrected HL for the entire interval $\phi \in \left[ 0, \, 2\pi \right]$, for $N \gg 1$. Note that this result is independent of the basis states $\{\ket{m}\}$. 

\section*{Supplementary Note 8. Overhead for $M=3$ Blocks with optimal measurements}

Here, we calculate the phase estimation error for the scheme where the initial state consists of GHZ states with sizes of $2^{k_i}$, where $k_i = 0, 1, \dots, k_\text{max}$, and each state has $M=3$ copies. For the readout, we assume optimal measurements are available. The total number of qubits is given by $N= 3(2^{k_\text{max}+1}-1)$. As mentioned in the main text, such a state can be written as
\begin{align}
\label{eq:fn_state}
    \ket{\psi} = \frac{1}{(N/3+1)^{3/2}} \sum_{n=0}^N \sqrt{f(n)} \ket{n}.
\end{align}
We can explicitly calculate the coefficients $f(n)$, with
\begin{align}
\label{eq:fn_defn}
    f(n) = \begin{cases}
        (n+1)\left( \frac{n}{2} + 1 \right) & \frac{1}{3} > \frac{n}{N} \geq 0 \\[2mm]
        \frac{(N+3)^2}{12} + \left(n - \frac{N-1}{2} \right) \left(n - \frac{N-3}{2} \right) & \frac{2}{3} \geq \frac{n}{N} \geq \frac{1}{3} \\[2mm]
        (N-n+1)\left( \frac{N-n}{2} + 1 \right) & 1 \geq \frac{n}{N} > \frac{2}{3}
    \end{cases}.
\end{align}
We assume a uniform prior in $[-\pi, \pi],$
and choose the commonly used cost function of $4 \sin^2{\left((\phi-\phi_{\text{est}})/2\right)} \approx (\phi-\phi_{\text{est}})^2$.
It was shown in Ref. \cite{kaftal2014usefulness}, that in this case, given the state in Eq. (\ref{eq:fn_state}) and the optimal measurements, $(\Delta \widetilde{\phi})^2$  is given by
\begin{align}
    (\Delta \widetilde{\phi})^2 = 2 \left[1 - \frac{S_f}{(N/3+1)^3} \right], \; S_f = \sum_{n=0}^{N-1} \sqrt{f(n)f(n+1)}.
\end{align}
Let us use the analytical expression for $f(n)$ in Eq. (\ref{eq:fn_defn}) to explicitly calculate the $(\Delta \widetilde{\phi})^2$ for large $N$. First, we can write $S_f$ as
\begin{align}
    S_f &= \sum_{n = 0}^{\frac{N}{3}-1} (n+2)\sqrt{(n+1)(n+3)} - \frac{(N+3)^2}{12} \nonumber \\
     &\quad+ \sum_{n=0}^{\frac{N-3}{6}} \sqrt{ \left(\frac{(N+3)^2}{12} - n^2 \right)^2 -n^2 }
\end{align}
The sum $S_f$ can be expanded using the Euler–Maclaurin formula, 
\begin{align}
    \sum_{m}^n g(i) &= \int_m^n g(x) dx + \sum_{k=1}^p \frac{B_k}{k!} \left(g^{(k-1)}(n) - g^{(k-1)}(m)\right) \nonumber \\
     &\quad + R_p,
\end{align}
where $B_k$ are the Bernoulli numbers, $R_p$ is an error term that depends on the order of the expansion, and $g^{(k)}(x)$ denotes the $k$th derivative of the function $g$ evaluated at $x$. The BMSE can be written as $\alpha/N + \beta/N^2 + O(1/N^3)$, where we want to calculate $\alpha$ and $\beta$. Since the sum $S_f$ is multiplied by a factor $\approx N^{-3}$, the terms that will contribute to $\alpha$ and $\beta$ in $S_f$ will scale as $N^2$ and $N$, respectively. Then, we do not need to find the constant terms in $S_f$, which reduces the number of terms that we need to calculate in the expansion significantly. Accordingly, we find
\begin{align}
    \sum_{n = 0}^{\frac{N}{3}-1} (n+2)\sqrt{(n+1)(n+3)} = \frac{N^3}{81} + \frac{N^2}{6} + \frac{5N}{9} + O(1)
\end{align}
and
\begin{align}
    &\sum_{n=0}^{\frac{N-3}{6}} \sqrt{ \left(\frac{(N+3)^2}{12} - n^2 \right)^2 -n^2 } \nonumber \\ & \quad = \frac{2N^3}{81} + \frac{N^2}{4} + \frac{(17 - \frac{3\sqrt{3}}{2} \text{ln}(2+\sqrt{3}))N}{18} + O(1)
\end{align}
Plugging these sums into $S_f$, we find that
\begin{align}
    (\Delta \widetilde{\phi})^2 = \frac{9\sqrt{3} \ln{(2+\sqrt{3})}}{2} \frac{1}{N^2} + O(N^{-3}).
\end{align}

\section*{Supplementary Note 9. Optimization of the adaptive measurement}
\label{app:adaptive_meas}

Here, we show the optimization over the single-qubit rotations in the adaptive measurement. The BMSE can be rewritten as
\begin{align}
\label{eqn_supp:bmse}
(\Delta \tilde{\phi})^2 = \int \sum_{\vec{n}} p(\vec{n}, \vec{\Phi} |\phi)(\phi-\phi_{est}(\vec{n}))^2 \mathcal{P}_{\delta\phi}(\phi)\, d\phi
\end{align}
where the optimal Bayesian estimator for each branch, $\phi_{est}^*(\vec{n})$, is given by
\begin{align}
\label{eqn_supp:optimal_est}
\phi_{est}^*(\vec{n}) = \frac{\int \phi \, p(\vec{n}, \vec{\Phi} |\phi) \mathcal{P}_{\delta\phi}(\phi) d\phi}{\int p(\vec{n}, \vec{\Phi} |\phi) \mathcal{P}_{\delta\phi}(\phi) d\phi}
\end{align}
which can be thought of as the expected value of $\phi$ given that the branch $\vec{n}$ has been sampled. Eq. (\ref{eqn_supp:bmse}) and (\ref{eqn_supp:optimal_est}) result in the minimum possible BMSE, given by
\begin{align}
\label{eqn_supp:bmse_2}
(\Delta \tilde{\phi}^*)^2 = (\delta\phi)^2 - \sum_{\vec{n}} \frac{\left(\int \phi \, p(\vec{n}, \vec{\Phi} |\phi) \mathcal{P}_{\delta\phi}(\phi) d\phi \right)^2}{\int p(\vec{n}, \vec{\Phi} |\phi) \mathcal{P}_{\delta\phi}(\phi) d\phi} 
\end{align}
$p(\vec{n}, \vec{\Phi} |\phi)$ contains the 
rotations angles (system phases) $\vec{\Phi}$ that will be performed after each measurement along the branch $\vec{n}$. For example, if the initial state contains one block of a 2-atom GHZ state and two blocks of 1-atom GHZ states, there are $2^3 = 8$ possible branches depending on measurement results, occurring with probabilities
\begin{align}
\label{eqn_supp:probabilities_example}
p(\left[ i, j, k \right], \vec{\Phi} |\phi) = &\theta_i{\left( \frac{2 (\phi - \Phi_1)}{2} \right)} \theta_j{\left( \frac{\phi - \Phi_{2+i}}{2}\right)} \nonumber \\ & \theta_k{\left( \frac{ \phi - \Phi_{4+2i+j}}{2}\right)}, \nonumber \\
\theta_i(x) =& \begin{cases}
     \cos^2{(x)}, & \text{if}\ i = 0 \\
    \sin^2{(x)}, & \text{if}\ i = 1
    \end{cases}
\end{align}
for $i, \,j, \,k \in \{ 0,\, 1\}$. The branch vector $\left[ i, j, k \right]$ indicates the measurement result of a parity measurement on a GHZ block with $m$ atoms: 0 (1) represents even (odd) parity. 

We now want to minimize $\Delta \Tilde{\phi}^*$ 
with respect to the 
rotation angles
$\vec{\Phi}$. For this purpose, one needs to obtain $\partial \Delta \Tilde{\phi}^*/ \partial \Phi_i = 0$ for every $i$. $\partial \Delta \Tilde{\phi}^*/ \partial \Phi_i$ is found as
\begin{align}
\label{eqn_supp:derivative_1}
\frac{\partial\Delta\Tilde{\phi}^{*}}{\partial\Phi_{i}}=\sum_{\vec{n}}\phi_{est}^{*}(\vec{n})^{2}\int\partial_{\Phi_{i}}p(\vec{n},\vec{\Phi}|\phi)\mathcal{P}_{\delta\phi}(\phi)d\phi \nonumber \\ -2\phi_{est}^{*}(\vec{n})\int\phi\,\partial_{\Phi_{i}}p(\vec{n},\vec{\Phi}|\phi)\mathcal{P}_{\delta\phi}(\phi)d\phi
\end{align}
Note that $\partial_{\Phi_i} p(\vec{n}, \vec{\Phi} |\phi) = 0 $ if the branch $\vec{n}$ does not contain $\Phi_i$. Assuming that $\Phi_i$ is contained in the branch $\vec{n}$, and that it is the $l^\text{th}$ measurement, performed on a GHZ state with m atoms, from Eq. (\ref{eqn_supp:probabilities_example}), we see that $\partial_{\Phi_i} p(\vec{n}, \vec{\Phi} |\phi)$ has two possible forms:
\begin{align}
\label{supp_eqn:derivative_cases}
\partial_{\Phi_i} p(\vec{n}, \vec{\Phi} |\phi) = \begin{cases}
    \;\; m\,p(\vec{n}, \vec{\Phi} |\phi) \tan{\left( \frac{m (\phi - \Phi_i)}{2} \right)}, & \text{if}\ n_l = 0 \\
     - m\,p(\vec{n}, \vec{\Phi} |\phi) \cot{\left( \frac{m (\phi - \Phi_i)}{2} \right)}, & \text{if}\ n_l = 1
\end{cases}
\end{align}
Plugging the expression for the derivative in Eq. (\ref{supp_eqn:derivative_cases}) into Eq. (\ref{eqn_supp:derivative_1}), we obtain
\begin{widetext}
\begin{align}
\label{eqn_supp:derivative_2}
\frac{\partial\Delta\Tilde{\phi}^{*}}{\partial\Phi_{i}}=\sum_{\vec{n},\Phi_{i}\ \text{in}\ \vec{n}}\phi_{est}^{*}(\vec{n})^{2}\int f(n_{l},\Phi_{i}|\phi)p(\vec{n},\vec{\Phi}|\phi)\mathcal{P}_{\delta\phi}(\phi)\,d\phi-2\phi_{est}^{*}(\vec{n})\int\phi\,f(n_{l},\Phi_{i}|\phi)p(\vec{n},\vec{\Phi}|\phi)\mathcal{P}_{\delta\phi}(\phi)\,d\phi
\end{align}
\end{widetext}
with 
\begin{align}
f(n_l, \Phi_i |\phi) = \begin{cases}
    \;\;\, m \tan{\left( \frac{m (\phi - \Phi_i)}{2} \right)}, & \text{if}\ n_l = 0 \\
    - m \cot{\left( \frac{m (\phi - \Phi_i)}{2} \right)}, & \text{if}\ n_l = 1
\end{cases}
\end{align}
and $m$ is the number of atoms contained in the GHZ block such that the feedback $\Phi_i$ is applied after its measurement.

To find the system phases that minimize the BMSE, we perform a gradient descent, where the system phases are updated according to
\begin{align}
\Phi_i \rightarrow \Phi_i - \alpha \frac{\partial \Delta \Tilde{\phi}*}{\partial \Phi_i}
\end{align}
where $\alpha$ is the step size and the derivative is given in Eq. (\ref{eqn_supp:derivative_2}). We implement the Adam optimizer \cite{kingma_adam_2017} to perform the gradient descent.

\section*{Supplementary Note 10. Proof of claim I (OQI and CSS Plateaus)}
\label{app:proof_of_phase_slip_bounds}

We provide here a detailed proof for the BMSE expressions in the limit of $N \rightarrow \infty$ given in the main text. For completeness, the statement of the claim is:

\textit{Claim.} The OQI
BMSE in the limit of $N \rightarrow \infty$ for any prior distribution $\mathcal{P}_{\delta \phi}(\phi)$ is \begin{align}
\label{eqn:plateau_hl_supp}
(\Delta \Tilde{\phi})^2_{\text{OQI}} =\left(\delta\phi\right)^{2}-\overset{\pi}{\underset{-\pi}{\int}}\frac{\left(\underset{k}{\sum}\phi_{k}\mathcal{P}_{\delta\phi}\left(\phi_{k}\right)\right)^{2}}{\underset{k}{\sum}\mathcal{P}_{\delta\phi}\left(\phi_{k}\right)}d\phi,
\end{align}
where $\phi_k = \phi + 2\pi k, \; k \in \mathbb{Z}.$\\
Assuming a $J_{\theta} \coloneqq \cos\left(\theta\right)J_{x}-\sin\left(\theta\right)J_{y}$ single quadrature readout, the CSS BMSE in the limit of $N \rightarrow \infty$ for any prior distribution $\mathcal{P}_{\delta \phi}(\phi)$ is
\begin{align}
\begin{split}
&(\Delta\Tilde{\phi})_{\text{SQL}}^{2}=\left(\delta\phi\right)^{2}-
\overset{\pi}{\underset{-\pi}{\int}}d\phi \, \cdot\\
&\frac{\left(\underset{k}{\sum}\left(\phi_{k}-\theta\right)\mathcal{P}_{\delta\phi}\left(\phi_{k}-\theta\right)-\left(\phi_{k}+\theta\right)\mathcal{P}_{\delta\phi}\left(-\phi_{k}-\theta\right)\right)^{2}}{\underset{k}{\sum}\mathcal{P}_{\delta\phi}\left(\phi_{k}-\theta\right)+\mathcal{P}_{\delta\phi}\left(-\phi_{k}-\theta\right)}.
\label{eqn:plateau_sql_supp}
\end{split} 
\end{align}

\textit{Proof.}
We first prove the OQI BMSE, given in Eq. (\ref{eqn:plateau_hl_supp}). 
To this end, we first prove that it is a lower bound to the BMSE for any possible scheme. Then, we show that there exists a scheme that saturates it.
Observe that a solution to the following minimization problem provides a lower bound for any scheme:
\begin{align}
&\underset{\phi_{\text{est}}\left(x\right)}{\text{min}}\underset{p\left(x|\phi\right)}{\text{min}}\sum_{x}\int_{-\infty}^{\infty}\left(\phi_{\text{est}}\left(x\right)-\phi\right)^{2}p\left(x|\phi\right)\mathcal{P}_{\delta\phi}\left(\phi\right)\;d\phi \nonumber \\
&\text{subject to: }\sum_{x}p\left(x|\phi\right)=1, \;p\left(x|\phi\right)=p\left(x|\phi+2\pi\right).
\end{align}
The first step is to show that we can map this into an equivalent minimization problem that involves $\phi_{\text{est}}$ only:
\begin{align}
\text{\ensuremath{\underset{\phi_{\text{est}}\left(\phi\right)}{\text{min}}}}\int_{-\pi}^{\pi}\sum_{k}\left(\phi_{\text{est}}\left(\phi\right)-\left(\phi+2\pi k\right)\right)^{2}\mathcal{P}_{\delta\phi}\left(\phi+2\pi k\right)\;d\phi.    
\end{align}
To see this, observe first that due to the constraint $p\left(x|\phi\right)=p\left(x|\phi+2\pi\right)$, we can write the BMSE as:
\begin{align}
&\int_{-\pi}^{\pi}\underset{k}{\sum}\sum_{x}\left(\phi_{\text{est}}\left(x\right)-\phi_k\right)^{2}p\left(x|\phi\right)\mathcal{P}_{\delta\phi}\left(\phi_k\right)\;d\phi    
\end{align}
where $\phi_k = \phi + 2\pi k, \; k \in \mathbb{Z}$. For each $\phi$, there exists an $x_{\text{min},\phi}$ for which $\sum_{k}\left(\phi_{\text{est}}\left(x_{\text{min},\phi}\right)-\phi_k\right)^{2}\mathcal{P}_{\delta\phi}\left(\phi_k\right)$ is minimal. Therefore, the optimal $p\left(x|\phi\right)$ is $p\left(x|\phi\right)=\delta\left(x-x_{\text{min},\phi}\right).$
Denoting $\phi_{\text{est}}\left(x_{\text{min},\phi}\right)$ as $\phi_{\text{est}}\left(\phi\right),$ we thus want to minimize $\sum_{k}\left(\phi_{\text{est}}\left(\phi\right)-\phi_k\right)^{2}\mathcal{P}_{\delta\phi}\left(\phi_k\right)$ over $\phi_{\text{est}}\left(\phi\right)$ for each $\phi$. This is a convex minimization problem with respect to $\phi_{\text{est}}\left(\phi\right)$, and the optimal $\phi_{\text{est}}\left(\phi\right)$ is given by: 
\begin{align}
\phi_{\text{est}}\left(\phi\right)=\underset{k}{\sum}\frac{\phi_k\mathcal{P}_{\delta\phi}\left(\phi_k \right)}{\sum_{j}\mathcal{P}_{\delta\phi}\left(\phi_j\right)}. 
\label{eqn:phi_est_opt_supp}
\end{align}
To get the expression of Eq. (\ref{eqn:plateau_hl_supp}), 
we use BMSE expression of Eq. (\ref{eqn:bmse_2}). 
For this lower bound, the set of outcomes is the continuous range $\left[-\pi,\pi\right]$, hence:
\begin{align}
&\left(\Delta\phi\right)^{2}=\left(\delta\phi\right)^{2}-\overset{\pi}{\underset{-\pi}{\int}}\phi_{\text{est}}\left(\phi\right)^{2}p\left(\phi\right)\;d\phi \nonumber \\
&=\left(\delta\phi\right)^{2}-\overset{\pi}{\underset{-\pi}{\int}}\frac{\left(\underset{k}{\sum}\phi_{k}\mathcal{P}_{\delta\phi}\left(\phi_{k}\right)\right)^{2}}{\underset{k}{\sum}\mathcal{P}_{\delta\phi}\left(\phi_{k}\right)}d\phi,   
\end{align}
which is the desired lower bound.

For tightness, we show that a CSS interrogation with a dual quadrature readout attains this bound in the limit of $N \rightarrow \infty.$
Let us first define a dual quadrature measurement: given a $2N$-qubits CSS, we split it into a product of two identical $N$-qubits CSS: $|\psi_{\phi,2N}\rangle=|\psi_{\phi,N}\rangle|\psi_{\phi,N}\rangle$.
We measure the $J_{x}=\frac{1}{2}\sum_{i=1}^{N}\sigma^{i}_x$ operator of the first $|\psi_{\phi,N}\rangle$  and the $J_{y}=\frac{1}{2}\sum_{i=N+1}^{2N}\sigma_y^{i}$ operator of the second $|\psi_{\phi,N}\rangle$ .
Let us denote
$x_{x}=\frac{2 N_x}{N}, x_{y}=\frac{2 N_y}{N},$
where $N_{\theta}\in\{-N/2,...,N/2-1,N/2\}$ is the outcome of $J_{\theta}$ measurement.
$x_{x},x_{y}$ have independent shifted binomial distributions.
In the limit of large $N$, the joint distribution $\left(x_{x},x_{y}\right)$ converges to the multivariate Gaussian distribution $p\left(x_{x},x_{y}|\phi\right)\rightarrow N\left(\left[\cos\left(\phi\right),\sin\left(\phi\right)\right]^T,\Sigma=\text{diag}\left(s_{x}^{2},s_{y}^{2}\right)\right)$, where $s_{x}^{2}=\sin\left(\phi\right)^{2}/N, \; s_{y}^{2}=\cos\left(\phi\right)^{2}/N.$
 Therefore, in the limit of $N\rightarrow\infty$  we obtain
 \begin{align}
&p\left(x_{x},x_{y}|\phi\right)\rightarrow\delta\left(x_{x}-\cos\left(\phi\right)\right)\delta\left(x_{y}-\sin\left(\phi\right)\right) \nonumber\\
&\Rightarrow p\left(r,\varphi|\phi\right)=\delta\left(r-1\right)\left(\sum_{k\in\mathbb{Z}}\delta\left(\varphi-\phi-2\pi k\right)\right), 
 \end{align}
where $r=\sqrt{x_{x}^{2}+x_{y}^{2}}$, $\varphi=\arg\left(x_{x}+ix_{y}\right).$
This $p\left(\varphi|\phi\right)$, yields the optimal $\phi_\text{est}$ of Eq. (\ref{eqn:phi_est_opt_supp}), hence the bound is saturated.
This proves the first part of the claim, the OQI BMSE.

Let us now compute the bound for a CSS interrogation with a $J_\theta$ single quadrature readout.
Similar to above, we denote $x_{\theta}=\frac{2N_{\theta}}{N}$.
$x_{\theta}$ has a shifted 
 binomial distribution with an average of $\cos\left(\phi+\theta\right)$ and a variance of $\sin\left(\phi+\theta\right)^{2}/N$.
 Therefore, in the limit of $N\rightarrow\infty$, $p\left(x_{\theta}|\phi\right)$ converges to
\begin{align}
\delta\left(x_{\theta}-\cos\left(\phi+\theta\right)\right) &= \underset{k\in\mathbb{Z}}{\sum}\delta\left(\phi_k-\arccos\left(x_{\theta}\right)+\theta\right) \nonumber \\
&\quad+\delta\left(\phi_k+\arccos\left(x_{\theta}\right)+\theta\right)
\end{align}
Inserting this into Eq. (\ref{eqn:bmse_2}) results in Eq. (\ref{eqn:plateau_sql_supp}).

\section*{Supplementary Note 11. Prior distribution when additional classical interrogations are introduced}
\label{app:classical_interrogations}

Phase unwinding with uncorrelated interrogations is equivalent to estimating the variable $P$ after a phase accumulation of $\phi = 2\pi P + \beta$ with Ramsey interferometry, where $-\pi < \beta < \pi$ and $\beta$ can be estimated with an STD that scales as $1/N$ if Heisenberg-limited protocols are used. Here, we show that for a wide prior distribution, i.e. $\delta\phi \gg \pi$, after $P$ is estimated, the posterior distribution of $\phi$ converges to a uniform distribution over the interval $\left[2\pi P -\pi, \, 2\pi P +\pi \right]$. The posterior distribution of $\phi$, after measuring $P = P_m$ with negligible error, is supported only on $2\pi P_m -\pi < \phi < 2\pi P_m +\pi$. It is given by
\begin{align}
    p(\phi |P_m) &= \frac{p(\phi )}{p(P = P_m)} \nonumber \\
    &=\frac{e^{-\phi^2/2(\delta\phi)^2}}{\int_{2\pi P_m -\pi}^{2\pi P_m +\pi} d \phi \, e^{-\phi^2/2(\delta\phi)^2}} \,.
\end{align}
This distribution is equivalent to
\begin{align}
    p(\theta|P_m) =\sqrt{\frac{2}{\pi(\delta\phi)^2}}\frac{e^{-(2\pi P_m + \theta)^2/2(\delta\phi)^2}}{ \text{erf}\left(\frac{2\pi P_m + \pi}{\sqrt{2(\delta\phi)^2}}\right) - \text{erf}\left(\frac{2\pi P_m - \pi}{\sqrt{2(\delta\phi)^2}}\right)}
\end{align}
for $-\pi \leq \theta \leq \pi$, where $\text{erf}(z)$ is the error function. In the limit of $\delta\phi \gg \pi$, we have
\begin{align}
    p(\theta|P_m) \approx \sqrt{\frac{2}{\pi(\delta\phi)^2}}\frac{e^{-2\pi^2 P_m^2/(\delta\phi)^2}\left(1-\frac{2\pi P_m\theta}{(\delta\phi)^2} \right)}{ \text{erf}\left(\frac{2\pi P_m + \pi}{\sqrt{2(\delta\phi)^2}}\right) - \text{erf}\left(\frac{2\pi P_m - \pi}{\sqrt{2(\delta\phi)^2}}\right)}
\end{align}
which reduces to a uniform distribution in $\theta \in \left[ -\pi, \, \pi \right]$ in the limit of $\delta\phi \rightarrow \infty$.

\section*{Supplementary Note 12. Performance of the schemes for a uniform prior distribution}
\label{app:flat_prior}

Here, we analyze the performance of the schemes for a uniform prior phase distribution defined in the interval $[-\pi/2, \, \pi/2]$. We choose not to extend the limits of the distribution to $-\pi$ and $\pi$, since we observed numerically for this case that the BMSE increases significantly due to the finite dynamic range of the schemes. The MSE approaches $4\pi^2$ as $|\phi| \rightarrow \pi$, and any gain in precision is masked by these errors. 
Therefore, we restrict the range of the prior distribution to a smaller interval.

\begin{figure}
    \centering
    \includegraphics[width=0.9\columnwidth]{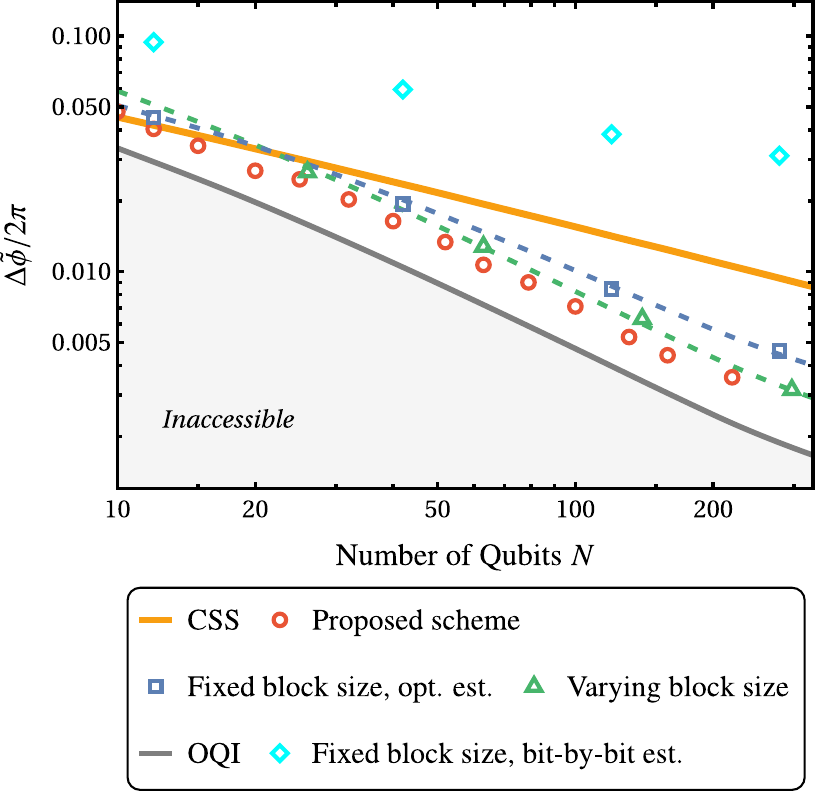}
        \caption{RBMSE $\Delta \tilde \phi$  normalized by $2\pi$, as a function of the qubit number $N$, for various schemes. Here, we assume a uniform prior distribution in the interval $[-\pi/2, \, \pi/2]$. We observe that the proposed scheme and the scheme with a varying block size \cite{higgins_demonstrating_2009} show HS with an overhead of 1.50 and 1.75, respectively. We also observe that the scheme with a fixed block size \cite{kessler_heisenberg-limited_2014} shows HS up to a logarithmic overhead, and achieves sub-SQL precision when optimal Bayesian estimators are used. However, this scheme fails to achieve sub-SQL precision when sub-optimal estimators, such as the bit-by-bit estimator, is used (see \supplukin\ for more details on the estimator). }
    \label{fig:flat_prior}
\end{figure}

The results can be found in Fig. \ref{fig:flat_prior}. For the scheme with a fixed block size, we plot the RBMSE for both the optimal Bayesian estimators, and the bit-by-bit estimator (see \supplukin\ for a detailed description of the estimators). We observe that this scheme shows HS up to a logarithmic overhead of $\sqrt{\text{log}(N)}$, where $N$ is the number of qubits, when optimal Bayesian estimators are used. However, it fails to achieve a sub-SQL performance when the bit-by-bit estimator is used, as it was the case for Gaussian prior distributions. Furthermore, the proposed scheme and the scheme with a varying block size both show HS, where the proposed scheme has a smaller overhead of 1.50, and the latter has an overhead of 1.75.

\section*{Supplementary Note 13. Effect of amplitude damping noise}
\label{app:amplitude_damping}

\begin{figure}[h]
\centering\includegraphics[width=0.9\columnwidth]{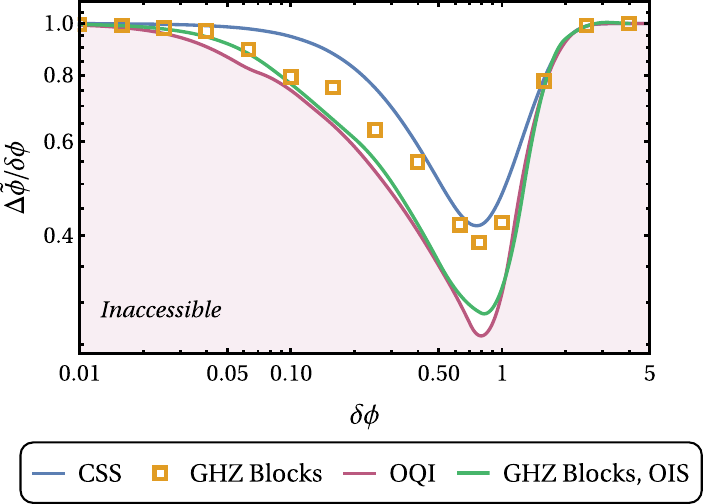}
    \caption{RBMSE $\Delta \tilde \phi$ as a function of the prior width $\delta\phi$, normalized by $\delta\phi$, for $N = 12$ and in the presence of amplitude damping with a probability of $p_a = 10^{-2}$. The benchmark is the noiseless OQI, plotted in purple. The 12-qubit CSS for $p_a = 10^{-2}$ is also shown in the plot in blue. The minimum RBMSE obtained from optimal partitioning into blocks of GHZ states---i.e., the optimal initial state (OIS) for blocks of GHZ states---for $p_a = 10^{-2}$, assuming that optimal measurements are available, is plotted in green. We notice that this curve is very close to the noiseless OQI. Finally, the RBMSE given optimal partitioning into blocks of GHZ states and local, adaptive measurements is plotted with yellow rectangles. We see that in the presence of amplitude damping, even though the optimal partitions perform very close to the OQI when optimal measurements are available, there is a large gap between the minimum achievable sensitivity and the sensitivity achieved when local, adaptive measurements are used.}
\label{fig:21_atom_bmse_amp_damp_app}
\end{figure}

Let us analyze the effect of amplitude damping noise on GHZ states and on the resulting probabilities.
A $k$-qubit GHZ state, $\ket{\psi} = \frac{1}{\sqrt{2}}\left(|0\rangle^{\otimes k}+e^{i k\phi}|1\rangle^{\otimes k}\right)$, undergoes a $k$-qubit amplitude damping channel: 
\begin{align}
    \rho&\rightarrow\varepsilon^{\otimes k}\left(\rho\right), \nonumber \\
    \varepsilon\left(\rho\right) &=p_a\,\sigma_{-}\rho\sigma_{+}+\sqrt{\mathbf{1}-p_a\,\sigma_{+}\sigma_{-}}\rho\sqrt{\mathbf{1}-p_a\,\sigma_{+}\sigma_{-}}\,, \nonumber \\
    \sigma_{+} &= \left[\begin{array}{cc}
        0 & 1 \\
        0 & 0
    \end{array} \right],\,\sigma_{-} = \left[\begin{array}{cc}
        0 & 0 \\
        1 & 0
    \end{array} \right],\,\mathbf{1} = \left[\begin{array}{cc}
        1 & 0 \\
        0 & 1
    \end{array} \right].
\label{eq:amp_damping_channel}    
\end{align}
where $p_a>0$ is the amplitude damping probability. The GHZ state is thus transformed to an ensemble $\sum_{\vec{v}}c(\vec{v})\rho_{\vec{v}},$ where $\vec{v}\in\left\{ 0,1\right\} ^{N}$, such that $v_{i}=0\left(1\right)$ indicates loss (no loss) in the $i^{th}$ qubit, and  $\rho_{\vec{v}} = |\vec{v}\rangle\langle\vec{v}|$. $c(\vec{v})$ are given by
\begin{align}
    c(\vec{0}) &= \frac{1}{2} \left(1 + (1-p_a)^k \right), \nonumber \\
    c(\vec{v}) &= \frac{1}{2} \left(1-p_a\right)^{\underset{i}{\sum}v_{i}}p_a^{1-\underset{i}{\sum}v_{i}},\quad\text{for}\;\;\vec{v} \neq \vec{0}.
\end{align}
The phase information is now encoded only in $|\vec{0}\rangle$ which equals to
\begin{align}
    |\vec{0}\rangle = \frac{1}{\sqrt{2 c(\vec{0})}}\left(|0\rangle^{\otimes k}+\left(1-p_a\right)^{k/2} e^{i k\phi}|1\rangle^{\otimes k}\right).
\end{align}
Let us derive the probabilities obtained from performing local Pauli $X$ measurements on the state.
We denote the $\pm 1$ eigenstates of Pauli $X$ as $|\pm\rangle$ respectively and the probability of obtaining $m$ times $|-\rangle$ as $p\left( m \right).$
The probability for an even parity is the probability of obtaining an even $m$ and is therefore equal to: $p\left(\text{even}\right)=\underset{m \text{ even}}{\sum}p\left(m\right).$
Note that
\begin{align}
    p(\text{even}) = \sum_{\vec{v}} c(\vec{v})\,p\left(\text{even}|\vec{v}\right).
\end{align}
For any $\vec{v}\neq\vec{0},$
$p\left(\text{odd}|\vec{v}\right)=p\left(\text{even}|\vec{v}\right)=\frac{1}{2}.$
Let us denote the contribution from $\sum_{\vec{v}\neq \vec{0}}\rho_{\vec{v}}$ as $p_{\vec{v}\neq \vec{0}}\left(\text{even}\right)$, $p_{\vec{v}\neq\vec{0}}\left(\text{odd}\right)$. We have 
\begin{align}
&\underset{\vec{v}\neq \vec{0}}{\sum}p_{\vec{v}\neq \vec{0}}\left(\text{even}\right)=\underset{\vec{v}\neq \vec{0}}{\sum}p_{\vec{v}\neq \vec{0}}\left(\text{odd}\right) = \sum_{\vec{v}\neq \vec{0}} c(\vec{v})\,p\left(\text{even}|\vec{v}\right) \nonumber \\&=\underset{\vec{v}\neq \vec{0}}{\sum}\left(1-p_a\right)^{\underset{i}{\sum}v_{i}} p_a^{1-\underset{i}{\sum}v_{i}}\frac{1}{4}\nonumber \\
&=\frac{1}{4}\left(1-\left(1-p_a\right)^{k}\right).
\end{align}
Now, let us calculate the contribution from $\vec{v}=\vec{0},$ denoted as $p_{\vec{0}}\left(\text{even}\right),\,p_{\vec{0}}\left(\text{odd}\right)$. 
For any state $\alpha|0\rangle^{\otimes k}+\beta|1\rangle^{\otimes k}$, we have:
\begin{align}
\alpha|0\rangle^{\otimes k}+\beta|1\rangle^{\otimes k} =\left(\frac{1}{\sqrt{2}}\right)^{k}\Bigl(&\alpha\left(|+\rangle+|-\rangle\right)^{k} + \nonumber \\ & \beta\left(|+\rangle-|-\rangle\right)^{k}\Bigr).    
\end{align}
Therefore,
the probability of measuring $m$ times $|-\rangle$ given this state is $p\left(m\right)=\left(\frac{1}{2}\right)^{k} \binom{k}{m}|\alpha+\left(-1\right)^{m}\beta|^{2},$
and thus the probability for odd/even $m$ is given by: $p_{\pm}=\frac{1}{2}|\alpha \pm \beta|^{2}.$
Applying this to our case, i.e. $|\vec{0}\rangle$), then $\alpha=\frac{1}{\sqrt{2c(\vec{0})}},\beta=\frac{\left(1-p\right)^{k/2}e^{ik\phi}}{\sqrt{2c(\vec{0})}},$ and thus  
\begin{align}
\begin{split}
p_{\vec{0}}\left(\text{even}\right) &=\frac{1}{4}\left(1+\left(1-p_a\right)^{k}+2\left(1-p_a\right)^{k/2}\cos\left(k\phi\right)\right),\\
p_{\vec{0}}\left(\text{odd}\right) &=\frac{1}{4}\left(1+\left(1-p_a\right)^{k}-2\left(1-p_a\right)^{k/2}\cos\left(k\phi\right)\right).
\end{split}
\label{eq:probs_ghz}
\end{align}

Summing up the contributions from all $\vec{v}$ 
we obtain that $p\left(\text{even}\right), \,p\left(\text{odd}\right)$ are:
\begin{align}
p\left(\text{even}\right)&=\frac{1}{2}\left(1+\left(1-p_a\right)^{k/2}\cos\left(k\phi\right)\right),\\
p\left(\text{odd}\right)&=\frac{1}{2}\left(1-\left(1-p_a\right)^{k/2}\cos\left(k\phi\right)\right).
\end{align}
We considered Pauli $X$ measurement, however any other Pauli operator in the $X-Y$ plane would give equivalent results: For a measurement of $\cos\left(\phi_{0}\right)X+\sin\left(\phi_{0}\right)Y$ the probabilities will take the same form and $\phi$ will be changed to:  
$\phi\rightarrow\phi-\phi_{0}.$

Note that given an initial GHZ state, amplitude damping noise commutes with the unitary phase encoding, $U\left(\phi\right)$.
We can therefore assume that amplitude damping acts first, followed by the phase encoding. 
This can be easily verified: given $\rho=|\psi\rangle\langle\psi|$ of a $k$-qubit GHZ state, we have
\begin{align}
\begin{split}
&\epsilon^{\otimes k}\left(U\left(\phi\right)\rho \,U\left(\phi\right)^{\dagger}\right)=U\left(\phi\right)\left(\epsilon^{\otimes k}\left(\rho\right)\right)U\left(\phi\right)^{\dagger}\\
&=|\psi'_{\phi}\rangle\langle\psi'_{\phi}|+\underset{\vec{v}\neq \vec{0}}{\sum}\rho_{\vec{v}},
\end{split}
\end{align}
where $|\psi'_{\phi}\rangle=\frac{1}{\sqrt{2}}\left(|0\rangle^{\otimes k}+\left(1-p_a\right)^{k/2}e^{-ik\phi}|1\rangle^{\otimes k}\right).$

\begin{figure*}[t]
    \raggedright
    \includegraphics[width=2.05\columnwidth]{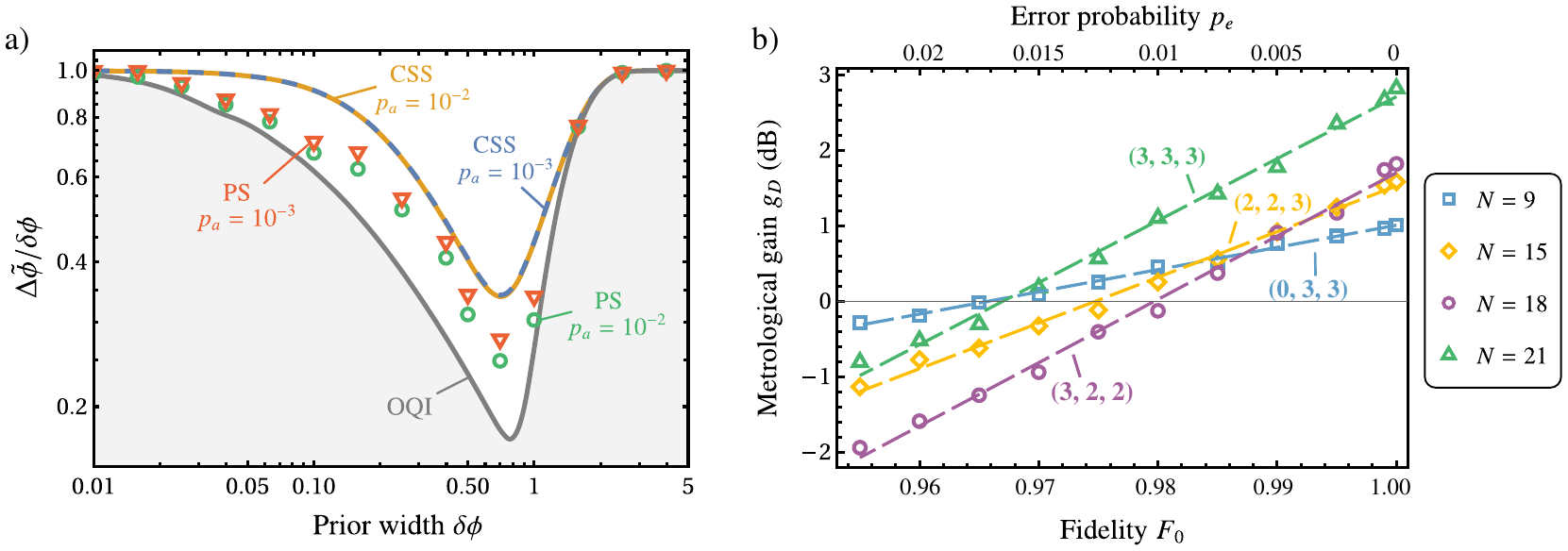} 
    \caption{Impact of noise on the proposed scheme. We consider amplitude damping, imperfect state preparation, and measurement errors to be the main processes limiting the sensitivity of the phase estimation. (a) The root Bayesian mean squared error (RBMSE) $\Delta\tilde\phi$, as a function of the prior width, $\delta\phi$, for $N = 21$ and various probabilities of decay $p_a$. As before, the RBMSE is normalized by the prior width. We plot the RBMSE of the 21-qubit CSS and the proposed scheme (PS on the Figure) for $p_a = 10^{-3}$ and $p_a = 10^{-2}$. Our benchmark is the noiseless OQI, plotted in gray. Increase in the RBMSE due to amplitude damping is most prominent around the prior width of $\delta\phi = 0.7$ rad. (b) The metrological gain as a function of the effective fidelity per qubit $F_0$, or the bit-flip measurement error probability $p_e$. We plot the performances of the optimal partitions for qubit numbers of $N = 9, 15, 18, 21$. The partitions are shown in the plot in the form of $(x, y, z)$, where $x, y$, and $z$ are the number of 4-qubit, 2-qubit, and 1-qubit GHZ states, respectively. We fit lines with different slopes to the performances of the scheme for different $N$, signifying an exponential decay in the metrological gain with decreasing $F_0$, or increasing $p_e$.
    We observe a correlation between the decay rate and the number of copies of the 4-qubit GHZ state: the partitions with $N = 21$ and $N=18$ both have three 4-qubit GHZ states, and they have similar decay rates.}
    \label{fig:21_atom_bmse_amp_damp}
\end{figure*}

Now, let us examine how amplitude damping affects the proposed scheme with and without optimal measurements. A case study is shown in Fig. \ref{fig:21_atom_bmse_amp_damp_app}. Here, we work with 12 qubits and assume an amplitude damping probability of $p_a = 10^{-2}$. The purple line represents the RBMSE of the noiseless OQI, i.e. without any amplitude damping effects. The green line is the RBMSE obtainable by using optimal partitions of GHZ states if optimal measurements are available. We see that the RBMSE of the optimal partitions, even in the presence of amplitude damping, can get very close to the RBMSE of the OQI. However, local, adaptive measurements are not enough to achieve this precision: we show the performance of the proposed scheme when such measurements are used with yellow rectangles, and see that there is a gap between that and the RBMSE achievable with optimal measurements. Intuitively, this gap can be explained by the following reasoning: since for the adaptive measurement, the single-qubit rotation applied to a given qubit relies so much on the previous measurement results, any error during the previous measurements changes the trajectory of the future measurements, resulting in a large error in phase estimation.
We conjecture that this issue can be mitigated
by using error detection scheme that checks whether the state remained in the GHZ code space.
Note that for every GHZ state with $k \geq 2$ atoms, decay errors can be detected by measuring if the GHZ state remains in the code space of $|0\rangle^{\otimes k},|1\rangle^{\otimes k}$, which can be achieved by e.g., measuring the stabilizers of this repetition code space.
We therefore detect a decay by measuring if the state is outside of this code space. Given a decay detection, the state has no phase information, and therefore we can ignore its measurement result. Otherwise, we can proceed with the same, adaptive, local measurements. This scheme converts amplitude damping noise into an erasure noise \cite{Kubica_erasure_qubits, Scholl2023} and involves only error detection and not error correction. This idea was recently proposed and analyzed also in Ref. \cite{kielinski2024ghz}.

Let us illustrate the advantage in using such error detection scheme in the frequentist case. Given a $k$-qubit GHZ state, the probabilities of even/odd parities are $\frac{1}{2}\left(1\pm\left(1-p_a\right)^{k/2}\cos\left(k\phi\right)\right),$ while with error detection the relevant probabilities are given by $p_{\vec{0}}\left(\text{even}\right),p_{\vec{0}}\left(\text{odd}\right).$
By direct calculation of the Cramer-Rao Bound (CRB), it can be observed that  in the limit of $\left(1-p_{a}\right)^{k}\ll1$, the MSE given by the  latter case is smaller by a factor of two than the former case. Hence, error detection leads to a factor of two improvement.

\section*{Supplementary Note 14. Impact of noise on the proposed scheme}
\label{sec:decoherence_effects}

In atomic clocks, beyond local oscillator noise, we consider the impact of amplitude damping \cite{yan_distinguishing_2018, burnett_decoherence_2019,levine_high-fidelity_2018,herbschleb_ultra-long_2019, balasubramanian_ultralong_2009}, imperfect state preparation due to finite gate fidelities, and measurement errors. To estimate the effect of such noise processes on the phase estimation, we again define the metrological gain $g_{\text{D}} = (\Delta\phi_{\text{CSS}})^2/(\Delta\Tilde{\phi})^2$ as our figure of merit, where $(\Delta\phi_{\text{CSS}})^2$ and $(\Delta\Tilde{\phi})^2$ are the BMSE of the phase estimation performed using a noiseless CSS, and the BMSE of the proposed scheme in the presence of the given noise process, respectively. 

\subsubsection*{Amplitude damping}

The effect of amplitude damping on a multi-qubit state is described in \suppamplitudedamping, and is given by the channel of Eq. (\ref{eq:amp_damping_channel}).
As shown in Supplementary Note 13, in the presence of amplitude damping, the probabilities are given by \mbox{$P(\pm) = (1 \pm (1-p_a)^{N/2}\cos{(N\phi)})/2$}, where $\{ +, -\}$ denote even, odd parity. 
The phase information thus drops exponentially 
with the number of qubits, $N$.
This implies that the optimal partition depends on the noise strength: as the noise gets stronger, GHZ states with smaller $N$ become more favorable. 

For simplicity, we choose to work with channels with small decay probabilities $p_a$, such that the optimal partition is unchanged (i.e. the BMSE obtained by this partition is relatively close to the BMSE obtainable by the optimal initial state of the noiseless case). In the context of atomic clocks, the decay probability is given by $p_a = 1-\text{exp}(-2\gamma_{\text{ind}}\tau)$ \cite{nielsen2001quantum}, where $\gamma_{\text{ind}}$ is the individual, uncorrelated, decay rate out of the clock state, due to radiative decay or scattering from trapping light. $\tau$ is the total interrogation time (see \suppclocks). Then, working with small decay probabilities corresponds to operating with a short total interrogation time, such that $p_a \approx 2\gamma_{\text{ind}}\tau$, $\tau \ll \gamma_{\text{ind}}^{-1}$. This is the relevant regime where protocols using entanglement can provide considerably better atomic clock stability \cite{huelga_improvement_1997}.

We simulate the performance of the proposed scheme for the amplitude damping probabilities $p_a = 10^{-3}$ and $p_a = 10^{-2}$ with $N=21$ qubits in Fig. \ref{fig:21_atom_bmse_amp_damp}. We compare the RBMSE obtained with the proposed scheme to the RBMSE obtained with a 21-qubit CSS and the noiseless OQI as a function of the prior width, $\delta\phi$. 
We see that for small $p_a$, the proposed scheme continues surpassing the performance of the 21-qubit CSS, and achieves a metrological gain of $g_D = 1.85$ (2.67 dB) for $p_a = 10^{-3}$, and $g_D = 1.51$ (1.78 dB) for $p_a = 10^{-2}$, at $\delta\phi=0.7$ rad.

Regarding the adaptive protocol, amplitude damping renders the optimization over the single-qubit rotations used during the adaptive measurement scheme challenging since it flattens the hypersurface over which the gradient descent is being performed.
Furthermore, due to the sequential nature of the measurement, errors performed during the beginning of the measurement sequence can alter the resulting estimate of the phase significantly, resulting in errors increasing with the probability of decay, $p_a$. This can be seen from Fig. \ref{fig:21_atom_bmse_amp_damp}, where the performance of the scheme degrades when $p_a$ is increased from $10^{-3}$ to $10^{-2}$.
As mentioned in Supplementary Note 13,
this issue can be mitigated
by using error detection measurements, instead of local measurements, for each GHZ state.

\subsubsection*{GHZ state preparation and single-qubit measurement errors}

State preparation and measurement errors result in a parity signal with a limited contrast, $C<1$, where the outcomes of a parity measurement performed on an $N$-qubit GHZ state with contrast $C(N)$ are written as \mbox{$P(\pm) = \left[1 \pm C(N)\cos{(N\phi)} \right]/2$}, where $\{ +, -\}$ denote even, odd parity. Let us show this for measurement errors: 
assuming the noisy measurement is given by a symmetric bit-flip channel that proceeds a perfect projective measurement in the $\ket{+}, \ket{-}$ basis, the measurement operators of each qubit are given by $M_{\pm}=\left(1-p_e\right)|\pm\rangle\langle \pm|+p_e|\mp\rangle\langle \mp|$ with $\ket{\pm} = (\ket{0}\pm \ket{1})/\sqrt{2}$, and $p_e$ being the bit-flip probability.
The probabilities of even/odd parities are then: \mbox{$P(\pm) = P(\pm)\, p_\text{even} + P(\mp)\, p_\text{odd}$}, where $p_\text{even}$ is the probability of an even number of bit-flip errors occurring, given by
\begin{align}
    p_\text{even} = \sum_{i = 0}^{\lfloor \frac{N}{2}\rfloor} \binom{N}{i} p_e^{2i} (1-p_e)^{N-2i}
\end{align}
and $p_\text{odd} = 1 - p_\text{even}$. These errors therefore result in a contrast of $C(N) = p_\text{even} -p_\text{odd} = \left(1-2p_{e}\right)^{N}.$

A bit-flip measurement error is equivalent to a phase-flip error just before the readout pulse. Since phase-flip errors commute with the $\phi$-encoding unitary $U(\phi)$, this is equivalent to a phase flip error occurring during state preparation. Hence, such preparation errors are completely equivalent to bit-flip measurement errors, and would lead to the same reduced parity. 
For imperfect state preparation, we therefore assume a model where the contrast scales with the qubit number as $C(N) = F_0^N$, where $F_0$ denotes the effective fidelity per qubit.

The metrological gain $g_D$ in dB for various qubit numbers $N$, as a function of the effective fidelity per qubit $F_0$, or the bit-flip probability $p_e$ can be found in Figure \ref{fig:21_atom_bmse_amp_damp}. Note that a metrological gain above 0 dB signifies that the proposed scheme surpasses the noiseless CSS interrogation for the respective qubit number. The partitions used for the proposed scheme are plotted in the Figure in the form of $(x, y, z)$, where $x, y$, and $z$ denote the number of repetitions of the 4-qubit, 2-qubit, and 1-qubit GHZ state. Since we aim to understand how the gain scales with the fidelity or the error probability as a function of the particular partitions, we choose to examine the case of (i) $N = 9$, which does not contain any 4-qubit GHZ states, (ii) $N = 15$, which contains two 4-qubit GHZ states, (iii) $N = 18$ and $N = 21$, which both contain three 4-qubit GHZ states. We fit functions in the form of $g_D = A \cdot \text{exp}(-B(1-F_0)) = A \cdot \text{exp}(-2 B p_e)$ to the metrological gains of different $N$, plotted with dashed lines. The decay rate $B$ for $N = 9, 15, 18$ and $21$ is calculated to be 6.78, 13.9, 19.3, and 18.9, respectively. 
Interestingly, the gain appears to be mainly a function of the number of repetitions of the largest GHZ state, the larger this number is, the quicker the gain decays with decreasing fidelity, or increasing bit-flip error probability. It is approximately independent of the number of repetitions of the smaller blocks of GHZ states (the 2-qubit and 1-qubit GHZ states in this case).

More sophisticated measurement schemes might be employed to overcome the decay in the metrological gain due to measurement errors. It has been shown that in the presence of noisy measurements (assuming no other noise source), attaining HS depends on the available control operations. For example, HS is attainable in this regime when global control operations and post-processing methods are used, however it is elusive when local control operations are available only \cite{Len2022,Zhou_noisy_meas}. We defer studying the effect of using more sophisticated measurement schemes on the proposed scheme to future work.

\begin{figure}[t]
    \centering
\includegraphics[width=0.8\linewidth]{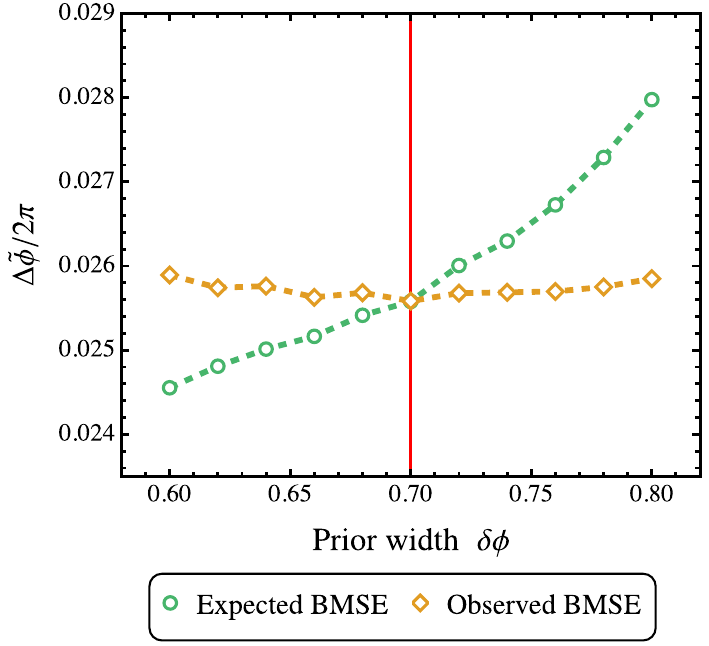}
    \caption{RBMSE $\Delta\tilde{\phi}$ obtained by the proposed scheme, normalized by $2\pi$, as a function of the prior width $\delta\phi$, for $N=25$ qubits. We plot in green the \textit{expected} RBMSE achieved by optimizing over the partition over GHZ blocks and the adaptive measurement. In yellow, we plot the \textit{observed} RBMSE when the true prior width ($\delta\phi^* = 0.7$ for this Figure) is unknown, therefore the scheme is run assuming various prior widths, and using the respective optimized measurements. The observed RBMSE is minimized at the true value of the prior width, where the expected and observed RBMSE intersect.}
    \label{fig:prior_width}
\end{figure}

\section*{Supplementary Note 15. Effect of Uncertain Prior Width}
\label{app:prior_width}

In the main text, we assume that the prior width $\delta\phi$ is known with complete certainty. This assumption relies on the fact that
the prior width can be estimated from measurements of the laser noise frequency spectrum \cite{bishof2013optical,Leroux_2017,shaw_multi-ensemble_2023}. However,
our knowledge of the prior can be partial due to, e.g. estimation errors.
We therefore provide here a case study for how the proposed scheme performs when the prior width is not perfectly known. In Figure \ref{fig:prior_width}, we plot the \textit{expected} and \textit{observed} RBMSE for the proposed scheme, for $N= 25$ qubits. The expected RBMSE is obtained by
applying our scheme (optimizing the partition and the adaptive measurement) for a given prior width.
For the observed RBMSE, there is a mismatch between the true and assumed prior widths: we apply our scheme assuming various prior widths while the true (unknown) prior width is $\delta\phi^* = 0.7.$ 
We observe that the two curves for the RBMSE intersect at the true value for the prior width. For smaller (larger) prior widths, the observed RBMSE is larger (smaller) than the expected RBMSE. Therefore, this case study demonstrates that if the prior width is not known certainly, we can estimate it by running the proposed scheme and comparing the expected and the observed RBMSEs. 

\bibliography{refs_2}

\end{document}